\documentclass[traditabstract]{aa} 
\usepackage{graphicx}
\usepackage{txfonts}

\begin{document}

\def\etal{{et al. \rm}}
\def\teff{$T_{\rm eff}$}
\def\logg{$\log g$}
\def\micro{$\xi$}
\def\kms{km s$^{-1}$}
\def\p{$\pm$}
\def\fe{[Fe/H]}
\def\loggf{$\log gf$}
\def\c{CoRoT}
\def\numax{$\nu_{\rm max}$}
\def\dnu{$\Delta \nu$}
\def\iso{\element[][12]{C}/\element[][13]{C}}

  \title{Atmospheric parameters and chemical properties of red giants in the \c \ asteroseismology fields
\thanks{Based on observations collected at La Silla Observatory, ESO (Chile) with the FEROS and HARPS spectrograph at the 2.2 and 3.6-m telescopes under programs LP178.D-0361, LP182.D-0356, and LP185.D-0056.}
\thanks{Tables \ref{tab_EWs} to \ref{tab_abundances_ratios} are only available in electronic form at the CDS via anonymous ftp to {\tt cdsarc.u-strasbg.fr (130.79.128.5)} or via {\tt http://cdsweb.u-strasbg.fr/cgi-bin/qcat?J/A+A/???/???}}}

  \titlerunning{Abundance study of CoRoT red-giant targets}
  \authorrunning{T. Morel et al.}

     \author{T. Morel
         \inst{1}
         \and
         A. Miglio
         \inst{2,3}
	 \and
	 N. Lagarde
         \inst{2,3}
         \and
         J. Montalb\'an
         \inst{1}
         \and
         M. Rainer
         \inst{4}
         \and
         E. Poretti
         \inst{4}
         \and
         P. Eggenberger
         \inst{5}
         \and
         S. Hekker
         \inst{6,7}
         \and
         T. Kallinger
         \inst{8,9}
         \and
         B. Mosser
         \inst{10}
         \and
         M. Valentini
         \inst{1,11}
         \and
         F. Carrier
         \inst{8}
         \and
         M. Hareter 
         \inst{9}
         \and
         L. Mantegazza
         \inst{4}
         }

  \offprints{Thierry Morel, \email{morel@astro.ulg.ac.be}.}
  \institute{
  Institut d'Astrophysique et de G\'eophysique, Universit\'e de Li\`ege, All\'ee du 6 Ao\^ut, B\^at. B5c, 4000 Li\`ege, Belgium
  \and
  School of Physics and Astronomy, University of Birmingham, Edgbaston, Birmingham, B15 2TT, UK
  \and
  Stellar Astrophysics Centre (SAC), Department of Physics and Astronomy, Aarhus University, Ny Munkegade 120, DK-8000, Aarhus C, Denmark 
  \and
  INAF -- Osservatorio Astronomico di Brera, via E. Bianchi 46, 23807 Merate (LC), Italy
  \and
  Geneva Observatory, University of Geneva, Chemin des Maillettes 51, 1290, Versoix, Switzerland
  \and
  Astronomical Institute 'Anton Pannekoek', University of Amsterdam, Science Park 904, 1098 XH Amsterdam, The Netherlands
  \and
  Max-Planck-Institut f\"ur Sonnensystemforschung, Justus-von-Liebig-Weg 3, 37077 G\"ottingen, Germany
  \and
  Katholieke Universiteit Leuven, Departement Natuurkunde en Sterrenkunde, Instituut voor Sterrenkunde, Celestijnenlaan 200D, B-3001 Leuven, Belgium
  \and
  Institute for Astrophysics, University of Vienna, T\"urkenschanzstrasse 17, 1180 Vienna, Austria
  \and
  LESIA, CNRS, Universit\'e Pierre et Marie Curie, Universit\'e Denis Diderot, Observatoire de Paris, 92195 Meudon Cedex, France
  \and
  Leibniz-Institut f\"ur Astrophysik Potsdam (AIP), An der Sternwarte 16, 14482, Potsdam, Germany
  }

  \date{Received 7 October 2013 / Accepted 19 February 2014}

  \abstract{A precise characterisation of the red giants in the seismology fields of the CoRoT satellite is a prerequisite for further in-depth seismic modelling. High-resolution FEROS and HARPS spectra were obtained as part of the ground-based follow-up campaigns for 19 targets holding great asteroseismic potential. These data are used to accurately estimate their fundamental parameters and the abundances of 16 chemical species in a self-consistent manner. Some powerful probes of mixing are investigated (the Li and CNO abundances, as well as the carbon isotopic ratio in a few cases). The information provided by the spectroscopic and seismic data is combined to provide more accurate physical parameters and abundances. The stars in our sample follow the general abundance trends as a function of the metallicity observed in stars of the Galactic disk. After an allowance is made for the chemical evolution of the interstellar medium, the observational signature of internal mixing phenomena is revealed through the detection at the stellar surface of the products of the CN cycle. A contamination by NeNa-cycled material in the most massive stars is also discussed. With the asteroseismic constraints, these data will pave the way for a detailed theoretical investigation of the physical processes responsible for the transport of chemical elements in evolved, low- and intermediate-mass stars.} 


  \maketitle
%

\section{Introduction} \label{sect_introduction}
Early observations by the MOST (e.g., Kallinger \etal \cite{kallinger08}) and WIRE (e.g., Stello \etal \cite{stello08}) satellites have demonstrated the tremendous potential of extremely precise and quasi-uninterrupted photometric observations from space for studies of red-giant stars. Breakthrough results are currently being made from observations collected by \c \ (Michel \etal \cite{michel08}) and {\it Kepler} (Borucki \etal \cite{borucki10}), which offer the opportunity for the first time to derive some fundamental properties of a vast number of red giants from the modelling of their solar-like oscillations (see the reviews by Christensen-Dalsgaard \cite{christensen_dalsgaard11}, Chaplin \& Miglio \cite{chaplin13}, and Hekker \cite{hekker13}). Amongst the most exciting results achievable by asteroseismology of red-giant stars are the possibility of inferring their evolutionary status (e.g., Montalb\'an \etal \cite{montalban10}; Bedding \etal \cite{bedding11}; Mosser \etal \cite{mosser11}), constraining their rotation profile (e.g., Beck \etal \cite{beck12}; Deheuvels \etal \cite{deheuvels12}), or determining the detailed properties of the core in He-burning stars (Mosser \etal \cite{mosser12}, Montalb\'an \etal \cite{montalban13}). In addition, their global seismic properties can provide a high level of accuracy of their fundamental properties, such as masses, radii, and distances, which may then be used to map and date stellar populations in our Galaxy (e.g., Miglio \etal \cite{miglio09}, \cite{miglio13}).

Carrying out an abundance analysis of red-giant pulsators is relevant for two closely related reasons. The most obvious one is that only accurate values of the effective temperature and chemical composition are independently derived from ground-based observations that permit a robust modelling of the seismic data (e.g., Gai \etal \cite{gai11}; Creevey \etal \cite{creevey12}). Conversely, asteroseismology can provide the fundamental quantities (e.g., mass, evolutionary status) that are needed to best interpret the abundance data. This would allow us, for instance, to better understand the physical processes controlling the amount of internal mixing in red giants. One issue of particular interest and is currently actively debated -- and this is one of the objectives of this project -- is to investigate the nature of the transport phenomena that are known to occur for low-mass stars after the first dredge-up but before the onset of the He-core flash (e.g., Gilroy \& Brown \cite{gilroy91}). 

Thanks to their brightness, a comprehensive study of the chemical properties of the red giants lying in the \c \ seismology fields is relatively easy to achieve. This can be compared with the case of the fainter stars observed in the exofields of \c \ (Gazzano \etal \cite{gazzano10}; Valentini \etal \cite{valentini13}) or in the {\it Kepler} field (Bruntt \etal \cite{bruntt11}; Thygesen \etal \cite{thygesen12}), for which the abundances of the key indicators of mixing (C, N, Li, and \iso) have not been systematically investigated to our knowledge. The most noticeable attempts in the case of the {\it Kepler} red giants are the low-precision (uncertainties of the order of 0.5 dex) carbon abundances derived for a dozen stars by Thygesen \etal (\cite{thygesen12}) and the study of lithium in the open cluster \object{NGC 6819} by Anthony-Twarog \etal (\cite{anthony_twarog13}).

This paper is organised as follows: Sect.~\ref{sect_targets} presents the targets observed, while Sect.~\ref{sect_observations} discusses the observations and data reduction. The determination of the seismic gravities is described in Sect.~\ref{sect_seismic_constraints}. The methodology implemented to derive the chemical abundances and stellar parameters is detailed in Sects.~\ref{sect_chemical_abundances} and \ref{sect_parameters}, respectively. The uncertainties and reliability of our results are examined in Sects.~\ref{sect_errors} and \ref{sect_validation}, respectively. We present the procedure used to correct the abundances of the mixing indicators from the effect of the chemical evolution of the interstellar medium (ISM) in Sect.~\ref{sect_correction_chemical_evolution}. Section \ref{sect_key_results} is devoted to a qualitative discussion of some key results. Finally, some future prospects are mentioned in Sect.~\ref{sect_conclusion}.

\section{The targets} \label{sect_targets}
Our sample is made up of 19 red-giant targets for the asteroseismology programme of the satellite. This includes four stars, which were initially considered as potential targets, but were eventually not observed (\object{HD 40726}, \object{HD 42911}, \object{HD 43023}, and \object{HD 175294}).\footnote{The \c \ satellite ceased operations on November 2nd, 2012.} The stars lie in either the \c \ eye pointing roughly towards the Galactic centre (around $\alpha$ = 18 h 50 min and $\delta$ = 0$\degr$) or the anticentre (around $\alpha$ = 6 h 50 min and $\delta$ = 0$\degr$) and were observed during an initial (IR), a short (SR), or a long run (LR) with a typical duration ranging from about 30 to 160 days. The white-light photometric measurements are quasi-uninterrupted (time gaps only occur under normal circumstances during the passage across the South Atlantic Anomaly, resulting in a duty cycle of about 90\%).

Three stars based on their radial velocities (\object{HD 170053}, \object{HD 170174}, and \object{HD 170231}) are likely members of the young open cluster \object{NGC 6633}. Although membership was also suspected for \object{HD 170031}, the radial velocity derived from our ground-based observations is discrepant with the values obtained for the three stars above (Barban \etal \cite{barban14}; Poretti et al., in preparation) and argues against this possibility unless this star is a runaway (an explanation in terms of binarity can be ruled out). For \object{HD 45398}, a possible member of \object{NGC 2232}, both our radial velocity and iron abundance are at odds with the values obtained for bona fide members of this metal-rich cluster (Monroe \& Pilachowski \cite{monroe_pilachowski10}). 

Thanks to seismic constraints, surface gravities are available for all but three of the stars observed by \c. This is discussed in more detail in Sect.~\ref{sect_seismic_constraints}. On the other hand, a detailed modelling of the \c \ data is described for HD 50890 by Baudin \etal (\cite{baudin12}) and for HD 181907 by Carrier \etal (\cite{carrier10}) and Miglio \etal (\cite{miglio10}).

Five bright, well-studied red giants (\object{$\alpha$ Boo}, \object{$\eta$ Ser}, \object{$\epsilon$ Oph}, \object{$\xi$ Hya}, and \object{$\beta$ Aql}) with less model-dependent estimates of the effective temperature and surface gravity compared to what can be provided by spectroscopy (from interferometric and seismic data, respectively) were also observed and analysed in exactly the same way as the main targets to validate the procedures implemented. 

\begin{figure*}
\centering
\includegraphics[width=18.5cm]{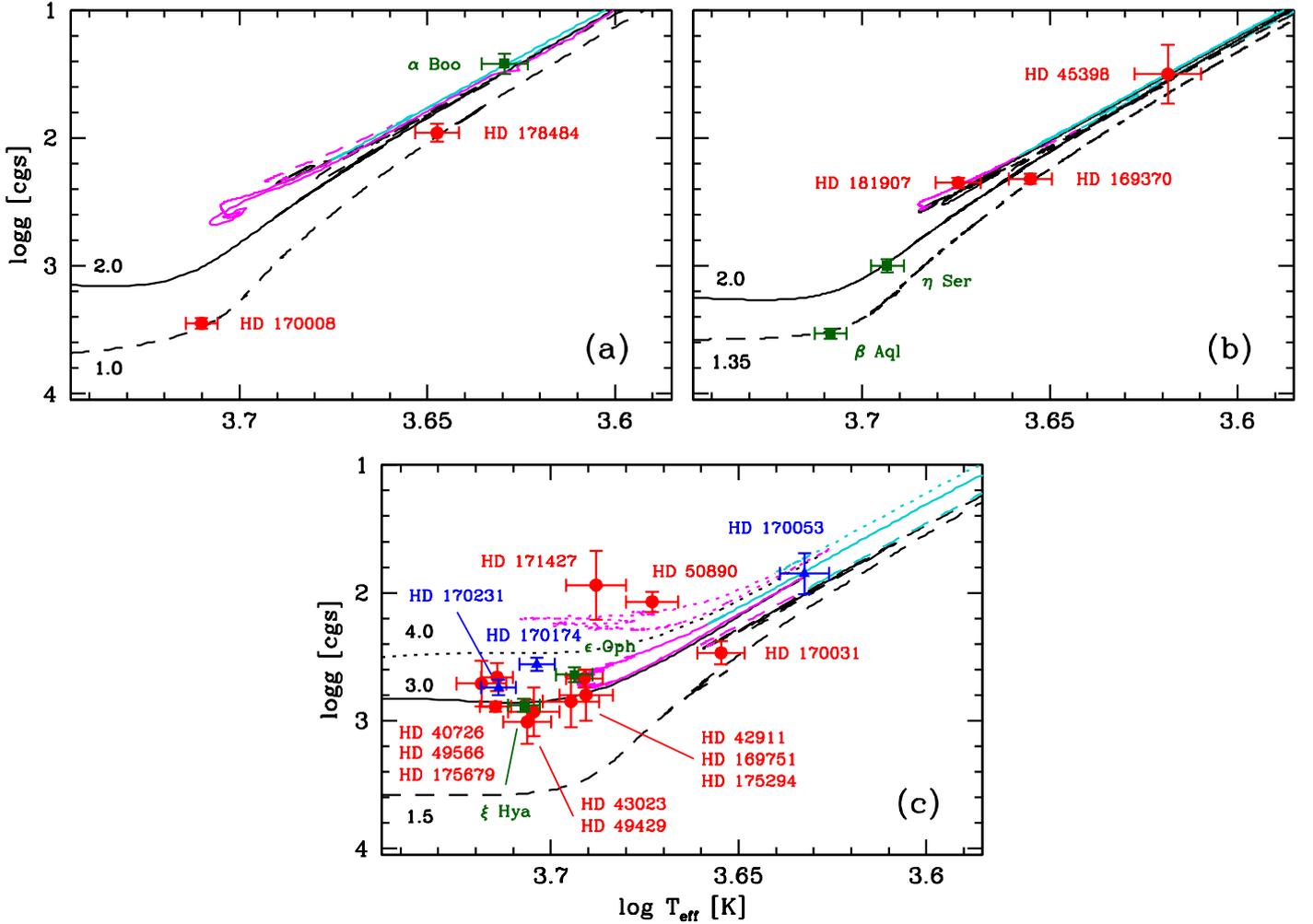}
\caption{Position of the targets in the $\log$\teff-\logg \ plane (red circles: \c \ targets, blue triangles: stars in \object{NGC 6633}, green squares: benchmark stars). The predictions of evolutionary models, which include rotation-induced mixing ($V/V_{\rm crit}$ = 0.45 on the zero-age main sequence; ZAMS) and thermohaline mixing, are overplotted (Lagarde \etal \cite{lagarde12}). The initial mass of the models in solar units is indicated to the left of each panel (The tracks are shown with a different linestyle depending on the mass.). The colour of the track indicates the evolutionary phase: red-giant branch (RGB; black), core-He burning (magenta), and asymptotic-giant branch (AGB; cyan). The data are separated into three metallicity domains: theoretical tracks for \fe \ = --0.56 and data for stars with \fe \ $<$ --0.26 ({\it panel a}); tracks for \fe \ = --0.25 and data for stars with --0.26 $\leq$ \fe \ $<$ --0.12 ({\it panel b}); and tracks at solar metallicity ($Z$ = 0.014) and data for stars with \fe \ $\geq$ --0.12 ({\it panel c}). When available, the stellar parameters are those obtained with the gravity constrained to the seismic value (Sect.~\ref{sect_parameters}).}
\label{fig_logg_logTeff_starevol}
\end{figure*}

Relatively inaccurate parallaxes, poorly-known interstellar extinction (many stars lying very close to the plane of the Galaxy)\footnote{Because of the lack of reliable reddening estimates in many cases, no attempts were made to derive the temperatures from colour indices. On the contrary, our spectroscopic results are completely free from these problems.} and unavailability of 2MASS data for the brightest targets conspire to often make the luminosities of the \c \ targets uncertain. Instead of placing the programme stars in a traditional Hertzsprung-Russell (HR) diagram, we therefore show their position in the $\log$\teff-\logg \ plane in Fig.~\ref{fig_logg_logTeff_starevol}. Solely based on evolutionary tracks, our sample appears to be made up of stars in various evolutionary stages and that are on average significantly more massive than the Sun. A more complete description of our sample based on asteroseismic diagnostics will be presented in a forthcoming publication (Lagarde et al., in preparation). 

\section{Observations and data reduction} \label{sect_observations}
The scientific rationale of the ESO large programmes devoted to the ground-based observations of the \c \ targets (LP178.D-0361, LP182.D-0356, and LP185.D-0056; Poretti \etal \cite{poretti13}) has continuously been adapted to the new results obtained by the satellite. The discovery of the solar-like oscillations in red giants (e.g., De Ridder \etal \cite{de_ridder09}) and their full exploitation as asteroseismic tools (e.g., Miglio \etal \cite{miglio09}, \cite{miglio13}) was one of the most relevant. Therefore, very high signal-to-noise ratio (S/N $\ga$ 200) spectra were obtained during the period 2007-2012 to perform an accurate determination of both the stellar parameters (\teff \ and \logg) and the chemical abundances. As a final improvement of the ground-based complement to \c \ data, dense spectroscopic timeseries were obtained on selected targets (among which HD 45398 and stars in NGC 6633). They are aimed at comparing amplitudes in photometric flux and radial velocity. Double-site, coordinated campaigns involving HARPS and SOPHIE (mounted at the 1.93-m telescope at the Observatoire de Haute Provence, France; OHP) were organised for this purpose (Poretti et al., in preparation).

Most of the observations were obtained with the HARPS spectrograph attached to the 3.6-m telescope at La Silla Observatory (Chile) in either the high-efficiency (EGGS) or the high-resolution (HAM) mode. The spectral range covered is 3780--6910 \AA \ with a resolving power $R$ $\sim$ 80\,000 and 115\,000, respectively. Spectra of three stars were acquired in 2007 at the ESO 2.2-m telescope with the fiber-fed, cross-dispersed \'echelle spectrograph FEROS in the object+sky configuration. The spectral range covered is 3600--9200 \AA \ and $R$ $\sim$ 48\,000. This wider spectral coverage compared to HARPS allowed a determination of the nitrogen abundance from a few \element[][12]{CN} lines in the range 8002-8004 \AA \ and the \iso \ isotopic ratio through the analysis of the $^{13}$CN feature at 8004.7 \AA. 

Four of the bright, benchmark stars were intensively monitored with the ELODIE (\object{$\epsilon$ Oph}) or the HARPS (\object{$\eta$ Ser}, \object{$\xi$ Hya}, and \object{$\beta$ Aql}) spectrographs to study their pulsational behaviour (Further details can be found in De Ridder \etal \cite{de_ridder06} and Kjeldsen \etal \cite{kjeldsen08} for \object{$\epsilon$ Oph} and \object{$\beta$ Aql}, respectively.). The time series were extracted from the instrument archives. A very high-quality spectral atlas was employed in the case of \object{$\alpha$ Boo} (Hinkle \etal \cite{hinkle00}). Further details about the observations are provided in Table \ref{tab_observations}.

\begin{table*}
\caption{Observations and basic parameters of the targets.}
\label{tab_observations}
\scriptsize
\centering
\begin{tabular}{rcccccclrccc} 
\hline\hline
\multicolumn{1}{c}{HD}     & HR                     & HIP                    & Other                  & Spectral  & Magnitude in    & Parallax            & Instrument    & \multicolumn{1}{c}{Resolving} & Instrumental & Number of & S/N at   \\
\multicolumn{1}{c}{number} & number                  & number                  & Name                   & type      & the $V$ band &  [mas]              &               & \multicolumn{1}{c}{power}     & broadening [\AA]   & spectra   & 5815 \AA \\
\hline
40726  & \multicolumn{1}{c}{...} & 28485                   & ...                    & G5 III    &   7.00 &  3.48\p0.56                  & HARPS EGGS     &  80\,000  & 0.089 &   1 & 252\\
42911  & \multicolumn{1}{c}{...} & 29526                   & ...                    & G7 III    &   7.38 &  7.36\p0.58                  & HARPS EGGS     &  80\,000  & 0.089 &   1 & 277\\
43023  & 2218                    & 29575                   & ...                    & G8 III    &   5.83 & 10.61\p0.38                  & HARPS EGGS     &  80\,000  & 0.089 &   1 & 342\\
45398  & \multicolumn{1}{c}{...} & 30691                   & ...                    & K0        &   6.90 &  4.58\p0.59                  & HARPS HAM      & 115\,000  & 0.062 &   1 & 249\\ 
49429  & \multicolumn{1}{c}{...} & 32659                   & ...                    & K0        &   6.91 &  6.19\p0.87                  & HARPS EGGS     &  80\,000  & 0.089 &   1 & 238\\
49566  & \multicolumn{1}{c}{...} & 32705                   & ...                    & G5        &   7.71 &  3.54\p0.67                  & HARPS EGGS     &  80\,000  & 0.089 &   1 & 301\\
50890  & 2582                    & 33243                   & ...                    & G6 III    &   6.03 &  2.99\p0.44                  & HARPS EGGS     &  80\,000  & 0.089 &   5 & 396\\
169370 & 6892                    & 90238                   & ...                    & K0        &   6.30 & 10.52\p0.57                  & HARPS EGGS     &  80\,000  & 0.089 &   1 & 225\\
169751 & \multicolumn{1}{c}{...} & 90379                   & ...                    & K2        &   8.37 &  7.26\p0.75                  & HARPS EGGS     &  80\,000  & 0.089 &   1 & 200\\
170008 & \multicolumn{1}{c}{...} & 90427                   & ...                    & G5        &   7.42 & 12.45\p0.71                  & HARPS EGGS     &  80\,000  & 0.089 &   1 & 240\\
170031 & \multicolumn{1}{c}{...} & \multicolumn{1}{c}{...} & ...                    & K5        &   8.20 &  ...                         & HARPS HAM      & 115\,000  & 0.062 & 108 & 141\\
171427 & \multicolumn{1}{c}{...} & 91063                   & ...                    & K2        &   7.22 &  2.06\p0.53                  & HARPS EGGS     &  80\,000  & 0.089 &   1 & 266\\
175294 & \multicolumn{1}{c}{...} & 92807                   & ...                    & K0        &   7.40 &  2.94\p0.66                  & FEROS          &  48\,000  & 0.151 &   1 & 330\\
175679 & 7144                    & 92968                   & ...                    & G8 III    &   6.14 &  6.39\p0.43                  & FEROS          &  48\,000  & 0.151 &   1 & 439\\
178484 & \multicolumn{1}{c}{...} & 94053                   & ...                    & K0        &   6.61 &  5.00\p0.46                  & HARPS HAM      & 115\,000  & 0.062 &   1 & 257\\ 
181907 & 7349                    & 95222                   & ...                    & G8 III    &   5.82 &  9.64\p0.34                  & FEROS          &  48\,000  & 0.151 &   1 & 490\\
170053 & \multicolumn{1}{c}{...} & \multicolumn{1}{c}{...} & ...                    & K2 II     &   7.30 &  2.67\p0.32\tablefootmark{a} & HARPS HAM      & 115\,000  & 0.062 &  99 & 154\\
170174 & \multicolumn{1}{c}{...} & \multicolumn{1}{c}{...} & ...                    & K2        &   8.31 &  2.67\p0.32\tablefootmark{a} & HARPS HAM      & 115\,000  & 0.062 &   1 & 158\\
170231 & \multicolumn{1}{c}{...} & \multicolumn{1}{c}{...} & ...                    & K2        &   8.69 &  2.67\p0.32\tablefootmark{a} & HARPS HAM      & 115\,000  & 0.062 &  93 & 137\\ 
124897 & 5340                    & 69673                   & $\alpha$ Boo, Arcturus & K1.5 III  & --0.04 & 88.83\p0.54                  & KPNO \'echelle & 150\,000  & 0.045 &   1 & $\sim$1000\tablefootmark{b}\\
168723 & 6869                    & 89962                   & $\eta$ Ser             & K0 III-IV &   3.26 & 53.93\p0.18                  & HARPS HAM      & 115\,000  & 0.062 & 129 & 246\\
146791 & 6075                    & 79882                   & $\epsilon$ Oph         & G9.5 III  &   3.24 & 30.64\p0.20                  & ELODIE         &  48\,000  & 0.151 & 181 & 184\tablefootmark{c}\\
100407 & 4450                    & 56343                   & $\xi$ Hya              & G7 III    &   3.54 & 25.16\p0.16                  & HARPS HAM      & 115\,000  & 0.062 &  59 & 284\\
188512 & 7602                    & 98036                   & $\beta$ Aql, Alshain   & G9.5 IV   &   3.71 & 73.00\p0.20                  & HARPS HAM      & 115\,000  & 0.062 & 135 & 285\\
\hline
\end{tabular}
\tablefoot{Spectral types and magnitudes in the $V$ band from SIMBAD database. {\it Hipparcos} parallaxes from van Leeuwen (\cite{van_leeuwen07}). The instrumental broadening is the full width at half-maximum of lines measured in calibration lamps at $\sim$6700 \AA. The quoted S/N is the typical value for one exposure. \tablefoottext{a}{Parallax of the NGC 6633 cluster (van Leeuwen \cite{van_leeuwen09}).} \tablefoottext{b}{Mean S/N over the wavelength range 3600--9300 \AA, as quoted by Hinkle \etal (\cite{hinkle00}).} \tablefoottext{c}{S/N at 5500 \AA.}}
\end{table*}

The data reduction (i.e., bias subtraction, flat-field correction, removal of scattered light, order extraction, merging of the orders, and wavelength calibration) was carried out for the \c \ red giants using dedicated tools developed at Brera observatory (Poretti \etal \cite{poretti13}). For the spectra of the benchmark stars extracted from the archives, the final data products provided by the reduction pipelines were used. As a final step, the spectra were put in the laboratory rest frame and the continuum was normalised by fitting low-order cubic spline or Legendre polynomials to the line-free regions using standard tasks implemented in the IRAF\footnote{{\tt IRAF} is distributed by the National Optical Astronomy Observatories, which is operated by the Association of Universities for Research in Astronomy, Inc., under cooperative agreement with the National Science Foundation.} software. In case of multiple observations, the individual exposures were co-added (weighted by the S/N ratio and ignoring the poor-quality spectra) to create an averaged spectrum, which was subsequently used for the abundance analysis. The only exception was HD 45398, for which we based our analysis on the exposure where \ion{[O}{i]} $\lambda$6300 was the least (in that case negligibly) affected by telluric features.

\section{Seismic constraints on the surface gravity}\label{sect_seismic_constraints}

Radii, masses, and surface gravities of solar-like oscillating stars can be estimated from the average seismic parameters that globally characterise their oscillation spectra: the average large frequency separation (\dnu) and the frequency corresponding to the maximum oscillation power (\numax).

Three methods described in Mosser \& Appourchaux (\cite{mosser_appourchaux09}), Hekker \etal (\cite{hekker10}), and Kallinger \etal (\cite{kallinger10}) were applied to the \c \ light curves to detect oscillations and measure the global oscillations parameters \dnu \ and \numax. We only consider stars for which 2 out of the 3 methods gave a positive detection of both quantities. This was not the case for \object{HD 45398}, \object{HD 49429}, and \object{HD 171427}. The seismic gravities of these three stars are therefore not discussed further. The final values for \dnu \ and \numax \ were adopted from the pipeline developed by Mosser \etal (\cite{mosser10}). We determined the uncertainties on \dnu \ and \numax \ by adding the formal uncertainty given by this pipeline and the scatter of the values obtained by the two other methods in quadrature. For \object{HD 170053}, the values of \dnu \ based on the methods of Mosser \& Appourchaux (\cite{mosser_appourchaux09}) and Hekker \etal (\cite{hekker10}) were only considered due to the larger (by a factor $\sim$5) uncertainty of the determination provided by the pipeline of Kallinger \etal (\cite{kallinger10}). For the benchmark stars for which oscillations were detected using sparse/ground-based data, we adopted an uncertainty of 2.5\% in \dnu \ and of 5\% in \numax, as suggested in Bruntt \etal (\cite{bruntt10}) and also adopted in Morel \& Miglio (\cite{morel_miglio12}). 

The frequency of maximum oscillation power is expected to scale as the acoustic cut-off frequency, and a straightforward relation has been proposed that links \numax \ to the surface gravity (e.g., Brown \etal \cite{brown91}):

\begin{equation}
\log g \simeq \log g_\odot + \log \left(
{{\nu_{\rm max} \over \nu_{{\rm max}, \, \odot}}}
\right)
+ {1 \over 2}
\log \left({T_{\rm eff}\over {\rm T}_{{\rm eff}, \, \odot}}\right) {\rm .}
\label{eq_numax}
\end{equation}
Theoretical support to this scaling law is provided by Belkacem \etal (\cite{belkacem11}). It is important to stress that this relation is largely insensitive to the effective temperature assumed ($\Delta T_{\rm eff}$ = 100 K leads to $\Delta \log g$ $\sim$ 0.005 dex only for a typical red-clump giant.). 
The average large frequency spacing, on the other hand, scales approximatively as the square root of the mean density of the star (e.g., Tassoul \cite{tassoul80}):
\begin{equation}
\Delta\nu \simeq \sqrt{\frac{M/M_\odot}{(R/R_\odot)^3}} \, \Delta\nu_{\odot}{\rm .}
\label{eq_dnu}
\end{equation}

We have considered several procedures to estimate \logg:
\begin{itemize}
\item \logg0: using Eq.\ref{eq_numax} directly and the spectroscopically determined \teff.
\item \logg1: using PARAM (da Silva \etal \cite{da_silva06}; Miglio \etal \cite{miglio13}), a Bayesian stellar parameter estimation method based on the Girardi \etal (\cite{girardi2000}) stellar evolutionary tracks, and considering \teff, \fe, \dnu, and \numax \ as observables . 
\item \logg2: using PARAM but taking \dnu \ as the only seismic constraint (see Ozel \etal \cite{ozel13}). 
\item \logg3: as \logg1 but adopting larger uncertainties in \dnu \ and \numax \ (as suggested in Huber \etal \cite{huber13}; see below).
\item \logg4: as \logg2 but artificially increasing/decreasing the observed \dnu \ to account for possible biases in Eq. \ref{eq_dnu} (see below).
\end{itemize}
When estimating \logg, we adopted the following in Eqs. \ref{eq_numax} and \ref{eq_dnu}: $\nu_{\rm max, \, \odot}$ = 3090 $\mu$Hz, $\Delta\nu_\odot=135.1$ $\mu$Hz (Huber \etal \cite{huber13}), and T$_{\rm eff, \, \odot}$ = 5777 K. 

\begin{figure*}
\centering
\includegraphics[width=14cm]{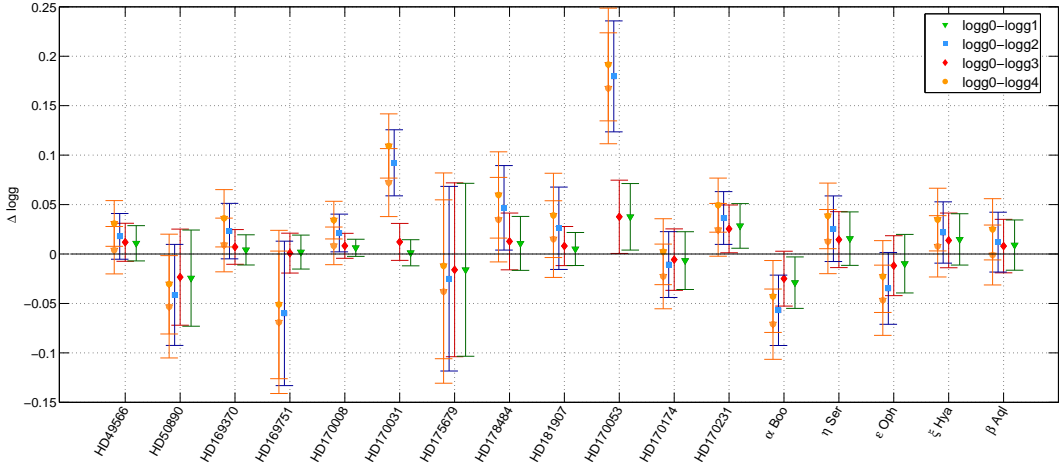}
\caption{Comparison of the \logg \ values determined under different assumptions. The two \logg4 values correspond to an increase/decrease of the observed \dnu \ by 2.5\% (see text).}
\label{fig_logg}
\end{figure*}

Given that both Eqs.\ref{eq_numax} and \ref{eq_dnu} are approximate expressions, we have considered the effect of possible biases in such relations. First, we increased the uncertainties by 2.5\% (see also Huber \etal \cite{huber13}) on both the observed \dnu \ and \numax \ (\logg3). Second, comparisons with stellar models suggest that Eq.\ref{eq_dnu} for stars similar to those in this study is accurate to $\sim$3\% (see White \etal \cite{white11}; Miglio \etal \cite{miglio12}, \cite{miglio13}, and the analysis presented in Mosser \etal \cite{mosser13}). To check the effect of such a systematic uncertainty on \logg, we have increased/decreased the observed \dnu \ by 2.5\%, while we consider it as the only seismic constraint. This led to a couple of gravity values (\logg4a,b). The comparison between the different estimates is presented in Fig.~\ref{fig_logg}. As can be seen, there is a good level of consistency between the values obtained with the exception of a few cases. The determination of the seismic gravity is therefore robust against the choice of the method used. 

However, the difference between \logg0 and \logg4 is particularly prominent for HD 170053, a likely member of the cluster NGC 6633. Using PARAM and all available constraints, we find a stellar mass $M$ $\sim$ 2.8 M$_{\odot}$, a radius $R$ $\sim$ 34 R$_{\odot}$, \logg \ $\sim$ 1.81, and $\log \,$($\rho$/$\rho_{\odot}$) $\sim$ --4.17. These estimates are compatible with a giant belonging to the cluster, although in a rather fast evolutionary phase (RGB or early AGB). If \numax \ is no longer considered as a constraint, then the only strong seismic constraint is on the stellar mean density, and PARAM then finds a low-mass star on the RGB with the values of $M$ $\sim$ 1.2 M$_{\odot}$, $R$ $\sim$ 26 R$_{\odot}$, and \logg \ $\sim$ 1.67, which still respects $\log \,$($\rho$/$\rho_{\odot}$) $\sim$ --4.17 and the spectroscopic constraints on \teff \ and \fe, as a more likely solution. The turn-off mass of NGC 6633 lies in the range 2.4-2.7 M$_{\odot}$ (Smiljanic \etal \cite{smiljanic09}), which excludes the latter possibility if this star is indeed a cluster member.

The final value of \logg \ we adopted resulted from using Eq.\ref{eq_numax} alone (\logg0). Once it was determined, the spectroscopic parameters were re-estimated (Sect.~\ref{sect_parameters}), and the procedure was repeated until convergence. The uncertainty was determined as the quadratic sum of the formal uncertainty in \logg0 and the scatter that is between \logg0 and the most discrepant value of the couple \logg4a,b (determined using \dnu \ only). This leads to the values listed in Table \ref{tab_seismic_gravities}.

\begin{table}
\caption{Seismic gravities.}
\label{tab_seismic_gravities}
\centering
\begin{tabular}{llc} 
\hline\hline
Name                     & \multicolumn{1}{l}{Source of data} & \logg \ seismology [cgs]\\ 
\hline
\object{HD 45398}        & \c \ SRa04               & Not available\\   
\object{HD 49429}        & \c \ SRa01               & Not available\\  
\object{HD 49566}        & \c \ SRa01               & 2.89\p0.04\\   
\object{HD 50890}        & \c \ IRa01               & 2.07\p0.08\\ 
\object{HD 169370}       & \c \ LRc03               & 2.32\p0.04\\ 
\object{HD 169751}       & \c \ LRc03               & 2.67\p0.07\\  
\object{HD 170008}       & \c \ LRc03               & 3.45\p0.04\\ 
\object{HD 170031}       & \c \ LRc07, LRc08        & 2.47\p0.09\\   
\object{HD 171427}       & \c \ LRc02               & Not available\\    
\object{HD 175679}       & \c \ SRc01               & 2.66\p0.11\\ 
\object{HD 178484}       & \c \ LRc09               & 1.96\p0.07\\
\object{HD 181907}       & \c \ LRc01               & 2.35\p0.04\\ 
\object{HD 170053}       & \c \ LRc07, LRc08        & 1.85\p0.16\\     
\object{HD 170174}       & \c \ LRc07, LRc08        & 2.56\p0.05\\    
\object{HD 170231}       & \c \ LRc07, LRc08        & 2.74\p0.06\\  
\object{$\alpha$ Boo}    & {\it Coriolis} satellite & 1.42\p0.08\\
\object{$\eta$ Ser}      & optical spectra          & 3.00\p0.05\\         
\object{$\epsilon$ Oph}  & MOST satellite           & 2.64\p0.06\\    
\object{$\xi$ Hya}       & optical  spectra         & 2.88\p0.05\\       
\object{$\beta$ Aql}     & optical  spectra         & 3.53\p0.04\\
\hline
\end{tabular}
\tablefoot{For the nomenclature of the \c \ runs, SRa04 is, for instance, the fourth short run in the anticentre direction. Further details about the seismic data for the stars used for validation can be found in Tarrant \etal (\cite{tarrant07}; \object{$\alpha$ Boo}), Barban \etal (\cite{barban04}; \object{$\eta$ Ser}), Kallinger \etal (\cite{kallinger08}; \object{$\epsilon$ Oph}), Frandsen \etal (\cite{frandsen02}; \object{$\xi$ Hya}), and Kjeldsen \etal (\cite{kjeldsen08}; \object{$\beta$ Aql}). For \object{HD 45398}, \object{HD 49429}, and \object{HD 171427}, the nature of the power spectrum hampered a robust determination of \dnu \ and/or \numax. The seismic gravities are therefore not quoted in those cases.}
\end{table}

\section{Determination of chemical abundances} \label{sect_chemical_abundances}
The atmospheric parameters (\teff, \logg, and microturbulence \micro) and abundances of 12 metals (Fe, Na, Mg, Al, Si, Ca, Sc, Ti, Cr, Co, Ni, and Ba) were self-consistently determined from the spectra using a classical curve-of-growth analysis. On the other hand, the abundances of Li, C, N, and O (as well as the \iso \ isotopic ratio for four stars) were determined from spectral synthesis. In each case, Kurucz plane-parallel atmospheric models computed with the ATLAS9 code ported under Linux (Sbordone \cite{sbordone05}) and the 2010 version of the line analysis software MOOG originally developed by Sneden (\cite{sneden73}) were used. Tests carried out using plane-parallel and spherical MARCS model atmospheres are briefly described in Sect.~\ref{sect_errors_parameters}. These calculations assume local thermodynamic equilibrium (LTE) and a solar helium abundance. A different atmospheric He content may be encountered within our sample, which is made up of stars in various evolutionary stages, but this is not expected to appreciably affect our results (see Pasquini \etal \cite{pasquini11}). Molecular equilibrium was achieved taking into account the 22 most common molecules. As for the solar analysis (see below), all models were computed with a length of the convective cell over the pressure scale height, $\alpha$ = $l/H_{\rm p}$ = 1.25, and without overshooting. Updated opacity distribution functions (ODFs) were employed. They incorporate the solar abundances of Grevesse \& Sauval (\cite{grevesse_sauval98}) and a more comprehensive treatment of molecules compared to the ODFs of Kurucz (\cite{kurucz90}). Further details are provided by Castelli \& Kurucz (\cite{castelli_kurucz04}).

\subsection{Curve-of-growth analysis} \label{sect_curve_of_growth}
The line list for the 12 metal species ($Z$ $>$ 8), whose abundances were directly determined from the equivalent width (EW) measurements, is made up of features selected to be unblended in a high-resolution atlas of the K1.5 III star \object{$\alpha$ Boo} (Hinkle \etal \cite{hinkle00}). Further details can be found in Morel \etal (\cite{morel03}, \cite{morel04}). As discussed in these papers, the selected transitions of the odd-$Z$ elements (Sc, Co, and Ba) are not significantly broadened by hyperfine structure. This list was completed by 13 \ion{Fe}{i} and 4 \ion{Fe}{ii} lines taken from Hekker \& Mel\'endez (\cite{hekker_melendez07}). These lines were carefully chosen to avoid blends with atomic or CN molecular features. Two lines were further added to our list: \ion{Na}{i} $\lambda$6160.75 and \ion{Al}{i} $\lambda$6696.02. 

The EWs were manually measured assuming Gaussian profiles, and only lines with a satisfactory fit were retained. Voigt profiles were used for the few lines with extended damping wings. Atomic lines significantly affected by telluric features were discarded from the analysis (the telluric atlas of Hinkle \etal \cite{hinkle00} was used). The EW measurements are provided in Table~\ref{tab_EWs}.

All the oscillator strengths were calibrated from an inverted solar analysis using a high S/N moonlight FEROS spectrum obtained during our first observing run. The oscillator strengths were tuned until the solar abundances of Grevesse \& Sauval (\cite{grevesse_sauval98}) were reproduced. They agree well with the laboratory measurements used by Hekker \& Mel\'endez (2007) with no evidence of systematic discrepancies ($\Delta \log gf$ = +0.002\p0.100 dex). Although such a differential analysis with respect to the Sun will not remove the systematic errors inherent to the modelling (e.g., inadequacies in the atmospheric structure) in view of the different fundamental parameters of our targets, such an approach is expected to minimise other sources of systematic errors, such as those related to the data reduction (e.g., continuum placement) or EW measurements. The solar oscillator strengths were derived using a plane-parallel LTE Kurucz solar model\footnote{This solar model is available online at: \newline {\tt http://wwwuser.oat.ts.astro.it/castelli/}.}  with \teff \ = 5777 K, \logg \ = 4.4377, and a depth-independent microturbulent velocity, \micro \ = 1.0 \kms.

Where possible (note that it is the case for the strongest \ion{Fe}{i} lines), the damping parameters for the van der Waals interaction were taken from Barklem \etal (\cite{barklem00}) and Barklem \& Aspelund-Johansson (\cite{barklem05}). The long interaction constant, $C_6$, was computed from the line broadening cross sections expressed in atomic units, $\sigma$, using:
\begin{equation}
C_6 = 6.46 \times 10^{-34} \, (\sigma/63.65)^{5/2}.
\end{equation}
A standard dependence of the cross sections on the temperature, as implemented in MOOG, was adopted (Uns\"old \cite{unsold55}). As discussed by Barklem \etal (\cite{barklem00}), the exact choice of the velocity parameter value, $\alpha$, does not usually lead to significant differences in the line profile. For lines without detailed calculations (about 35\%), we applied an enhancement factor, $E_\gamma$, to the classical Uns\"old line width parameter. No attempt was made to estimate this quantity on a line-to-line basis, and we assumed a typical value for a given ion. This was based on a comparison between our computed line widths and those assuming the classical van der Waals theory (for Fe and Ni) or a compilation of empirical results in the literature (Chen \etal \cite{chen00}; Feltzing \& Gonzalez \cite{feltzing_gonzalez01}; Bensby \etal \cite{bensby03}; Reddy \etal \cite{reddy03}; Thor\'en \etal \cite{thoren04}), which are based on the fitting of strong lines (for Na, Mg, Al, Si, Sc, and Ti). A value $E_\gamma$ $\sim$ 1.6 was adopted for both the \ion{Fe}{i} and \ion{Fe}{ii} lines. At least for \ion{Fe}{ii}, the value $E_\gamma$ = 2.5 that is often adopted in the literature seems too high (see Barklem \& Aspelund-Johansson \cite{barklem05}). A standard treatment of radiative and Stark broadening was used. The line list adopted in our study is provided in Table \ref{tab_atomic_data}.

\subsection{Spectral synthesis}\label{spectral_synthesis}
\subsubsection{Line selection and atomic data}\label{spectral_synthesis_atomic_data}
\begin{figure*}
\centering
\includegraphics[width=12cm,clip]{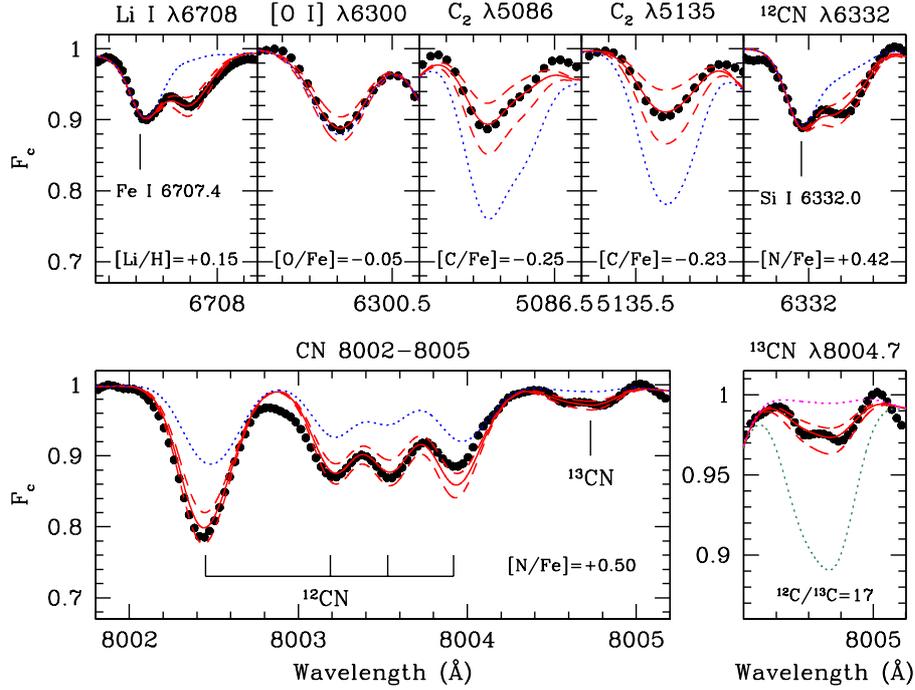}
\caption{Example of the fits to the spectral features used as mixing indicators in the case of \object{HD 175679}. The solid red line is the best-fitting synthetic profile, while the two dashed lines show the profiles for an abundance deviating by \p0.1 dex (and deviating by $\Delta$\iso = 5 in the case of $^{13}$CN 8004.7). The blue, dotted lines show the profiles for no Li present (\ion{Li}{i} $\lambda$6708) and solar abundance ratios with respect to iron for the CNO features. In the case of $^{13}$CN 8004.7, the magenta and dark green lines show the profiles for \iso = 89.4 and 3.5, which correspond to the terrestrial and CNO-equilibrium values, respectively.}
\label{fig_synthesis}
\end{figure*}

\begin{figure}
\centering
\includegraphics[width=9.0cm,clip]{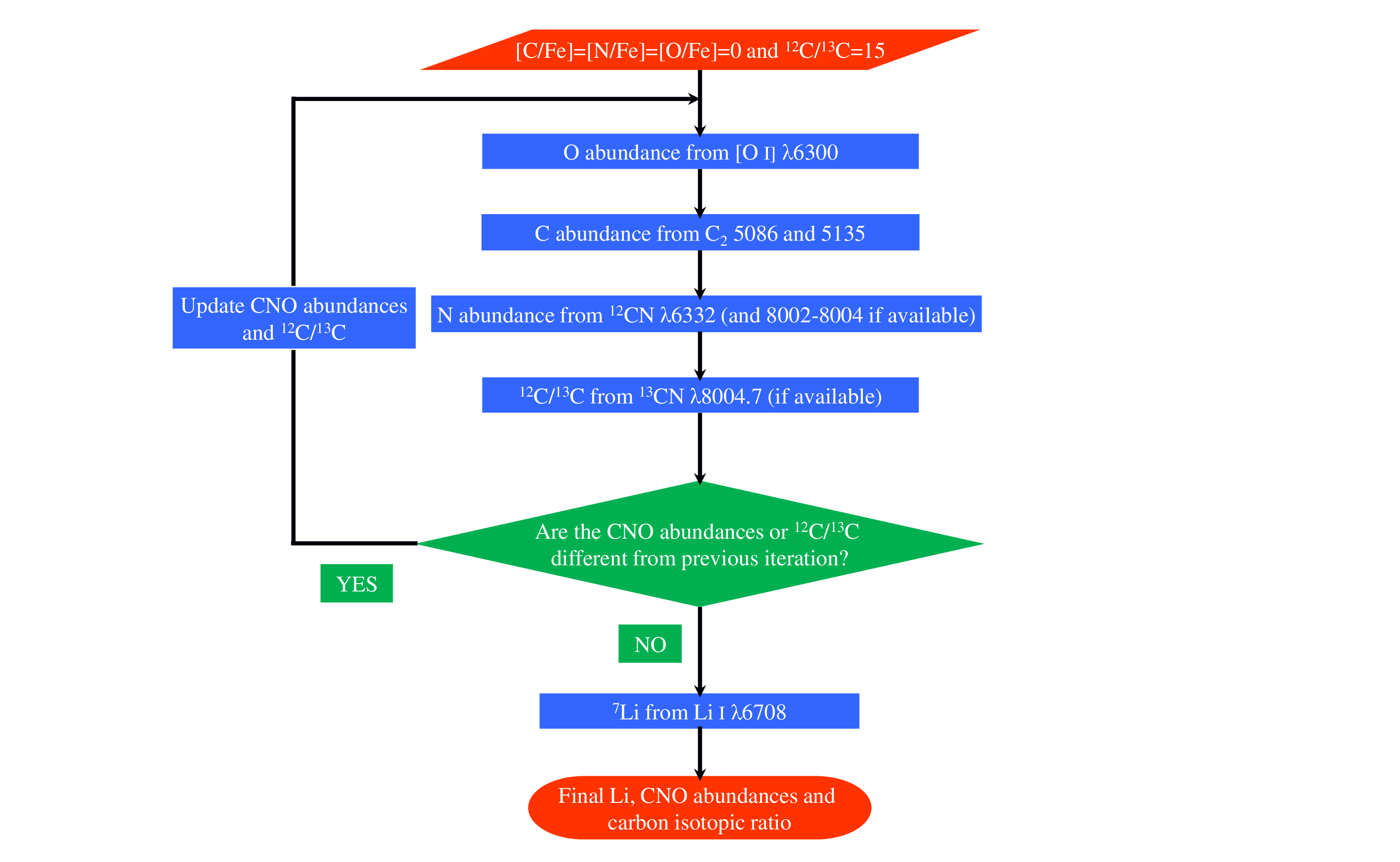}
\caption{Iterative scheme used for the spectral syntheses.}
\label{fig_flow_chart}
\end{figure}

The determination of the Li and CNO abundances relied on a spectral synthesis of the following atomic and molecular species: \ion{Li}{i} $\lambda$6708 (lithium), the C$_2$ lines at 5086 and 5135 \AA \ (carbon), \element[][12]{CN} $\lambda$6332 (nitrogen), and \ion{[O}{i]} $\lambda$6300 (oxygen). For only four stars, a number of \element[][12]{CN} lines in the range 8002-8004 \AA \ were also used for nitrogen. Some examples of the fits to the Li and CNO features are shown in the case of \object{HD 175679} in Fig.~\ref{fig_synthesis}. To ensure molecular equilibrium of the CNO-bearing molecules, the abundances of these three elements were iteratively varied until the values used were eventually the same in each synthesis. The scheme used is sketched in Fig.~\ref{fig_flow_chart}. In all cases, the list of atomic lines of other elements in the spectral ranges of interest was created using the data tabulated in the VALD-2 database assuming the mean abundances based on the EWs. The broadening parameters were estimated as in Sect.~\ref{sect_curve_of_growth}. The linear limb-darkening coefficients in the appropriate photometric band ($V$, $R$, or $I$) were interpolated from the tables of Claret (\cite{claret00}). The following dissociation energies were assumed for the molecular species: $D_0$ = 6.21 (C$_2$; Huber \& Herzberg \cite{huber_herzberg79}), 7.65 (CN; Bauschlicher \etal \cite{bauschlicher88}), 6.87 (TiO; Cox \cite{cox00}), 1.34 (MgH; Cox \cite{cox00}), and 3.06 eV (SiH; Cox \cite{cox00}). The effect of adopting other values for C$_2$ and CN is examined in Sect.~\ref{sect_errors_abundances}.

The lithium abundance was determined using the accurate laboratory atomic data quoted in Smith \etal (\cite{smith98}). The contribution of the $^6$Li isotope is expected to be negligible and was therefore ignored. The van der Waals damping parameters for the lithium components were taken from Barklem \etal (\cite{barklem00}). Spectral features of the diatomic molecules CN, TiO, MgH, and SiH were considered. An extensive CN line list was taken from Mandell \etal (\cite{mandell04}), who used a carbon arc spectrum to accurately estimate the wavelengths and oscillator strengths of the \element[][12]{CN} transitions in the spectral region of interest. A list of $^{48}$TiO lines from the $\gamma$ system was taken from Luck (\cite{luck77}). The relevant atomic data were retrieved from the 2006 catalogue version of the TiO database of Plez (\cite{plez98}).\footnote{Available online at: \newline {\tt http://www.graal.univ-montp2.fr/hosted/plez/}.} We carried out test calculations for HD 45398 using the updated and more extensive TiO line list implemented in the VALD-3 database, but this led to negligible differences in the Li abundance. It has to be noted that all the transitions considered have accurate laboratory wavenumbers (Davis \etal \cite{davis86}). A few MgH and SiH lines of significant strength were also taken from the Kurucz atomic database and incorporated.\footnote{Available online at: \newline {\tt http://kurucz.harvard.edu/molecules.html}.} An isotopic ratio 1.000:0.127:0.139 was assumed for $^{24}$MgH:$^{25}$MgH:$^{26}$MgH (Asplund \etal \cite{asplund09}). The iron and lithium abundances were adjusted until a satisfactory fit of the blend primarily formed by \ion{Fe}{i} $\lambda$6707.4 and the Li doublet was achieved (\loggf \ = --2.21 was adopted for the Fe line based on an inverted solar analysis.). A close agreement was found in all cases between the abundance yielded by this weak iron line and the mean values found with the EWs. A very small velocity shift ($\la$ 1 km s$^{-1}$) was occasionally applied to account for an imperfect correction of the stellar radial velocity. Owing to the weakness of the \ion{Li}{i} $\lambda$6708 feature in some objects or its absence thereof, only an upper limit could be determined. A fit of this feature in our solar spectrum yields $\log \epsilon$(Li) = 1.09, which we take as reference thereafter. We also provide non-LTE (NLTE) abundances (which we recommend to use) using corrections interpolated from the tables of Lind \etal (\cite{lind09}) in the following. These NLTE corrections are systematically positive and range from +0.11 to +0.36 dex. For the Sun, it amounts to +0.04 dex. It should be noted that abundances lower at the $\sim$0.15 dex level at solar metallicity could be expected if the formation of \ion{Li}{i} $\lambda$6708 is modelled using hydrodynamical simulations that take surface convection into account (Collet \etal \cite{collet07}). However, no correction for granulation effects was applied here because of the unavailability of detailed predictions on a star-to-star basis. 

The contribution of \ion{Ni}{i} $\lambda$6300.34 to \ion{[O}{i]} $\lambda$6300.30 was estimated adopting the oscillator strength determined by Johansson \etal (\cite{johansson03}) from laboratory measurements: \loggf \ = --2.11. The oscillator strength of the oxygen line, \loggf \ = --9.723, was inferred from an inverted solar analysis, and $E_\gamma$ = 1.8 was assumed. The oxygen abundance could also be determined for four stars using the EWs of the \ion{O}{i} triplet at about 7774 \AA. The values found systematically appear larger than those yielded by \ion{[O}{i]} $\lambda$6300 with differences ranging from +0.07 (\object{HD 175679}) to +0.26 dex (\object{HD 175294} and \object{$\alpha$ Boo}). Such a discrepancy is commonly observed in red giants and likely arises from the different sensitivity to NLTE effects (e.g., Schuler \etal \cite{schuler06}). For this reason, the triplet abundances are not discussed in the following. One should note that our results based on the forbidden line are also immune to the neglect of surface inhomogeneities related to time-dependent convection phenomena (Collet \etal \cite{collet07}).

\begin{figure}
\centering
\includegraphics[width=9cm]{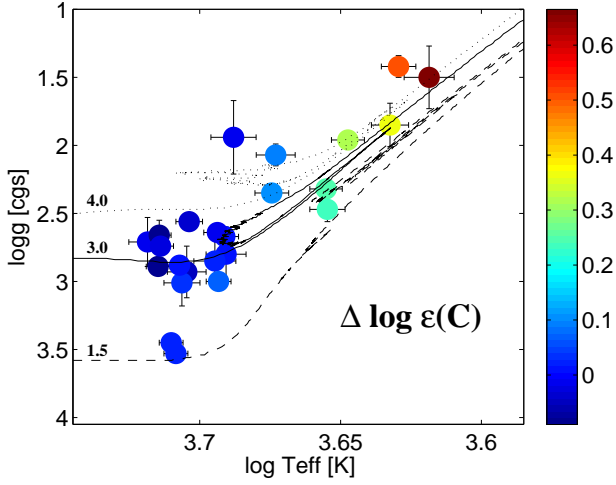}
\caption{Difference between the abundances yielded by \ion{C}{i} $\lambda$5380 and the mean values using the C$_2$ features. The predictions of evolutionary models at solar metallicity and for masses of 1.5, 3, and 4 M$_{\odot}$ are overplotted for illustrative purposes. Same tracks as in Fig.~\ref{fig_logg_logTeff_starevol}, except that the evolutionary phase, is not colour coded.}
\label{fig_diff_C2_5380}
\end{figure}

The atomic data for the lines of the C$_2$ Swan system at 5086 and 5135 \AA \ were taken from Lambert \& Ries (\cite{lambert_ries81}). In the case of C$_2$ $\lambda$5135, however, small adjustments were applied based on a fit of this feature in the Sun. An enhancement factor, $E_\gamma$ = 2, was assumed (Asplund \etal \cite{asplundetal05}). Both lines could be used for all stars but \object{$\epsilon$ Oph} (C$_2$ $\lambda$5086 is affected by a telluric feature), and these two indicators agree closely: $\langle$$\log \epsilon$ (C$_2$ $\lambda$5086) -- $\log \epsilon$ (C$_2$ $\lambda$5135)$\rangle$ = --0.02$\pm$0.05 dex (1$\sigma$, 23 stars). The C abundance was also estimated from the high-excitation \ion{C}{i} $\lambda$5380.3 line assuming \loggf \ = --1.704 (derived from an inverted solar analysis) and $E_\gamma$ = 2. However, these abundances were found to be discrepant in the six coolest objects with those yielded by the C$_2$ features (see Fig.~\ref{fig_diff_C2_5380}). For the remaining stars, there is a near-perfect agreement: $\langle$$\log \epsilon$ (\ion{C}{i} $\lambda$5380) -- $\log \epsilon$ (C$_2$)$\rangle$ = +0.01$\pm$0.05 dex (1$\sigma$, 18 stars). The origin of the discrepancy at low \teff \ is unclear: departures from LTE (Fabbian \etal \cite{fabbian06}), unrecognized blends (Luck \& Heiter \cite{luck_heiter06}, \cite{luck_heiter07}), and/or granulation effects. In view of the problems plaguing the \ion{C}{i} $\lambda$5380 abundances, they are not discussed further. Although the departures from LTE affecting the C$_2$ features are largely unknown (Asplund \cite{asplund05}), the neglect of granulation should not lead to large errors (Dobrovolskas \etal \cite{dobrovolskas13}).

For the molecular feature at 6332.2 \AA, seven CN(5,1) components were considered (with wavelengths from R. Smiljanic, private communication). The \loggf \ values were taken from J\o rgensen \& Larsson (\cite{jorgensen_larsson90}) with some adjustments based on the fit of the solar spectrum (see Table~\ref{tab_CN}). Two \ion{Si}{i} and \ion{Fe}{ii} lines at about 6331.95 \AA \ lie in close vicinity of the CN feature. The oscillator strength of the stronger Si line (\loggf \ = --2.64) was inferred  from an inverted solar analysis, while the value in VALD-2 was assumed for the Fe line (\loggf \ = --2.071; Raassen \& Uylings \cite{raassen_uylings98}). Both the Si and the N abundances were adjusted during the fit. The Si abundances found systematically agreed within error with the mean values derived from the EWs. The atomic data for the lines of the A$^2$$\Pi$-X$^2$$\Sigma^+$ system of CN(2,0) in the range 8002-8005 \AA \ were taken from J\o rgensen \& Larsson (\cite{jorgensen_larsson90}), but the wavelengths come from other sources (Wyller \cite{wyller66}; Tomkin \etal \cite{tomkin75}; Boyarchuk \etal \cite{boyarchuk91}; Barbuy \etal \cite{barbuy92}). The \loggf \ value for \ion{Fe}{i} $\lambda$8002.58 was taken from Boyarchuk \etal (\cite{boyarchuk91}). Although the two indicators can only be used in a few stars, the N abundances agree well: $\langle$$\log \epsilon$ (CN 6332) -- $\log \epsilon$ (CN 8002-8004)$\rangle$ = --0.07$\pm$0.07 dex (1$\sigma$, 4 stars). As for the C$_2$ features, the magnitude of the NLTE effects for CN is poorly known (Asplund \cite{asplund05}), and granulation is expected to have a limited impact (Dobrovolskas \etal \cite{dobrovolskas13}).

\begin{table}
\centering
\caption{Atomic data for the CN(5,1) components included in the modelling of the CN $\lambda$6332 feature.}
\label{tab_CN}
\begin{tabular}{cccc} 
\hline\hline
Component & $\lambda$ & LEP    & \loggf \\ 
          & [\AA]     & [eV]   &        \\
\hline
R$_2$(4)  & 6332.18   & 0.258  & --2.258\\
R$_2$(3)  & 6332.18   & 0.256  & --2.387\\ 
R$_2$(5)  & 6332.34   & 0.260  & --2.433\\ 
R$_2$(2)  & 6332.34   & 0.255  & --2.843\\ 
R$_2$(6)  & 6332.68   & 0.263  & --2.804\\ 
R$_2$(1)  & 6332.68   & 0.254  & --3.576\\
R$_2$(7)  & 6333.19   & 0.266  & --2.879\\ 
\hline
\end{tabular}
\end{table}

The \iso \ isotopic ratio could be determined for four stars by fitting the $^{13}$CN feature at 8004.7 \AA. For the other stars, we fixed this ratio to a value of 15, which may be regarded -- as judged from the values obtained for these few stars and previous results in the literature (e.g., Smiljanic \etal \cite{smiljanic09}) -- as representative of our sample. This choice has a very limited impact on the final results.

\subsubsection{Line broadening parameters}\label{spectral_synthesis_broadening}
The knowledge of the line broadening is needed to perform the spectral synthesis. The lines in red giants are broadened by stellar rotation and macroturbulent motions by a comparable amount. These two phenomena imprint a distinct, yet subtle, signature in the shape of the line profile that is best revealed through Fourier transforms applied to very high spectral resolution and S/N data (e.g., Gray \& Brown \cite{gray_brown06}). In most cases, it is difficult with our observations to clearly disentangle the contribution of these two processes. The radial-tangential macroturbulence, $\zeta_{\rm RT}$, was therefore estimated from a calibration of this parameter across the HR diagram (Massarotti \etal \cite{massarotti08}). The stellar luminosities were computed from the {\it Hipparcos} parallaxes (Table \ref{tab_observations}) and the bolometric corrections from the calibrations of Alonso \etal (\cite{alonso99}). The \c \ targets are relatively nearby ($d$ $\la$ 500 pc), and $A_V$ = 0.2 mag was assumed for all stars (no reddening was taken into account for the bright benchmark stars). For the \object{NGC 6633} members, we assumed a distance of 375 pc (van Leeuwen \cite{van_leeuwen09}). For \object{HD 170031}, we adopted the value determined for \object{HD 169370} ($\zeta_{\rm RT}$ = 3.4 \kms) in view of the similar physical parameters. Although these estimates of the stellar luminosity are uncertain (Sect.~\ref{sect_targets}), they are sufficient for our purpose. 

The projected rotational velocity, $v \sin i$, was subsequently derived by fitting a set of six relatively unblended \ion{Fe}{i} lines in the vicinity of the Li doublet. The other free parameter was \fe \ (For all stars, the iron abundances found are identical within error of the mean values derived from the curve-of-growth analysis.). The only exception was \object{$\alpha$ Boo}, for which we used the values $\zeta_{\rm RT}$ = 5.2 and $v \sin i$ = 1.5 \kms, as derived by Gray \& Brown (\cite{gray_brown06}) from Fourier techniques. The results we found ($\zeta_{\rm RT}$ = 4.2 and $v \sin i$ = 2.4 \kms) lead to nearly identical line profile shapes. Our values for $\eta$ Ser ($\zeta_{\rm RT}$ = 3.8 and $v \sin i$ = 1.5 \kms) agree within error with those of Carney \etal (\cite{carney08}), which are based on a Fourier transform analysis. The microturbulent velocity was fixed to the values derived as described in Sect.~\ref{sect_parameters} and listed in Table \ref{tab_parameters}. Instrumental broadening at the wavelength of the feature of interest was assumed to be Gaussian and estimated from calibration lamps (As an illustration, see Table \ref{tab_observations} for the values used for \ion{Li}{i} $\lambda$6708.). Finally, the linear limb-darkening coefficients in the $R$ band were taken from Claret (\cite{claret00}). 

\begin{figure}[h]
\centering
\includegraphics[width=9.0cm,clip]{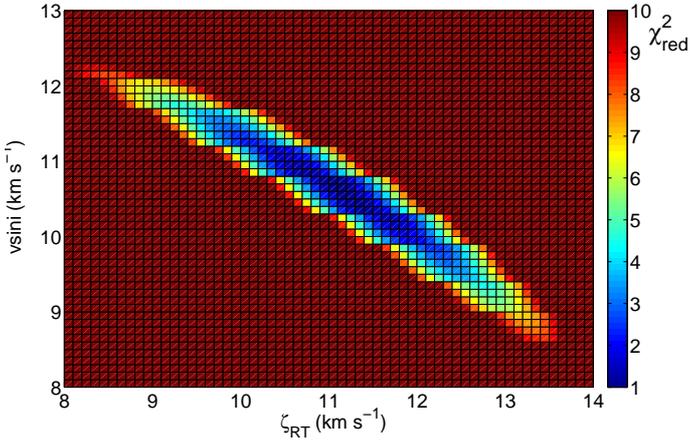}
\caption{Variation for the \ion{Fe}{i} $\lambda$6705 line in HD 50890 of the fit quality (colour-coded as a function of $\chi^2_{\rm red}$) for different combinations of $v\sin i$ and $\zeta_\mathrm{RT}$. The best fit is found for $\zeta_\mathrm{RT}$ = 11.1 and $v\sin i$ = 10.6 \kms \ (and \fe \ = +0.05). Similar results are obtained for \ion{Fe}{i} $\lambda$6703. Note that the fit quality of the analysis that made use of a calibrated $\zeta_{\rm RT}$ of 7.6 \kms \ (and which led to $v \sin i$ = 12.5 \kms) cannot be judged from this figure because the adopted metallicity is different.}
\label{fig_vsini_vmacro}
\end{figure}

The $v \sin i$ values we derive are strongly tied to the choice of the adopted macroturbulence and are only meant to provide a good fit to the features synthesised. They are, therefore, surrounded by a large uncertainty, and these are not quoted. It is worth mentioning, however, that the value adopted for HD 50890 ($\sim$ 12.5 \kms \ using a calibrated macroturbulence of 7.6 \kms) is much larger than that found for the other targets ($\lesssim$ 5 \kms). This suggests an unusually high rotation rate for a giant, although the crudeness of the calibration used to infer $\zeta_{\rm RT}$ should be kept in mind. In view of the well-resolved nature of the line profiles in that particular case, an attempt was made to separate the contribution of each broadening mechanism through fitting two iron lines in the vicinity of the Li doublet (\ion{Fe}{i} $\lambda$6703.6 and 6705.1 \AA) with a grid of synthetic spectra computed for a wide range of $v \sin i$ and $\zeta_{\rm RT}$ values. The other free parameter was the iron abundance (see Morel \etal \cite{morel13}). Although there is a clear degeneracy in the determination of these two quantities, this analysis indeed supports a high rotational velocity of the order of 10 \kms \ (Fig.~\ref{fig_vsini_vmacro}). The CNO and Li abundances were derived for this star using these best-fitting broadening parameters.

\section{Determination of stellar parameters}\label{sect_parameters}
\begin{figure}
\centering
\includegraphics[width=9cm]{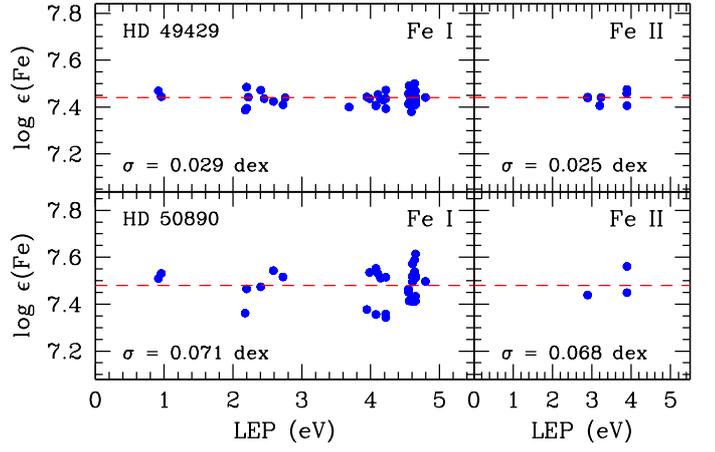}
\caption{Abundances derived from the \ion{Fe}{i} and \ion{Fe}{ii} lines for \object{HD 49429} and \object{HD 50890} as a function of the LEP. The horizontal dashed line indicates the mean iron abundance.}
\label{fig_Fe_EP}
\end{figure}

The model parameters (\teff, \logg, \micro, \fe, and abundances of the $\alpha$ elements) were iteratively modified until all the following conditions were simultaneously fulfilled: (1) the \ion{Fe}{i} abundances exhibit no trend with lower excitation potential (LEP) or reduced EW (the logarithm of the EW divided by the wavelength of the transition). Our selected neutral iron lines span a wide range in LEP and strength and are therefore well suited to constrain both the temperature and the microturbulence; (2) the abundances derived from the \ion{Fe}{i} and \ion{Fe}{ii} lines are identical; and (3) the Fe and $\alpha$-element abundances are consistent with the values adopted for the model atmosphere. The number of iron lines used ranges from 33 to 63 for \ion{Fe}{i} and from 3 to 9 for \ion{Fe}{ii}. The typical line-to-line scatter of the Fe abundances (either for \ion{Fe}{i} or \ion{Fe}{ii}) is 0.05 dex. Figure \ref{fig_Fe_EP} shows the variation of the Fe abundances as a function of the LEP for two stars showing amongst the lowest and largest scatters. 

The ODFs and Rosseland opacity tables were chosen according to the microturbulence and Fe abundance (rounded to the nearest 0.1 dex), as derived from the spectral analysis. Furthermore, the $\alpha$-element abundance of the model varied depending on [Fe/H], which follows the same convention as adopted for the MARCS suite of models (e.g., [$\alpha$/Fe] = +0.2 for \fe \ = --0.4; Gustafsson \etal \cite{gustafsson08}). If necessary, the ODFs for the appropriate Fe and $\alpha$-element abundances were linearly interpolated from a pre-calculated grid available online.\footnote{For more details, see: \newline {\tt http://wwwuser.oat.ts.astro.it/castelli/odfnew.html}.}

As discussed in Sect.~\ref{sect_seismic_constraints}, the seismic gravities are likely not only more precise but also more accurate than the values derived from the ionisation balance of iron. We therefore repeated the analysis after fixing the gravity of the models to this value. Such an approach is now routinely implemented in spectroscopic studies of seismic targets (e.g., Batalha \etal \cite{batalha11}; Carter \etal \cite{carter12}; Thygesen \etal \cite{thygesen12}; Huber \etal \cite{huber13}) and is expected to provide more robust estimates of the physical parameters and ultimately chemical abundances. The temperature was also determined from Fe excitation balance. A change in \logg \ of 0.1 dex typically leads to variations in \teff \ of 15 K and in \fe \ of 0.04 dex. Similar figures are obtained for solar-like stars (Torres \etal \cite{torres12}; Huber \etal \cite{huber13}). The good agreement in our case between the two sets of \logg \ values (see below) only implies small adjustments for \teff \ ($\lesssim$ 30 K) and the abundances ($\lesssim$ 0.1 dex). A more general discussion including results for {\it Kepler} targets is presented in Morel (\cite{morel14}).

Although ionisation balance of iron is usually fulfilled within the errors, the formal mean iron abundance yielded by the \ion{Fe}{i} and \ion{Fe}{ii} lines may differ by up to 0.18 dex (and on average by 0.07 dex) when the gravity is held fixed to the seismic value. There is therefore an ambiguity as to which metallicity value should be eventually adopted. As the \ion{Fe}{i} lines are known to be more prone to departures from LTE, it may be argued that the mean \ion{Fe}{ii}-based abundance is a better proxy of the stellar metallicity (e.g., Thygesen \etal \cite{thygesen12}). However, the abundances yielded by the \ion{Fe}{ii} lines are also affected by a number of caveats in red giants: (1) the features are only usually a few, difficult to measure, and potentially more affected by blends; (2) they are very sensitive to errors in the effective temperature (varying \teff \ by 50 K while keeping the gravity fixed typically changes the \ion{Fe}{i} and \ion{Fe}{ii} abundances by 0.01-0.02 and 0.06 dex, respectively; see also Ram{\'{\i}}rez \& Allende Prieto \cite{ramirez11}); (3) they suffer, as with the \ion{Fe}{i} lines (and perhaps even more according to models), from the neglect of granulation effects (Collet \etal \cite{collet07}; Ku\v{c}inskas \etal \cite{kucinskas13}; see also fig.15 of Dobrovolskas \etal \cite{dobrovolskas13}). In view of the uncertainties plaguing both the \ion{Fe}{i} and \ion{Fe}{ii} abundances, we consider in the following that the iron content is given by the average of the values yielded by these two ions.

\section{Computation of uncertainties}\label{sect_errors}
The uncertainties affecting our results are schematically two kinds: statistical and systematic. We tried to incorporate both in the total uncertainty budget, assuming that the choices of the line list and the set of model atmospheres were the main sources of systematic uncertainties. 

\subsection{Physical parameters}\label{sect_errors_parameters}
To assess the uncertainties in the physical parameters associated to the choice of the diagnostic lines, we repeated the analysis using the line lists of Morel \etal (\cite{morel03}, \cite{morel04}) and Hekker \& Mel\'endez (\cite{hekker_melendez07}) described in Sect.~\ref{sect_curve_of_growth}. The standard deviation of the results was about 40 K for \teff, 0.10 dex for \logg, and 0.04 \kms \ for \micro. Four stars were not observed with HARPS or ELODIE, but with instruments offering a wider wavelength coverage (FEROS and KPNO \'echelle). However, re-analysing these stars by only considering the lines covered by HARPS leads to very small differences. 

The uncertainties arising from the choice of the model atmosphere were estimated by analysing a number of stars using plane-parallel MARCS (Gustafsson \etal \cite{gustafsson08}) and Kurucz models computed with different assumptions regarding their metal content. Namely, we repeated the analysis by varying the input metallicity and $\alpha$ enhancement of the models by their typical uncertainty ($\Delta$\fe \ = 0.1 and $\Delta$[$\alpha$/Fe] = 0.2 dex). Furthermore, we assumed different treatments for the convection (adopting various ratios of the mixing length to scale height or incorporating overshooting). A few stars ($\alpha$ Boo, $\epsilon$ Oph, and $\xi$ Hya) were analysed with spherical MARCS models, but small differences in terms of atmospheric parameters were found with respect to plane-parallel Kurucz models (see also Carlberg \etal \cite{carlberg12}). As expected, the largest changes by far were found for the low-gravity star $\alpha$ Boo with \teff \ and \logg \ larger by 55 K and 0.13 dex, respectively. In contrast, extremely similar results were found for $\epsilon$ Oph and $\xi$ Hya that are more representative of our sample. These relatively small differences, which remain within the uncertainties, can be explained by the similar atmospheric structure for the range of parameters spanned by our targets (Gustafsson \etal \cite{gustafsson08}). Very small variations in the iron abundances were also found in accordance with previous studies (Heiter \& Eriksson \cite{heiter_eriksson06}). Once again, $\alpha$ Boo deviated the most, but [Fe/H] was only 0.04 dex larger. In addition, we experimented with the MARCS models that were moderately contaminated by CN-cycled material (as is the case for most of our targets as discussed below) but found negligible differences with respect to the models that assume a scaled-solar mixture. 

We finally obtain the following figures for the systematic uncertainties: 70 K for \teff, 0.15 dex for \logg, and 0.05 \kms \ for \micro. With the gravity fixed to the seismic value, this reduces to 35 K for \teff \ and 0.025 \kms \ for \micro.

The statistical uncertainties are first related to the errors made when fulfilling excitation and ionisation equilibrium of the iron lines (\teff \ and \logg) or when constraining the \ion{Fe}{i} abundances to be independent of the line strength (\micro). To estimate the uncertainty in \teff, for instance, we considered the range over which the slope of the relation between the \ion{Fe}{i} abundances and the LEP is consistent with zero within the uncertainties. As the parameters of the model (\teff, \logg, and \micro) are interdependent, changes in one of these parameters are necessarily accompanied by variations in the other two. Two of these parameters were therefore adjusted, while the third one was varied by the relevant uncertainty. In addition, to assess the uncertainties associated to the placement of the continuum level, we re-estimated the parameters of HD 175679 after shifting the continuum upwards by 1\%. This led to negligible variations for \teff \ and \micro, while \logg \ was 0.03 dex lower. These figures were considered as being representative and adopted for all the stars in our sample.

The final uncertainty was taken as the quadratic sum of the statistical and systematic errors (Because of the small scatter of the Fe abundances, the latter are often found to dominate.). 

\subsection{Chemical abundances}\label{sect_errors_abundances}
To investigate the sensitivity of the abundances obtained from curve-of-growth techniques to changes in the physical parameters, we repeated the analysis by varying each parameter by its global (systematic and statistical) uncertainty defined above. We proceeded as above to estimate the uncertainties related to the placement of the continuum level. As expected, this has a noticeable effect on \fe \ but a much lower impact on the abundance ratios with respect to iron. Finally, the line-to-line scatter, $\sigma_{\rm int}$, was quadratically summed to these values to obtain the final uncertainty (We assumed a rather generous value of 0.07 dex when the abundance was based on a single line.). The uncertainty budget is described in the case of \object{HD 175679} in Table \ref{tab_errors}.

\begin{table}
\centering
\caption{Uncertainty budget for the abundances derived from the EWs in the case of \object{HD 175679}.}
\scriptsize
\label{tab_errors}
\begin{tabular}{llccccc} \hline\hline
                                        & $\sigma_{\rm int}$    & $\sigma_{\rm Teff}$ & $\sigma_{\rm logg}$ & $\sigma_{\xi}$ & $\sigma_{\rm normalisation}$ & $\sigma_{\rm total}$ \\
\hline
$\Delta$${\rm [Fe/H]}$                  & 0.05                  &  +0.05 &  +0.00 & --0.01 &  +0.07 & 0.10\\
$\Delta$[\ion{O}{i} triplet/Fe]         & 0.05                  & --0.07 &  +0.11 & --0.01 &  +0.04 & 0.15\\ 
$\Delta$[Na/Fe]                         & 0.03                  & --0.02 & --0.02 &  +0.01 & --0.01 & 0.05\\
$\Delta$[Mg/Fe]                         & 0.07\tablefootmark{a} & --0.04 & --0.03 & --0.01 & --0.01 & 0.09\\
$\Delta$[Al/Fe]                         & 0.07                  & --0.02 & --0.02 &  +0.02 &  +0.01 & 0.08\\
$\Delta$[Si/Fe]                         & 0.08                  & --0.04 &  +0.03 &  +0.01 &  +0.01 & 0.10\\
$\Delta$[Ca/Fe]                         & 0.05                  & --0.01 & --0.02 & --0.01 &  +0.00 & 0.06\\
$\Delta$[Sc/Fe]                         & 0.07\tablefootmark{a} &  +0.05 &  +0.09 &  +0.03 &  +0.00 & 0.13\\
$\Delta$[Ti/Fe]                         & 0.07\tablefootmark{a} &  +0.01 & --0.01 &  +0.03 &  +0.03 & 0.09\\
$\Delta$[Cr/Fe]                         & 0.14                  &  +0.00 & --0.02 &  +0.01 & --0.01 & 0.15\\
$\Delta$[Co/Fe]                         & 0.07\tablefootmark{a} &  +0.01 &  +0.01 &  +0.02 &  +0.00 & 0.08\\
$\Delta$[Ni/Fe]                         & 0.04                  & --0.01 &  +0.02 &  +0.00 &  +0.00 & 0.05\\
$\Delta$[Ba/Fe]                         & 0.07\tablefootmark{a} &  +0.02 &  +0.08 & --0.02 &  +0.00 & 0.11\\
\hline
\end{tabular}
\tablefoot{The first column gives the line-to-line scatter. The quantities $\sigma_{\rm Teff}$, $\sigma_{\rm logg}$, and $\sigma_{\xi}$ give the uncertainties associated to the following changes in the atmospheric parameters: $\Delta$\teff \ = +80 K, $\Delta$\logg \ = +0.18 dex, and $\Delta$\micro \ = +0.07 \kms. Note that the two other parameters were simultaneously adjusted to fulfil excitation/ionisation equilibrium of iron or to have the \ion{Fe}{i} abundances that are independent of the line strength. Finally, $\sigma_{\rm normalisation}$ provides the changes associated to a continuum level shifted upwards by 1\% (These values were adopted for all the stars in our sample.). \tablefoottext{a}{Arbitrary value.}}
\end{table}

For the abundances obtained through spectral synthesis, the same procedure as above was applied to \object{HD 175679}, and the uncertainties were taken as representative of the whole sample. The only exception was \iso \ for which the sensitivity against the placement of the continuum was estimated on a star-to-star basis. It should be noted that this ratio is largely insensitive to the exact choice of the parameters (see Smiljanic \etal \cite{smiljanic09}). The abundances of all elements other than Li and CNO were updated and fixed in the synthesis to the values obtained with the new set of parameters. As the fit quality was evaluated by eye, we incorporated a typical error associated to this procedure. For \ion{[O}{i]} $\lambda$6300, varying the nickel abundance (and therefore the contamination of the blended Ni feature) within its uncertainty led to negligible changes. Finally, a typical uncertainty of 0.1 eV in the dissociation energy of the C$_2$ and CN molecules was also taken into account in the total uncertainty budget. Once again, the final uncertainty was taken as the quadratic sum of all these sources of errors (see Table \ref{tab_errors_mixing}). The uncertainties in the Li and CNO abundances associated to the use of 1D LTE models are discussed in Sect.~\ref{spectral_synthesis_atomic_data}.

\begin{table*}
\centering
\caption{Uncertainty budget for the abundances of the mixing indicators in the case of \object{HD 175679}.}
\label{tab_errors_mixing}
\begin{tabular}{llcccccccc} \hline\hline
& $\sigma_{\rm int}$ & $\sigma_{\rm Teff}$ & $\sigma_{\rm logg}$ & $\sigma_{\xi}$ & $\sigma_{\rm normalisation}$ & $\sigma_{\rm fit}$ & $\sigma_{\rm Ni}$  & $\sigma_{\rm D_0}$ & $\sigma_{\rm total}$\\
\hline
$\Delta$$\log \epsilon$(O)             & 0.05\tablefootmark{a}   &  +0.11 &  +0.08 &  +0.04 &  +0.06 & 0.03 &  +0.01 & +0.00 & 0.17\\
$\Delta$[O/Fe]                         & 0.05\tablefootmark{a}   &  +0.06 &  +0.08 &  +0.05 &  +0.06 & 0.03 &  +0.01 & +0.00 & 0.14\\\hline
$\Delta$$\log \epsilon$(C$_2$ 5086)    & \multicolumn{1}{c}{...} &  +0.06 &  +0.03 &  +0.02 &  +0.04 & 0.03 &  +0.00 & +0.04 & 0.10\\
$\Delta$$\log \epsilon$(C$_2$ 5135)    & \multicolumn{1}{c}{...} &  +0.06 &  +0.04 &  +0.02 &  +0.04 & 0.03 &  +0.00 & +0.04 & 0.10\\
$\Delta$$\log \epsilon$(C$_2$)         & 0.03                    &  +0.06 &  +0.04 &  +0.02 &  +0.04 & 0.03 &  +0.00 & +0.04 & 0.11\\
$\Delta$[C/Fe]                         & 0.03                    &  +0.01 &  +0.04 &  +0.03 &  +0.04 & 0.03 &  +0.00 & +0.04 & 0.09\\\hline
$\Delta$$\log \epsilon$(CN 6332)       & \multicolumn{1}{c}{...} &  +0.08 & --0.01 &  +0.03 &  +0.04 & 0.03 &  +0.00 & +0.08 & 0.13\\
$\Delta$$\log \epsilon$(CN 8003)       & \multicolumn{1}{c}{...} &  +0.08 &  +0.00 &  +0.03 &  +0.06 & 0.03 &  +0.00 & +0.09 & 0.14\\
$\Delta$$\log \epsilon$(N)             & 0.05                    &  +0.08 & --0.01 &  +0.03 &  +0.05 & 0.03 &  +0.00 & +0.08 & 0.14\\
$\Delta$[N/Fe]                         & 0.05                    &  +0.03 & --0.01 &  +0.04 &  +0.05 & 0.03 &  +0.00 & +0.08 & 0.13\\\hline
$\Delta$[N/C]                          & 0.06                    &  +0.02 & --0.04 &  +0.01 &  +0.01 & 0.04 &  +0.00 & +0.04 & 0.10\\
$\Delta$[N/O]                          & 0.07                    & --0.03 & --0.09 & --0.01 & --0.01 & 0.04 & --0.01 & +0.08 & 0.15\\
$\Delta$[C/O]                          & 0.06                    & --0.05 & --0.05 & --0.02 & --0.02 & 0.04 & --0.01 & +0.04 & 0.12\\\hline
$\Delta$\iso                           & \multicolumn{1}{c}{...} & --0.3  & --0.2  &  +0.00 & --5    &    1 &  +0.00 & +0.00 &  5.1\\\hline
$\Delta$[Li/H]                         & 0.05\tablefootmark{a}   &  +0.09 & --0.04 &  +0.03 &  +0.05 & 0.05 &  +0.00 & +0.00 & 0.13\\ 
\hline
\end{tabular}
\tablefoot{The first column gives the line-to-line scatter. The quantities $\sigma_{\rm Teff}$, $\sigma_{\rm logg}$, and $\sigma_{\xi}$ give the uncertainties associated to the following changes in the atmospheric parameters: $\Delta$\teff \ = +80 K, $\Delta$\logg \ = +0.18 dex, and $\Delta$\micro \ = +0.07 \kms. Note that the two other parameters were simultaneously adjusted to fulfil excitation/ionisation equilibrium of iron or to have the \ion{Fe}{i} abundances that are independent of the line strength. The uncertainty associated to a continuum level shifted upwards by 1\% is provided by $\sigma_{\rm normalisation}$. As the fit quality was evaluated by eye, $\sigma_{\rm fit}$ gives the rough uncertainty associated to this procedure. The error resulting from a lowering of the Ni abundance by its uncertainty (0.05 dex) is given by $\sigma_{\rm Ni}$. Finally, the effect of lowering the adopted dissociation energy of the C$_2$ and CN molecules by 0.1 eV is given by $\sigma_{\rm D_0}$. \tablefoottext{a}{Arbitrary value.}}
\end{table*}

The physical parameters are provided in Table \ref{tab_parameters}, while the chemical abundances are given in Tables \ref{tab_abundances} and \ref{tab_abundances_mixing}. Table \ref{tab_abundances_ratios} presents the logarithmic CNO abundance ratios. 

\begin{table}
\caption{Atmospheric parameters of the targets.}
\centering
\label{tab_parameters}
\begin{tabular}{lcccr} 
\hline\hline
                       & \teff \ [K] & \logg \ [cgs] & \micro \ [\kms] & \multicolumn{1}{c}{\fe} \\
\hline
\object{HD 40726}       & 5230\p80 &  2.71\p0.18  & 1.67\p0.07 &  +0.05\p0.11\\
\object{HD 42911}       & 4905\p80 &  2.80\p0.20  & 1.42\p0.07 &  +0.12\p0.11\\
\object{HD 43023}       & 5065\p80 &  2.93\p0.19  & 1.38\p0.07 & --0.05\p0.10\\
\object{HD 45398}       & 4155\p85 &  1.50\p0.23  & 1.50\p0.08 & --0.15\p0.14\\ 
\object{HD 49429}       & 5085\p75 &  3.01\p0.17  & 1.33\p0.06 & --0.06\p0.10\\
\object{HD 49566}       & 5170\p75 &  3.01\p0.17  & 1.39\p0.06 & --0.04\p0.10\\
                        & 5185\p50 & [2.89\p0.04] & 1.42\p0.04 & --0.08\p0.09\\ 
\object{HD 50890}       & 4730\p95 &  1.85\p0.26  & 1.98\p0.10 & --0.02\p0.13\\
                        & 4710\p75 & [2.07\p0.08] & 1.98\p0.09 &  +0.06\p0.12\\ 
\object{HD 169370}      & 4520\p85 &  2.31\p0.22  & 1.42\p0.07 & --0.27\p0.13\\
                        & 4520\p60 & [2.32\p0.04] & 1.42\p0.06 & --0.26\p0.10\\ 
\object{HD 169751}      & 4900\p80 &  2.72\p0.19  & 1.23\p0.07 &  +0.00\p0.11\\
                        & 4910\p55 & [2.67\p0.07] & 1.24\p0.05 & --0.02\p0.10\\ 
\object{HD 170008}      & 5130\p75 &  3.43\p0.17  & 1.04\p0.07 & --0.35\p0.10\\
                        & 5130\p50 & [3.45\p0.04] & 1.04\p0.05 & --0.34\p0.09\\ 
\object{HD 170031}      & 4535\p85 &  2.41\p0.21  & 1.43\p0.08 & --0.01\p0.14\\
                        & 4515\p65 & [2.47\p0.09] & 1.41\p0.07 &  +0.04\p0.11\\ 
\object{HD 171427}      & 4875\p90 &  1.94\p0.27  & 2.29\p0.10 & --0.02\p0.13\\
\object{HD 175294}      & 4950\p85 &  2.85\p0.20  & 1.60\p0.08 &  +0.25\p0.12\\
\object{HD 175679}      & 5150\p80 &  2.94\p0.18  & 1.58\p0.07 &  +0.11\p0.10\\
                        & 5180\p50 & [2.66\p0.11] & 1.63\p0.06 &  +0.02\p0.10\\ 
\object{HD 178484}      & 4450\p85 &  1.90\p0.23  & 1.58\p0.07 & --0.32\p0.12\\        
                        & 4440\p60 & [1.96\p0.07] & 1.58\p0.06 & --0.29\p0.10\\        
\object{HD 181907}      & 4705\p90 &  2.44\p0.23  & 1.59\p0.08 & --0.11\p0.13\\
                        & 4725\p65 & [2.35\p0.04] & 1.61\p0.06 & --0.15\p0.12\\ 
\object{HD 170053}      & 4315\p90 &  1.72\p0.23  & 1.69\p0.08 & --0.14\p0.13\\
                        & 4290\p65 & [1.85\p0.16] & 1.68\p0.07 & --0.03\p0.12\\ 
\object{HD 170174}      & 5035\p80 &  2.74\p0.19  & 1.55\p0.07 & --0.01\p0.11\\
                        & 5055\p55 & [2.56\p0.05] & 1.58\p0.06 & --0.07\p0.10\\ 
\object{HD 170231}      & 5150\p80 &  2.96\p0.19  & 1.45\p0.08 &  +0.04\p0.11\\ 
                        & 5175\p55 & [2.74\p0.06] & 1.49\p0.06 & --0.03\p0.10\\ 
\object{$\alpha$ Boo}   & 4255\p85 &  1.45\p0.23  & 1.77\p0.09 & --0.67\p0.13\\
                        & 4260\p60 & [1.42\p0.08] & 1.77\p0.08 & --0.69\p0.11\\ 
\object{$\eta$ Ser}     & 4915\p80 &  3.07\p0.18  & 1.14\p0.07 & --0.21\p0.10\\
                        & 4935\p50 & [3.00\p0.05] & 1.17\p0.05 & --0.24\p0.09\\ 
\object{$\epsilon$ Oph} & 4935\p85 &  2.66\p0.21  & 1.42\p0.07 & --0.03\p0.11\\
                        & 4940\p55 & [2.64\p0.06] & 1.43\p0.06 & --0.04\p0.10\\ 
\object{$\xi$ Hya}      & 5080\p75 &  2.96\p0.17  & 1.32\p0.06 &  +0.13\p0.10\\
                        & 5095\p50 & [2.88\p0.05] & 1.34\p0.05 &  +0.10\p0.09\\ 
\object{$\beta$ Aql}    & 5100\p80 &  3.56\p0.17  & 0.97\p0.07 & --0.21\p0.11\\
                        & 5110\p50 & [3.53\p0.04] & 0.99\p0.06 & --0.22\p0.09\\ 
\hline
\end{tabular}
\tablefoot{When available, the second row shows the results with the surface gravity fixed to the seismic value for each star (given in square brackets).}
\end{table}

\section{Validation of results} \label{sect_validation} 
\subsection{Physical parameters} \label{sect_validation_parameters}
\begin{figure}
\centering
\includegraphics[width=6.5cm]{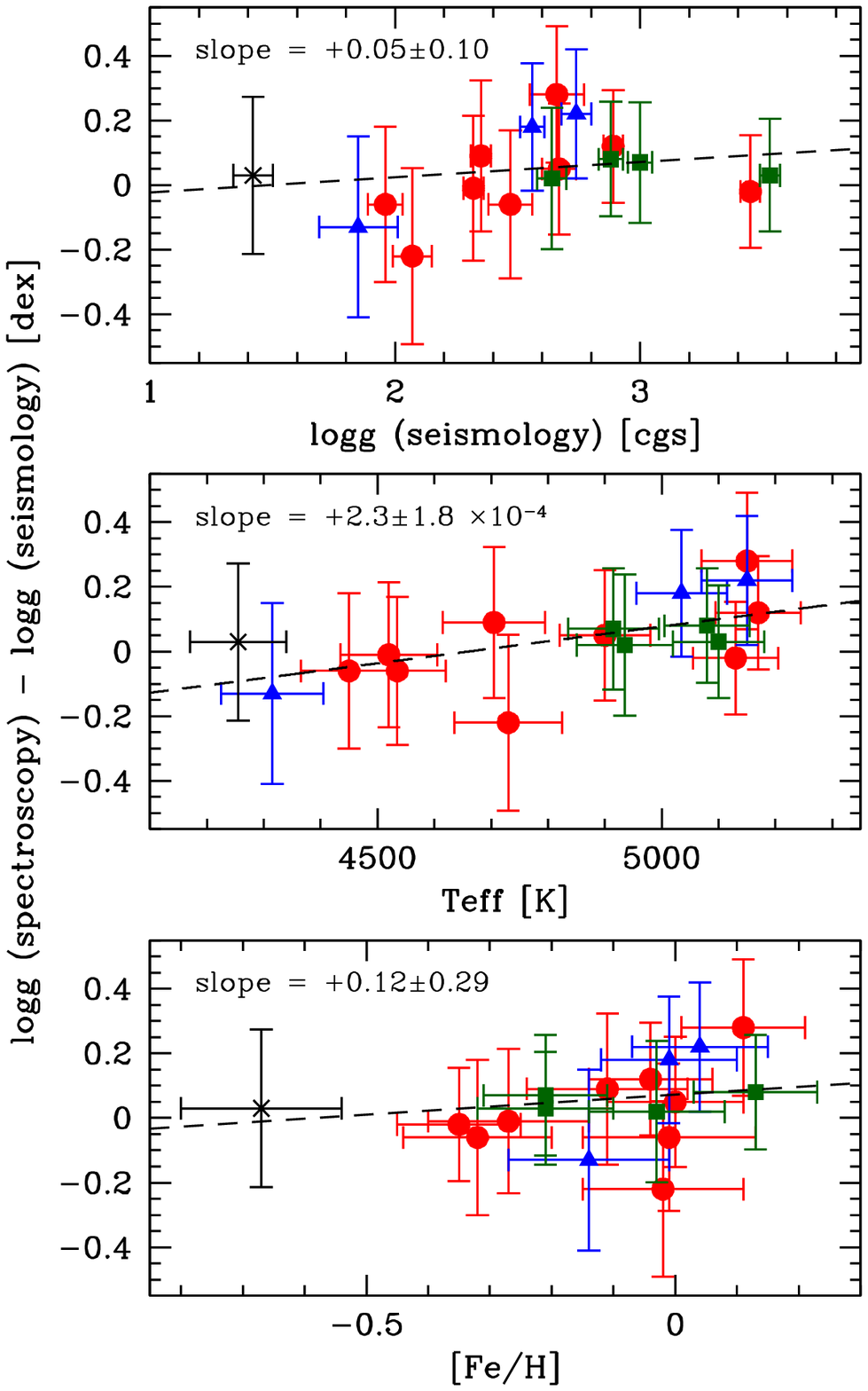}
\caption{Comparison between the surface gravities derived from ionisation balance of iron and from seismic data, as a function of the seismic \logg, \teff, and \fe \ (red dots: \c \ targets, blue triangles: stars in \object{NGC 6633}, black cross: \object{$\alpha$ Boo}, green squares: other stars used for validation). The fits weighted by the inverse variance are shown as dashed lines, and the slopes are indicated.} 
\label{fig_teff_logg_vs_loggsismo}
\end{figure}

The reliability of our spectroscopic gravities can be investigated for 17 stars in our sample by comparing with the independent estimates provided by asteroseismology (As discussed in Sect.~\ref{sect_seismic_constraints}, these values are mainly a function of the seismic observables and are only very weakly dependent on the \teff \ assumed.). As shown in Fig.~\ref{fig_teff_logg_vs_loggsismo}, these two sets of values agree well: $\langle$\logg \ [spectroscopy] \ -- \logg \ [seismology]$\rangle$ = +0.04\p0.13 dex. There is also no clear evidence of a trend as a function of the seismic gravity, effective temperature, or metallicity. None of the slopes is statistically different from zero.

\begin{figure*}
\centering
\includegraphics[width=14cm]{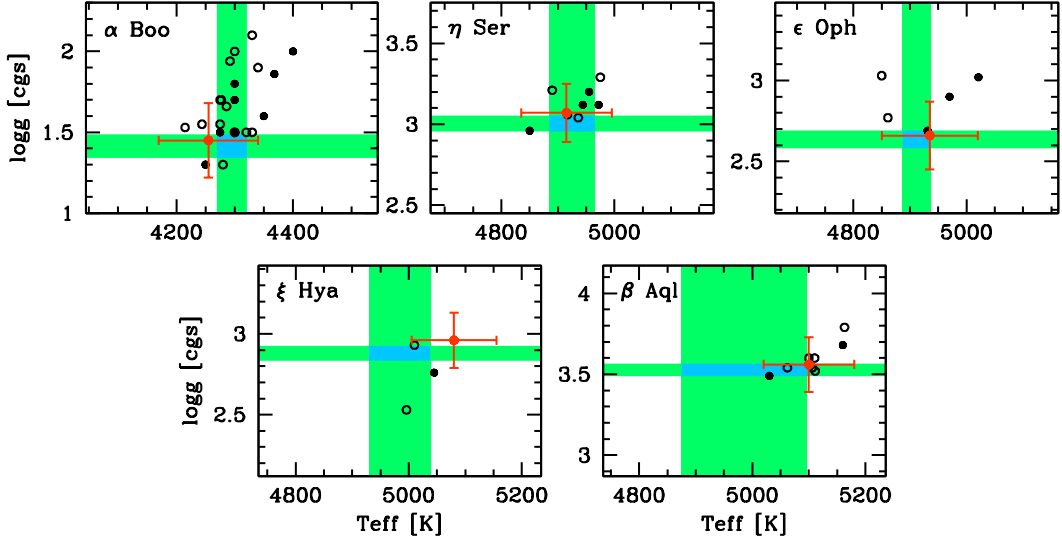}
\caption{Comparison for the stars used for validation between our \teff \ and \logg \ values (red) and those derived from interferometric and seismic data (green areas delimiting the 1-$\sigma$ error bars). Black dots: previous results from the literature (filled and open symbols: parameters derived as in the present study or using different techniques, respectively). Further details about the literature data can be found in the appendix.}
\label{fig_teff_logg_calibrators}
\end{figure*}

The five bright, well-studied red giants offer an opportunity to compare our results to the numerous ones already available in the literature. More importantly, it also allows us to investigate possible differences between our \teff \ values and the completely independent (and more likely accurate) ones derived using interferometric techniques. For \object{$\alpha$ Boo}, these measurements show a very high level of consistency, where \teff \ = 4303\p47 (Quirrenbach \etal \cite{quirrenbach96}), 4290\p30 (Griffin \& Lynas-Gray \cite{griffin_lynas_gray99}), and 4295\p26 K (Lacour \etal \cite{lacour08}). We adopt the last value in the following. The other stars only have a single value available in the literature: \object{$\eta$ Ser} (4925\p40 K; M\'erand \etal \cite{merand10}), \object{$\epsilon$ Oph} (4912\p25 K; Mazumdar \etal \cite{mazumdar09}), \object{$\xi$ Hya} (4984\p54 K; Bruntt \etal \cite{bruntt10}), and \object{$\beta$ Aql} (4986\p111 K; Bruntt \etal \cite{bruntt10}). Figure \ref{fig_teff_logg_calibrators} shows a comparison between our \logg \ and \teff \ values and those derived from seismic and interferometric data, respectively. Also shown are the previous results in the literature, which are summarised in Table \ref{tab_results_literature} (certainly not exhaustive for $\alpha$ Boo). We restrict ourselves to studies carried out over the past $\sim$25 years to select analyses based on higher-quality observational material and improved analysis techniques. We also make the distinction between fundamental parameters that are derived using similar methods as used here (excitation and ionisation equilibrium of Fe lines) or determined by other means. Our effective temperatures agree to the interferometric values within the uncertainties in all cases. Observations with the CHARA Array interferometer recently yielded \teff \ = 4577\p60 K for HD 50890 (Baines \etal \cite{baines13}). This is about 130 K cooler than our estimate.

Our results may suffer from the neglect of granulation and departures from LTE, but the checks described above suggest that the parameters determined from spectroscopy are reasonably accurate. For the range of parameters spanned by our targets (especially since they have a near-solar metallicity), departures from LTE are expected to have a limited impact on the temperature and gravity derived from excitation and ionisation balance (Lind \etal \cite{lind12}). Detailed calculations (Bergemann \etal \cite{bergemann12}; Lind \etal \cite{lind12}) available through the INSPECT database\footnote{See {\tt www.inspect-stars.net}.} indeed show that the NLTE corrections only amount to an average of 0.03 dex for a number of \ion{Fe}{i} and \ion{Fe}{ii} lines in our list that span a relatively wide range in strength and LEP. As shown by state-of-the-art 3D hydrodynamical simulations, the use of classical model atmospheres may be a more questionable assumption, although, once again, the problem is much more acute at low metallicities according to the calculations presented by Collet \etal (\cite{collet07}). The good level of consistency achieved between our spectroscopic parameters and independent estimates might indicate that the effect of granulation is not as severe at near-solar metallicity as anticipated by these particular models (see Dobrovolskas \etal \cite{dobrovolskas13}).

We show the few temperatures and gravities previously obtained for the \c \ targets in Table \ref{tab_results_literature_corot}. Our estimates and the mean values in the literature agree well. Significant differences are, however, found with respect to Valenti \& Fischer (\cite{valenti_fischer05}) and Liu \etal (\cite{liu10}) for \object{HD 170174} and \object{HD 175679}, respectively. The results of Valenti \& Fischer (\cite{valenti_fischer05}) are based on spectral synthesis techniques, while Liu \etal (\cite{liu10}) used photometric indices and isochrone fitting. 

\begin{table*}
\caption{Previous results obtained in the literature for the \c \ targets.}
\tiny
\centering
\label{tab_results_literature_corot}
\begin{tabular}{lcccccccccccc} \hline\hline
                        & \multicolumn{2}{c}{\teff \ [K]}   & \multicolumn{2}{c}{\logg \ [cgs]} & \micro \ [\kms] & \fe & [Li/H] & [C/Fe] & [N/Fe] & [O/Fe] & [Na/Fe] & Ref. \\
Star                    & Value                 & Method    & Value                 & Method   &                         & & LTE & & & &      & \\
\hline
\object{HD 43023}  & \bf{5065\p80} & \bf{E} & \bf{2.93\p0.19}     & \bf{I} & \bf{1.38\p0.07} & \bf{--0.05\p0.10} & \bf{$<$--1.40} & \bf{--0.23} & \bf{+0.39} & \bf{--0.02} & \bf{+0.10} &  \\
                   & 5140\p80      & E      & 3.10\p0.20          & I      & 1.41\p0.20      & +0.04\p0.15       & ...       & ...    & ...   & ...    & ...   & 1 \\
                   & 5105\p100     & E      & 3.08\p0.10          & I      & 1.50\p0.30      & --0.06\p0.12      & $<$--0.46 & --0.22 & +0.22 & +0.17  & +0.11 & 2 \\
                   & 5005\p75      & E      & 2.71\p0.15          & I      & 1.30\p0.08      & --0.10\p0.08      & ...       & --0.20 & ...   & ...    & +0.06 & 3\tablefootmark{a} \\
                   & 5027\p100     & P      & 2.97\p0.15          & LMT    & ...             & +0.06\p0.15       & ...       & ...    & ...   & ...    & ...   & 4 \\
                   & 4994\p100     & LDR    & 2.40\p0.20          & I      & 1.3\p0.2        & --0.13\p0.12      & $<$--0.59 & --0.27 & +0.28 & +0.00  & +0.02\tablefootmark{b} & 5 \\
                   & 5005          & S      & 2.68                & S      & ...             & --0.04            & ...       & ...    & ...   & ...    & ...   & 6 \\
\hline
\object{HD 50890}  & \bf{4730\p95} & \bf{E} & \bf{1.85\p0.26}     & \bf{I} & \bf{1.98\p0.10} & \bf{--0.02\p0.13} & \bf{--0.07}    & \bf{--0.40} & \bf{+0.58} & \bf{--0.24} & \bf{+0.51} &   \\
                   & \bf{4710\p75} & \bf{E} & \bf{[2.07\p0.08]}   & \bf{A} & \bf{1.98\p0.09} & \bf{+0.06\p0.12}  & \bf{--0.13}    & \bf{--0.43} & \bf{+0.57} & \bf{--0.20} & \bf{+0.41} &   \\ 
                   & 4665\p200     & E      & 1.4\p0.3            & I      & ...             & --0.18\p0.14      & ...       & ...    & ...   & ...    & ...   & 7 \\
\hline
\object{HD 169370} & \bf{4520\p85} & \bf{E} & \bf{2.31\p0.22}     & \bf{I} & \bf{1.42\p0.07} & \bf{--0.27\p0.13} & \bf{--1.41}    & \bf{--0.03} & \bf{+0.06} & \bf{+0.08}  & \bf{+0.03} &   \\
                   & \bf{4520\p60} & \bf{E} & \bf{[2.32\p0.04]}   & \bf{A} & \bf{1.42\p0.06} & \bf{--0.26\p0.10} & \bf{--1.41}    & \bf{--0.04} & \bf{+0.05} & \bf{+0.07}  & \bf{+0.02} &   \\
                   & 4460\p70      & E      & 2.3\p0.2            & I      & 1.3\p0.2        & --0.17\p0.10      & ...       & ...    & ...   & ...    & ...   & 8 \\
                   & 4547          & LDR    & ...                 & ...    & ...             & ...               & ...       & ...    & ...   & ...    & ...   & 9 \\
\hline
\object{HD 170053} & \bf{4315\p90} & \bf{E} & \bf{1.72\p0.23}     & \bf{I} & \bf{1.69\p0.08} & \bf{--0.14\p0.13} & \bf{+0.11}     & \bf{--0.15} & \bf{+0.42} & \bf{--0.04} & \bf{+0.27} &   \\
                   & \bf{4290\p65} & \bf{E} & \bf{[1.85\p0.16]}   & \bf{A} & \bf{1.68\p0.07} & \bf{--0.03\p0.12} & \bf{+0.06}     & \bf{--0.18} & \bf{+0.43} & \bf{--0.06} & \bf{+0.13} &   \\
                   & 4370\p60      & E      & 1.80\p0.26          & I      & 1.51\p0.08      & +0.04\p0.10       & ...       & --0.17 & +0.38 & --0.17 & --0.01\tablefootmark{b} & 10\\
\hline
\object{HD 170174} & \bf{5035\p80} & \bf{E} & \bf{2.74\p0.19}     & \bf{I} & \bf{1.55\p0.07} & \bf{--0.01\p0.11} & \bf{--0.48}    & \bf{--0.20} & \bf{+0.46} & \bf{--0.03} & \bf{+0.20} &   \\
                   & \bf{5055\p55} & \bf{E} & \bf{[2.56\p0.05]}   & \bf{A} & \bf{1.58\p0.06} & \bf{--0.07\p0.10} & \bf{--0.45}    & \bf{--0.17} & \bf{+0.52} & \bf{--0.05} & \bf{+0.28} &   \\
                   & 5015\p60      & E      & 2.85\p0.26          & I      & 1.44\p0.08      &  +0.11\p0.11      & ...       & --0.19 & +0.45 & --0.12 & --0.03\tablefootmark{b} & 10\\
                   & 4979\p72      & E      & 2.75\p0.12          & I      & 1.58\p0.10      & --0.08\p0.10      & ...       & ...    & ...   & ...    & ...   & 11\\
                   & 5245\p44      & S      & 3.11\p0.06          & S      & 0.85            &  +0.35\p0.03      & ...       & ...    & ...   & ...    & +0.07 & 12\\
\hline
\object{HD 175679} & \bf{5150\p80} & \bf{E} & \bf{2.94\p0.18}     & \bf{I} & \bf{1.58\p0.07} &  \bf{+0.11\p0.10} & \bf{+0.15}     & \bf{--0.24} & \bf{+0.46} & \bf{--0.05} & \bf{+0.17} &   \\
                   & \bf{5180\p50} & \bf{E} & \bf{[2.66\p0.11]}   & \bf{A} & \bf{1.63\p0.06} &  \bf{+0.02\p0.10} & \bf{+0.20}     & \bf{--0.19} & \bf{+0.56} & \bf{--0.09} & \bf{+0.29} &   \\
                   & 4844\p100     & P      & 2.59\p0.10          & LMT    & 1.4\p0.2        & --0.15\p0.10      & ...       & +0.04  & ...   & --0.02 & +0.17 & 13\\
\hline
\end{tabular}
\tablefoot{The rows in boldface show the results of this study (When available, the second one shows the results with the surface gravity fixed to the seismic value, which is given in square brackets for each star.). The abundances were rescaled to our adopted solar values when appropriate and whenever these were not quoted in the original paper. {\bf E}: from excitation equilibrium of the \ion{Fe}{i} lines; {\bf I}: from ionisation equilibrium of Fe; {\bf P}: from photometric data; {\bf LMT}: from estimates of the luminosity, mass, and effective temperature; {\bf S}: from spectral synthesis; {\bf LDR}: from line-depth ratios; {\bf A}: from asteroseismology. References. (1) Hekker \& Mel\'endez \cite{hekker_melendez07}; (2) Luck \& Heiter \cite{luck_heiter07}; (3) Takeda \etal \cite{takeda08}; (4) Zhao \etal \cite{zhao01}; (5) Mishenina \etal \cite{mishenina06}; (6) Soubiran \etal \cite{soubiran08}; (7) Baudin \etal \cite{baudin12}; (8) da Silva \etal \cite{da_silva06}; (9) Biazzo \etal \cite{biazzo07}; (10) Smiljanic \etal \cite{smiljanic09}; (11) Santos \etal \cite{santos09} (based on the line list of Hekker \& Mel\'endez \cite{hekker_melendez07}); (12) Valenti \& Fischer \cite{valenti_fischer05}; (13) Liu \etal \cite{liu10}. \tablefootmark{a}{As Takeda \etal caution, their oxygen abundances may not be reliable.} \tablefoottext{b}{NLTE value.}}
\end{table*}

As a final validation test, our spectrum of \object{HD 181907} was analysed with an automated tool by making use of the EWs of iron lines and MARCS model atmospheres (see Valentini \etal \cite{valentini13}). The following results were obtained: \teff \ = 4679\p54 K, \logg \ = 2.28\p0.11, \micro \ = 1.35\p0.09 \kms, and \fe \ = --0.17\p0.07. Fixing the gravity to the seismic value does not significantly change the results: \teff \ = 4735\p46 K, \micro \ = 1.48\p0.06 \ \kms, and \fe \ = --0.15\p0.05. These results are close to ours and indicate that the parameters obtained for this star are robust.

\subsection{Chemical abundances} \label{sect_validation_abundances}
For the sake of brevity, we restrict ourselves here to only discuss the metallicity scale and the chemical species used as a probe of mixing (Li, CNO, and Na). 

As seen in Table \ref{tab_results_literature}, our metallicities for the benchmark stars and those in the literature agree generally well. The cause of the rather low value we obtain for \object{$\alpha$ Boo} is unclear, but it is not attributable to a grossly overestimated microturbulence. 

Not surprisingly in view of the different parameters adopted (see above), our metallicities for two \c \ targets (Table \ref{tab_results_literature_corot}) are at odds with those of Valenti \& Fischer (\cite{valenti_fischer05}) and Liu \etal (\cite{liu10}). As is the case for $\eta$ Ser (Table \ref{tab_results_literature}), the higher \fe \ of Valenti \& Fischer (\cite{valenti_fischer05}) may stem for the most part from an adopted microturbulence that is too low. We obtain a metallicity that is higher at the $\sim$0.2 dex level for HD 50890 compared to Baudin \etal (\cite{baudin12}), but their gravity was poorly constrained.

Stars in the young open cluster \object{NGC 6633} offer an opportunity to assess the reliability of our metallicities because the values we obtain for the three likely members should be identical within the uncertainties. Santos \etal (\cite{santos09}) and Smiljanic \etal (\cite{smiljanic09}) obtained --0.08 and +0.07 for the mean metallicity of the cluster based on the analysis of three and two red giants, respectively. Table \ref{tab_results_literature_corot} shows a comparison between our results and those they obtained for the two stars in common: \object{HD 170053} and \object{HD 170174}.\footnote{We consider the results of Santos \etal (\cite{santos09}) based on the line list of Hekker \& Mel\'endez (\cite{hekker_melendez07}), which, as they discuss, is more appropriate for red giants. Their metallicities were also rescaled to the value they inferred from the analysis of a solar reflection spectrum.} On the other hand, Jeffries \etal (\cite{jeffries02}) obtained \fe \ = --0.10\p0.08 for 10 FG dwarfs and determined that NGC 6633 is significantly more metal poor than the Hyades (by $\sim$0.21 dex) or the Pleiades (by $\sim$0.07 dex), which should be regarded as a more robust result. We obtain --0.04\p0.04 for the mean metallicity of the cluster when considering the seismic constraints for the three likely members. It could be noted that HD 170053 is no longer discrepant in terms of iron content when the gravity is fixed to the more accurate seismic value (see Table \ref{tab_parameters}).

With regard to the Li, CNO, or Na abundances, there is a good overall agreement for the benchmark stars with respect to previous studies (listed in Table \ref{tab_results_literature}). There is, however, some evidence of slightly larger C abundances in our case compared to Bruntt \etal (\cite{bruntt10}) for three stars in common. Our upper limits for the Li abundance in two stars (\object{$\eta$ Ser} and \object{$\beta$ Aql}) also appear inconsistent with their report of a detection. Our value for $\xi$ Hya is also $\sim$0.3 dex lower.

To the best of our knowledge, only four \c \ targets have previous determinations of the abundances of these elements (\object{HD 43023}, \object{HD 170053}, \object{HD 170174}, and \object{HD 175679}). Although there are generally only small differences compared to the literature data (Table \ref{tab_results_literature_corot}), three discrepancies are worth pointing out: (1) the larger C abundance found in \object{HD 175679} by Liu \etal (\cite{liu10}); (2) the differences at the $\sim$0.2 dex level with the N and O abundances of Luck \& Heiter (\cite{luck_heiter07}) for HD 43023; and (3) the lower [Na/Fe] ratios reported by Smiljanic \etal (\cite{smiljanic09}) for two stars in NGC 6633. This may not arise from differences in the adopted parameters (see Table \ref{tab_errors}), and the cause of this disagreement is unclear. Our abundances can be revised downwards by $\sim$0.08 dex if we adopt their \loggf \ values and take into account their NLTE corrections. However, using their EWs would lead to abundances higher by $\sim$0.09 dex.

Smiljanic \etal (\cite{smiljanic09}) report a \iso \ isotopic ratio of 18\p8 and 21\p7 for \object{HD 170053} and \object{HD 170174}, respectively, which we are unable to confirm because no \element[][13]{C} lines are covered by our observations. The only point of comparison for this quantity is provided by \object{$\alpha$ Boo}. Our value (\iso \ = 8\p1) closely agrees with those in the literature, which range from 6 to 10 (e.g., Pilachowski \etal \cite{pilachowski97}; Pavlenko \cite{pavlenko08}; Abia \etal \cite{abia12}).

\section{Correction for the chemical evolution of the Galaxy} \label{sect_correction_chemical_evolution}
\begin{figure*}
\centering
\includegraphics[width=18.5cm]{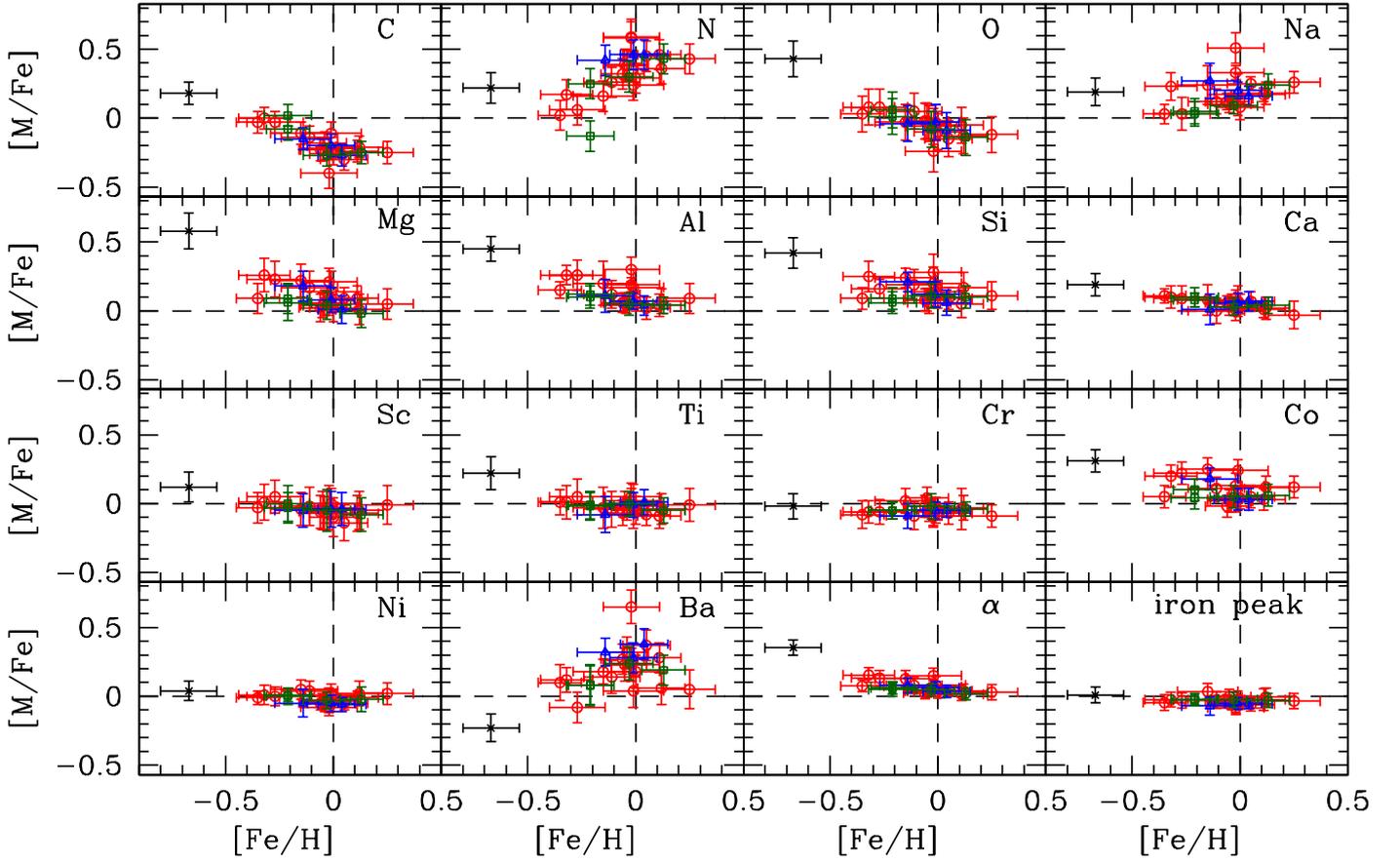}
\caption{Abundance ratios with respect to iron as a function of \fe. The results have been obtained using the spectroscopic gravities. Same symbols as in Fig.~\ref{fig_teff_logg_vs_loggsismo}. The mean abundance ratio of the $\alpha$-synthesised elements is defined as the unweighted mean of the Mg, Si, Ca, and Ti abundances. For the mean abundance of the iron-peak elements, we considered Cr and Ni.}
\label{fig_abundances_vs_Fe}
\end{figure*}

As shown in Fig.~\ref{fig_abundances_vs_Fe}, the abundance ratios with respect to iron of several elements (e.g., Ca) exhibit a clear trend with \fe. The larger abundance ratios, which are observed as \fe \ decreases, is a well-known feature of Galactic disk stars and arises from the different relative proportion of Type Ia/II supernovae yields and the various amounts of material lost by stellar winds from AGB or massive stars along the history of the Galaxy (In contrast, the iron-peak elements closely follow Fe as expected.). This behaviour is also observed in our sample for some mixing indicators and indicates that these abundances are not only affected by mixing processes but also -- and perhaps to a larger extent -- by the chemical evolution of the Galaxy. To disentangle the contribution of these two phenomena to first order, we corrected these abundances by removing the metallicity trend found in dwarfs of the Galactic thin disk by Ecuvillon \etal (\cite{ecuvillon04a}, \cite{ecuvillon06}) for C and O and by Reddy \etal (\cite{reddy03}) for Na, as follows:

\begin{eqnarray}
\rm [C/Fe]_{\rm corr} = [C/Fe] + 0.39 \fe,\\
\rm [O/Fe]_{\rm corr} = [O/Fe] + 0.50 \fe,\\
\rm [Na/Fe]_{\rm corr} = [Na/Fe] + 0.13 \fe.
\label{equ5}
\end{eqnarray}

\begin{figure}
\centering
\includegraphics[width=6.5cm]{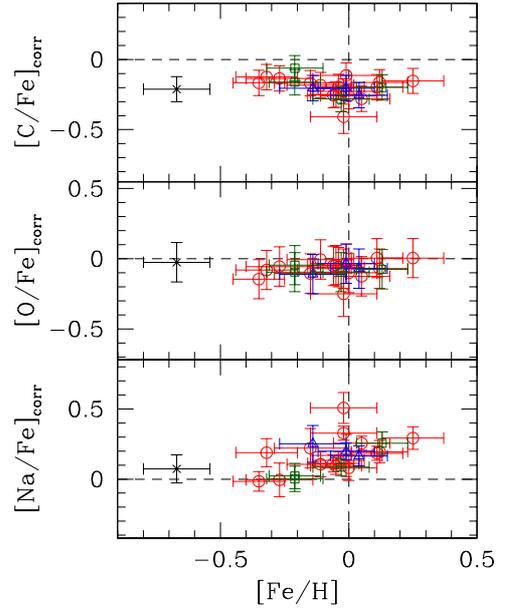}
\caption{Corrected abundance ratios with respect to iron for C, O, and Na, as a function of \fe. The results have been obtained using the spectroscopic gravities. Same symbols as in Fig.~\ref{fig_teff_logg_vs_loggsismo}.}
\label{fig_abundances_vs_Fe_corr}
\end{figure}

\begin{figure*}
\centering
\includegraphics[width=13cm]{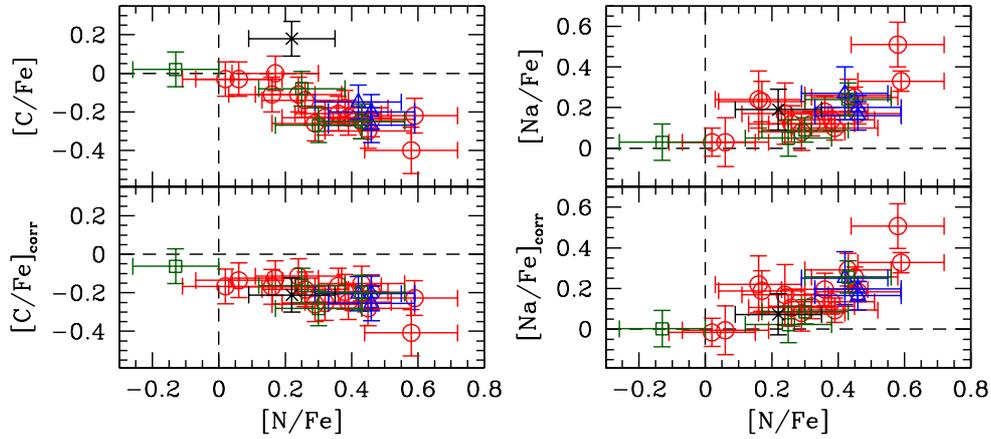}
\caption{{\it Top and bottom left panels:} [C/Fe] as a function of [N/Fe] prior and after correction for the effects of the chemical evolution of the Galaxy, respectively. {\it Top and bottom right panels:} [Na/Fe] as a function of [N/Fe]. The results have been obtained using the spectroscopic gravities. Same symbols as in Fig.~\ref{fig_teff_logg_vs_loggsismo}.}
\label{fig_NC_NNa_corr_not_corr}
\end{figure*}

Very similar slopes have been reported in the literature (see Reddy \etal \cite{reddy03}, Luck \& Heiter \cite{luck_heiter06}, and da Silva \etal \cite{da_silva11} in the case of C). For oxygen, we only considered the results based on the \ion{[O}{i]} $\lambda$6300 line (Ecuvillon \etal \cite{ecuvillon06}). For Na, we assumed that the trend found by Reddy \etal (\cite{reddy03}) extends to supersolar metallicities. No corrections are applied to the N abundances because no trend with \fe \ is discernible for this element (Reddy \etal \cite{reddy03}; Ecuvillon \etal \cite{ecuvillon04b}). For the thick disk star \object{$\alpha$ Boo}, we use the abundance offsets found by Reddy \etal (\cite{reddy06}) between kinematically-selected samples of thin and thick disk stars (The $\Delta$[X/Fe] values in their Table 7 were subtracted from the corrected ratios defined above.). As expected, this procedure efficiently erases the trends previously found for C and O (Fig.~\ref{fig_abundances_vs_Fe_corr}), as well as for [N/C] and [N/O] (not shown). Figure \ref{fig_NC_NNa_corr_not_corr} shows that the relation between [N/Fe] and [C/Fe] tightens after correction and that $\alpha$ Boo behaves as the other red giants. To first approximation, these corrected abundances (Tables \ref{tab_abundances_mixing} and \ref{tab_abundances_ratios}) are now free of the effects related to the nucleosynthesis history of the ISM and are, hence, more reliable probes of the mixing processes operating in our targets. As for nitrogen, note that there is no clear trend between the barium abundances (discussed below) and the metallicity (Reddy \etal \cite{reddy03}).

\section{Discussion of some key observational results} \label{sect_key_results}
\begin{figure*}
\begin{minipage}[t]{0.48\textwidth}
\includegraphics[width=\textwidth]{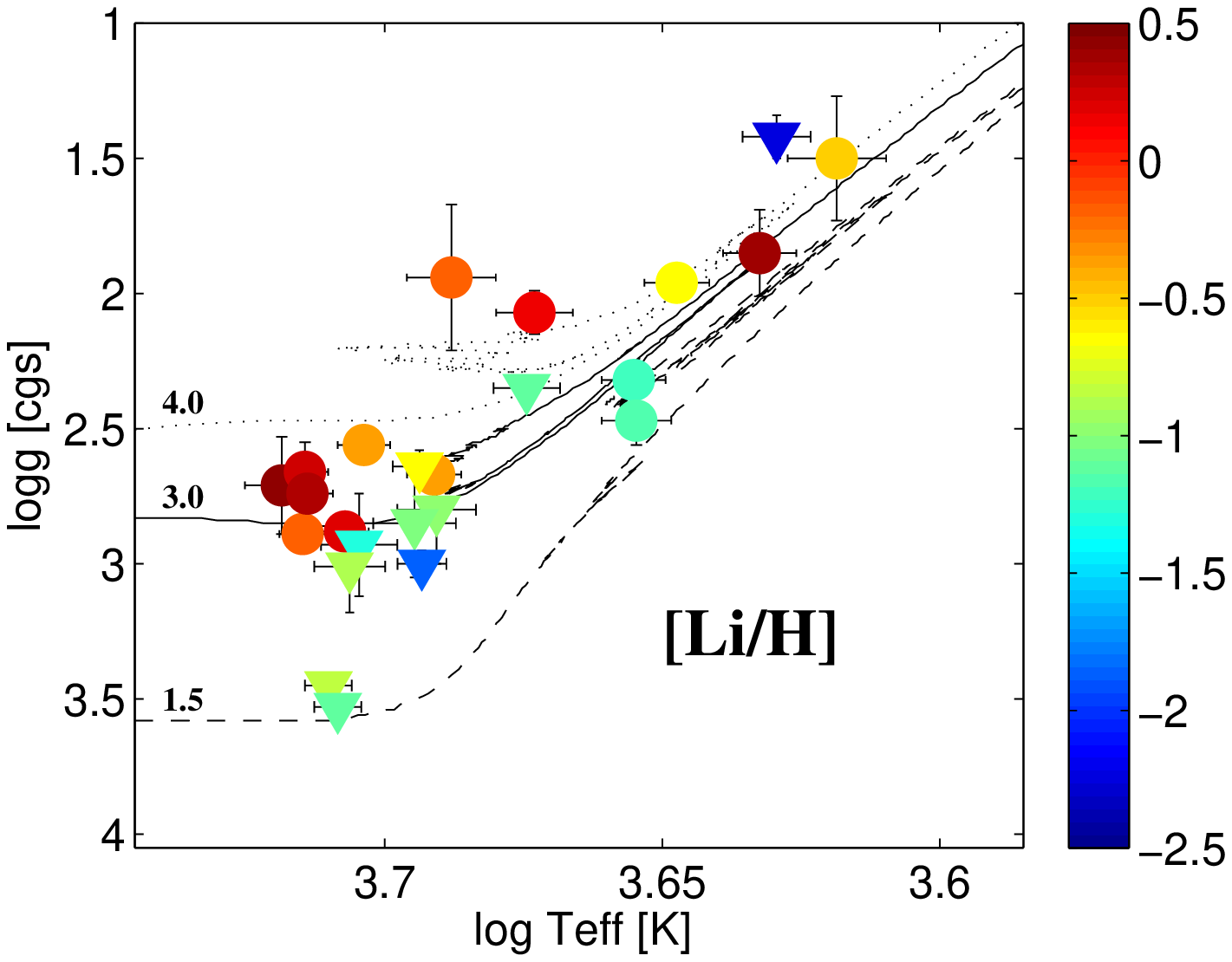}
\end{minipage}
\begin{minipage}[t]{0.48\textwidth}
\includegraphics[width=\textwidth]{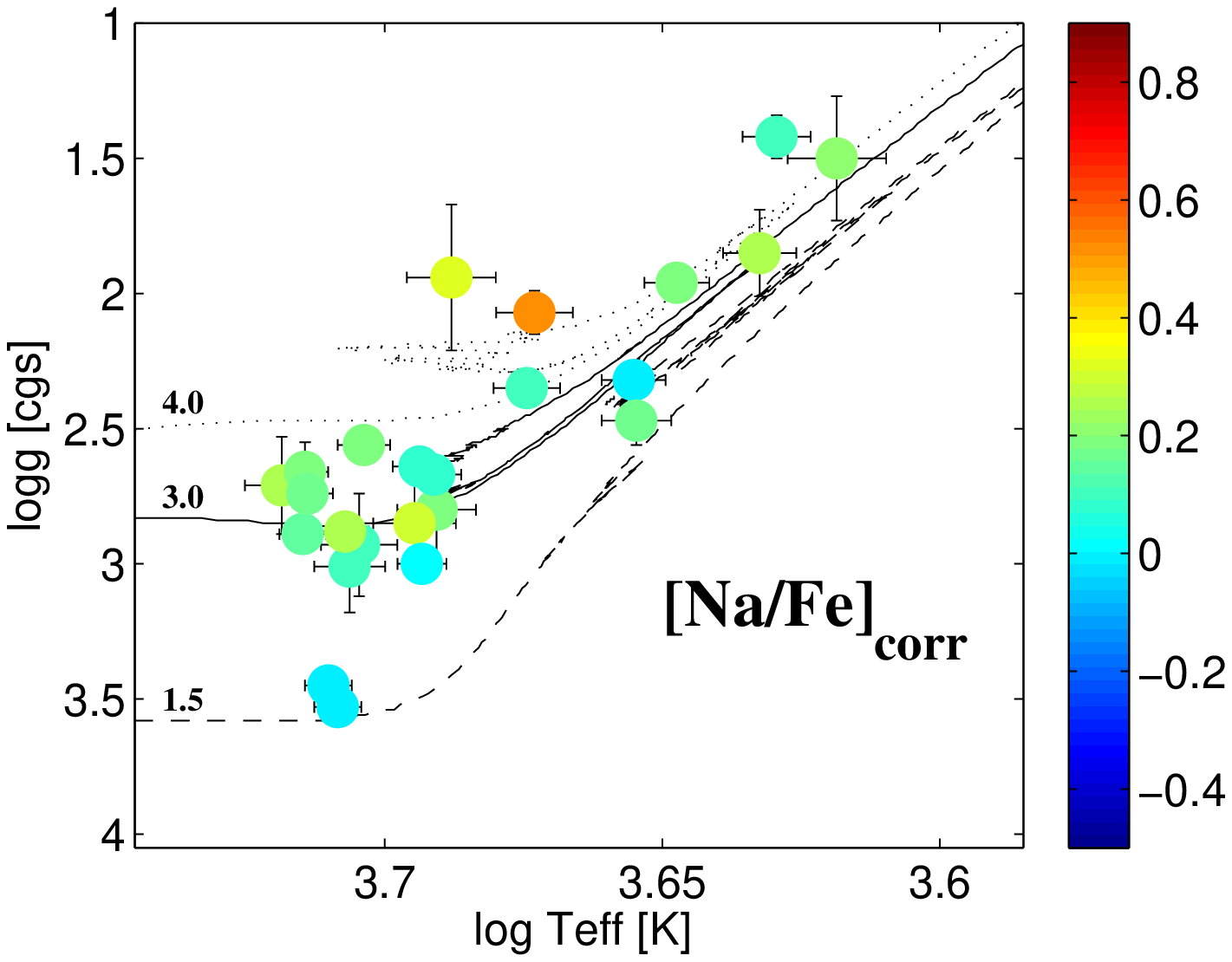}
\end{minipage}
\begin{minipage}[b]{0.48\textwidth}
\includegraphics[width=\textwidth]{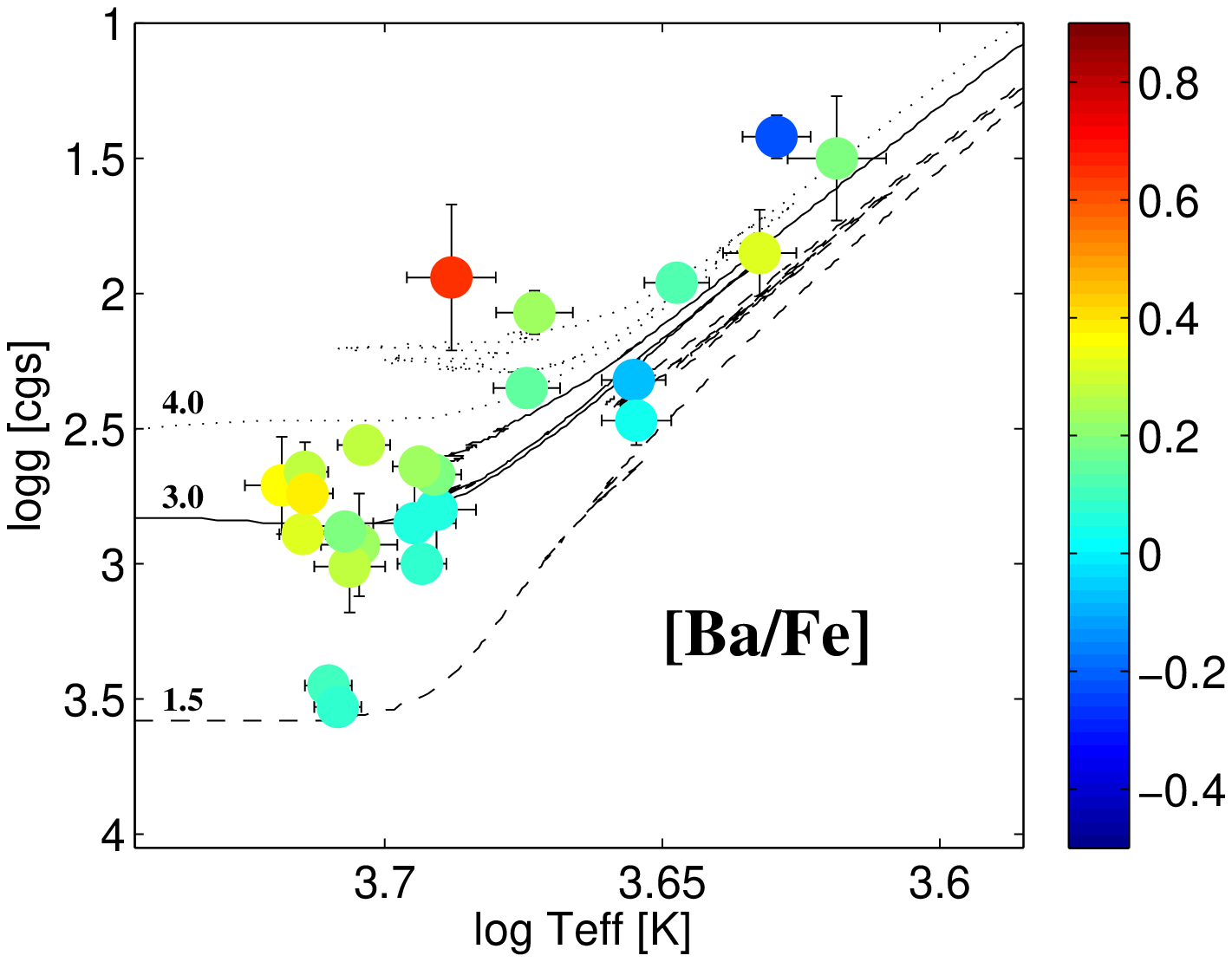}
\end{minipage}
\hspace*{0.6cm}
\begin{minipage}[b]{0.48\textwidth}
\includegraphics[width=\textwidth]{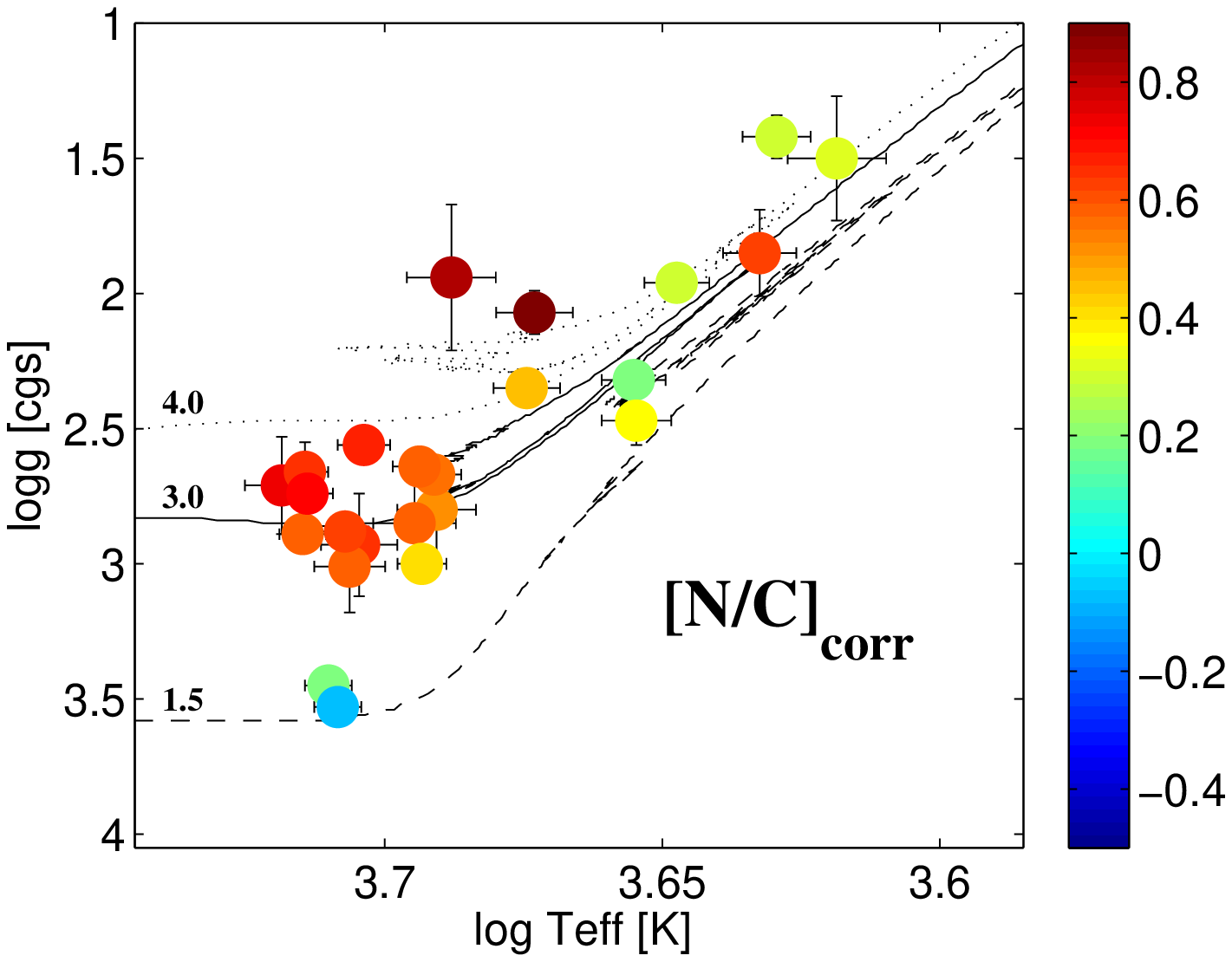}
\end{minipage}
\caption{Variations of some key abundance ratios across the $\log$\teff-\logg \ plane. For the NLTE lithium abundances, the downward-pointing triangles denote upper limits. The predictions at solar metallicity of evolutionary models for masses of 1.5, 3, and 4 M$_{\odot}$ are overplotted for illustrative purposes. Same tracks as in Fig.~\ref{fig_logg_logTeff_starevol}, except that the evolutionary phase, is not colour coded.} 
\label{fig_Li_Na_Ba_NC}
\end{figure*}

We defer to a detailed comparison between our Li and \iso \ abundance data and the predictions of theoretical models that incorporate rotational mixing and thermohaline instabilities to a forthcoming paper. However, let us discuss some salient results obtained for other species here. Adopting the seismic gravities leads to variations in the abundances of all elements that are comparable to the uncertainties (Tables \ref{tab_abundances} to \ref{tab_abundances_ratios}). In the general discussion which follows, we therefore only consider the results obtained in a consistent manner for all the stars using the spectroscopic gravities. Figure \ref{fig_Li_Na_Ba_NC} illustrates the complex behaviour of some key abundance ratios during the red-giant phase.

As has been known for a long time, ordinary red giants are C-poor and N-rich objects (e.g., Lambert \& Ries \cite{lambert_ries81}). Figure \ref{fig_NC_NNa_corr_not_corr} shows that carbon is increasingly depleted as nitrogen is enhanced, which illustrates the differing degrees of CN-cycled material transported at the surface of our targets. This relation is quantitatively very similar for other red-giant samples (Luck \& Heiter \cite{luck_heiter07}; Smiljanic \etal \cite{smiljanic09}; Tautvai\v sien\.e \etal \cite{tautvaisiene10}, \cite{tautvaisiene13}) once the corrections discussed above (Sect.~\ref{sect_correction_chemical_evolution}) are applied. The corrected oxygen abundances are identical within the uncertainties for all the targets (Fig.~\ref{fig_abundances_vs_Fe_corr}). There is therefore no observational evidence of ON-cycled material transported to the surface despite the detection of the products of the NeNa cycle in some stars (see below). The abundance differences indicative of an oxygen depletion might be buried in the noise.

The sodium abundance can be altered during the red-giant phase, as material processed by the NeNa cycle at temperatures in excess of $\sim$ 2.5 $\times$ 10$^7$ K is transported to the surface (e.g., Langer \etal \cite{langer93}). However, this is only expected to occur prior to the AGB phase for intermediate-mass ($M$ $\ga$ 2 M$_{\odot}$) stars in accordance with the solar abundance ratios found in stars with lower masses (e.g., Gratton \etal \cite{gratton00}). Evolutionary models predict a sodium overabundance of the order of 0.2 dex due to the first dredge-up but that can reach up to 0.8 dex for very high rotational velocities on the ZAMS (Charbonnel \& Lagarde \cite{charbonnel_lagarde10}). Smiljanic \etal (\cite{smiljanic09}) investigated the Na abundance in giants belonging to young open clusters and found a slight increase as a function of the turn-off mass (see also Takeda \etal \cite{takeda08} and Liu \etal \cite{liu10} for field giants). The largest value in our sample ([Na/Fe]$_{\rm corr}$ $\sim$ +0.5 dex) is found for HD 50890, which is a young (155-180 Myrs) and massive (3-5 M$_{\odot}$) star, according to Baudin \etal (\cite{baudin12}). Interestingly, there is also suggestive evidence that this star is spinning fast (Sect.~\ref{spectral_synthesis_broadening}). There has been some debate in the recent literature (e.g., Smiljanic \cite{smiljanic12}, and references therein) concerning the possible existence of large (up to 0.6 dex) sodium overabundances in red giants. Although models can accommodate such high values, two results make us believe that the large excesses we observe are real. First, solar values are found for some low-mass subgiants (e.g., HD 170008 or $\beta$ Aql; Fig.~\ref{fig_logg_logTeff_starevol}) for which no enrichment is expected. Second, there is clear evidence that the Na and N abundances increase in parallel (Fig.~\ref{fig_NC_NNa_corr_not_corr}). To our knowledge, this is the first time that such a trend is so clearly uncovered (see Mishenina \etal \cite{mishenina06}). It should be noted that the predicted NLTE corrections for Na (according to the calculations of Lind \etal \cite{lind11} and as quoted in the INSPECT database) are fairly uniform within our sample and similar to that in the Sun (about --0.1 dex). Our abundances (that are relative to solar) should hence be little affected by the neglect of NLTE effects. In the same vein, granulation effects are anticipated to have a limited impact on the strength of the \ion{Na}{i} lines used (Collet \etal \cite{collet07}).

The MgAl cycle operates above such high temperatures ($\sim$ 7 $\times$ 10$^7$ K; Langer \etal \cite{langer97}) that Al is not expected to be produced at the expense of Mg prior to the AGB phase. The variation of the Al abundance within our sample is comparable to the uncertainties, and there is a fortiori no evidence of an Al excess accompanied by an Mg depletion when similar corrections as described in Sect.~\ref{sect_correction_chemical_evolution} are applied using the data of Reddy \etal (\cite{reddy03}).

It is well known that a strong barium overabundance in giant stars can result from a previous episode of mass transfer with a formerly thermally pulsing AGB (TP-AGB) companion that is now a white dwarf (Alves-Brito \etal \cite{alves_brito11}, and references therein). Barium stars are C-rich (e.g., Barbuy \etal \cite{barbuy92}), and it is therefore very unlikely that this phenomenon plays a significant role in our sample. The Ba excess we occasionally observe may instead be attributable to a young age, as found in field dwarfs of the solar neighbourhood or members of open clusters (e.g., Bensby \etal \cite{bensby07}; D'Orazi \etal \cite{dorazi09}; da Silva \etal \cite{da_silva12}). In this respect, it may be regarded as significant that high abundances are found in the young, massive star HD 50890 and in the likely members of NGC 6633 with an estimated age in the range 450--575 Myrs according to Smiljanic \etal (\cite{smiljanic09}) and van Leeuwen (\cite{van_leeuwen09}). It is interesting to note that HD 170031 has a Ba abundance lower by about a factor 2 than these three stars, which strengthens the case for a non-membership (Sect.~\ref{sect_targets}).\footnote{In the context of the binary scenario mentioned above, very high precision and near-continuous radial-velocity measurements collected over about one week with SOPHIE and HARPS do not reveal the presence of a companion for the stars in NGC 6633 (Poretti et al., in preparation).} The largest Ba abundance is found in the massive star HD 171427 (see Fig.~\ref{fig_logg_logTeff_starevol}). Once again, this can be interpreted as arising from a time evolution along the history of the Galaxy, which is of relative proportion to the yields of the various stellar populations. The combined effect of departures from LTE and time-dependent/spatial variations in the atmospheric structure due to convection is relatively small in very low-metallicity environments such as globular clusters ($\Delta$[Ba/Fe] $\sim$ 0.15 dex compared to a 1D LTE analysis; Dobrovolskas \etal \cite{dobrovolskas12}), and the magnitude is likely much lower in our sample. 

We conclude this discussion by a word of caution. The discrepancies between different abundance indicators discussed in Sect.~\ref{spectral_synthesis_atomic_data} (see in particular Fig.~\ref{fig_diff_C2_5380}) warn us that some effects, which are not incorporated in our analysis (e.g., departures from LTE), may bias our results, especially in the objects with the most extended and diluted atmospheres. Although we attempted to evaluate the impact for some key chemical species as far as possible and concluded that the trends observed (e.g., between [N/Fe] and [Na/Fe]) are likely of physical origin, it should be kept in mind that these arguments rest on heterogeneous and often fragmentary calculations in the literature. 

\section{Concluding remarks and perspectives} \label{sect_conclusion}
We are entering a new era where spectroscopic and asteroseismic data of superb quality can be combined to provide a global view of red giants in unprecedented detail. Astrometric data from the {\it Gaia} space mission and new long-baseline interferometric facilities will soon also open new perspectives. On the other hand, major advances are being made on various theoretical aspects (e.g., Charbonnel \& Lagarde \cite{charbonnel_lagarde10}; Ludwig \& Ku\v{c}inskas \cite{ludwig12}). 

Our study is an effort to ultimately fully characterise the stars in our sample. This may be achieved for those for which detailed seismic information is available, such as HD 50890 (Baudin \etal \cite{baudin12}) or HD 181907 (Carrier \etal \cite{carrier10}; Miglio \etal \cite{miglio10}). The modelling of the \c \ data for other stars in the seismology fields is underway (e.g., Barban \etal \cite{barban14}).

The extent of mixing experienced by each of our targets results from the combined action of different physical processes (convective and rotational mixing, as well as thermohaline instabilities) whose relative efficiency (or merely occurrence) is a complex function of their evolutionary status, mass, metallicity, and rotational history. A preliminary comparison with evolutionary models supports the widespread occurrence of mixing processes other than convection in our sample. We will take advantage of the asteroseismic constraints to provide in a forthcoming paper (Lagarde et al., in preparation) a thorough interpretation of our abundance data based on theoretical models incorporating the three mechanisms mentioned above (Charbonnel \& Lagarde \cite{charbonnel_lagarde10}).

Finally, dramatic advances may be expected from the analysis of the large population of red-giant stars monitored by the {\it Kepler} satellite. The various evolutionary sequences can clearly be distinguished from asteroseismic diagnostics (e.g., Stello \etal \cite{stello13}; Montalb\'an \etal \cite{montalban13}), which opens up the possibility of mapping out the evolution of the mixing indicators during the shell-hydrogen and core-helium burning phases for a very large number of stars. An inspection of the spectra obtained by Thygesen \etal (\cite{thygesen12}) for 82 red giants in the {\it Kepler} field (mostly obtained with FIES installed on the Nordic Optical Telescope; NOT) shows that these data are not of sufficient quality to confidently measure the generally very weak Li and $^{13}$CN features. Although demanding in terms of telescope time, such a study is amenable for the brightest targets, which can be observed on larger facilities, and may be particularly rewarding.

\begin{acknowledgements}
T.M. acknowledges financial support from Belspo for contract PRODEX GAIA-DPAC. A.M. and N.L. acknowledge fund from the Stellar Astrophysics Centre provided by The Danish National Research Foundation (Grant agreement no.: DNRF106). N.L. acknowledges financial support from the Swiss National Fund. J.M. and M.V. acknowledge financial support from Belspo for contract PRODEX CoRoT. M.R. acknowledges financial support from the FP7 project {\it SPACEINN: Exploitation of Space Data for Innovative Helio- and Asteroseismology}. E.P. and M.L. acknowledge financial support from the PRIN-INAF 2010 ({\it Asteroseismology: looking inside the stars with space- and ground-based observations}). S.H. has received funding from the European Research Council under the European Communities Seventh Framework programme (FP7/2007-2013)/ERC grant agreement Stellar Ages \#338251. T.K. acknowledges financial support from the Austrian Science Fund (FWF P23608). We would like to thank the referee, U. Heiter, for a careful reading of the manuscript and valuable comments. We are grateful to N. Grevesse for enlightening discussions, F. Castelli for her assistance with the interpolation of the ODF tables, K. Lind for the program to interpolate the NLTE Li corrections, as well as C. Pereira and R. Smiljanic for their help with the molecular data. This research made use of the INSPECT database (version 1.0), NASA's Astrophysics Data System Bibliographic Services, the SIMBAD database operated at CDS, Strasbourg (France), as well as the Vienna Atomic Line Database (VALD) and the WEBDA database, both operated at the Institute for Astronomy of the University of Vienna.
\end{acknowledgements}

\appendix

\section{Literature results for the benchmark stars.}\label{sect_appendix}

Table \ref{tab_results_literature} provides the literature data for the benchmark stars.

\begin{table*}
\centering
\caption{Atmospheric parameters, iron content, and abundances of mixing indicators found in the literature for the benchmark stars.}
\label{tab_results_literature}
\begin{tabular}{lcccccccccccc} \hline\hline
                        & \multicolumn{2}{c}{\teff \ [K]}   & \multicolumn{2}{c}{\logg \ [cgs]} & \micro \ [\kms] & \fe & [Li/H] & [C/Fe] & [N/Fe] & [O/Fe] & [Na/Fe] & Ref. \\
Star                    & Value                 & Method    & Value                 & Method   &                         & & LTE & & & &      & \\
\hline 
\object{$\alpha$ Boo}   & \bf{4255}             & \bf{E}    & \bf{1.45}             & \bf{I}   & \bf{1.77} & \bf{--0.67}             & \bf{$<$--2.50} & \bf{+0.18} & \bf{+0.22} & \bf{+0.43} & \bf{+0.19} &\\
                        & \bf{4260}             & \bf{E}    & \bf{[1.42]}           & \bf{A}   & \bf{1.77} & \bf{--0.69}             & \bf{$<$--2.45} & \bf{+0.19} & \bf{+0.22} & \bf{+0.44} & \bf{+0.21} & \\
                        & 4400                  & E         & 2.0                   & I        & 1.5       & --0.51                  & ... & ... & ... & ... & +0.18 & 1\\
                        & 4250                  & E         & 1.3                   & I        & 1.7       & --0.68                  & ... & +0.00 & +0.28 & +0.45 & +0.20 & 2\\
                        & 4300                  & E         & 1.5                   & I        & 1.7       & --0.63                  & ... & ... & ... & ... & ... & 3\\
                        & 4300                  & E         & 1.7                   & I        & 1.6       & --0.72                  & ... & ... & ... & +0.42 & +0.32 & 4\\
                        & 4350                  & E         & 1.6                   & I        & 1.6       & --0.58                  & ... & ... & ... & ... & ... & 5\\
                        & 4275                  & E         & 1.5                   & I        & 1.6       & --0.58                  & ... & ... & ... & ... & ... & 6\tablefootmark{a}\\
                        & 4300                  & E         & 1.8                   & I        & 1.6       & --0.57                  & ... & ... & ... & ... & +0.05 & 7\\
                        & 4368                  & E         & 1.86                  & I        & 1.86      & --0.62                  & ... & ... & ... & ... & ... & 8\\
\cline{2-13}                                                                                                
                        & 4330                  & P         & 1.5                   & I        & 1.5       & --0.38                  & ... & ... & ... & ... & ... & 9\\
                        & 4340                  & P         & 1.9                   & LMT      & ...       & --0.37                  & $<$--1.89 & ... & ... & ... & ... & 10\\
                        & 4300                  & P         & 2.0                   & P        & 1.5       & --0.69                  & ... & ... & ... & ... & ... & 11\\
                        & 4330                  & P         & 2.1                   & LMT      & 1.6       & --0.58                  & ... & ... & ... & ... & --0.10 & 12\\
                        & 4300                  & S         & 1.5                   & S        & 1.7       & --0.5                   & ... & +0.04 & +0.43 & +0.50 & +0.30 & 13\\
                        & 4280                  & P         & 1.3                   & I        & 1.4       & --0.54                  & $<$--1.91 & ... & ... & ... & +0.10 & 14\\
                        & 4292                  & F         & 1.94                  & F        & ...       & --0.51                  & ... & ... & ... & ... & ... & 15\\
                        & 4320                  & F         & 1.50                  & F        & 1.7       & --0.5                   & ... & --0.06 & +0.19 & +0.35 & ... & 16\\
                        & 4275                  & P         & 1.55                  & LMT      & 1.65      & --0.54                  & ... & --0.05 & +0.35 & +0.47 & ... & 17\\
                        & 4277                  & S         & 1.7                   & S        & ...       & --0.47                  & ... & ... & ... & ... & ... & 18\\
                        & 4300                  & E         & 1.50                  & LMT      & 1.5       & --0.49                  & ... & ... & ... & ... & +0.04 & 19\\
                        & 4286                  & F         & 1.66                  & LMT      & 1.74      & --0.52                  & ... & +0.43 & ... & +0.50 & +0.11 & 20\\
                        & 4215                  & E         & 1.53                  & LMT      & 1.65      & --0.60                  & ... & ... & ... & +0.67\tablefootmark{b} & ... & 21\\
                        & 4275                  & P         & 1.7                   & LMT      & 1.85      & --0.52                  & ... & --0.04 & +0.24 & +0.33 & ... & 22\\
                        & 4244                  & E         & 1.55                  & LMT      & 1.61      & --0.55                  & ... & ... & ... & ... & ... & 23\\
\hline                                                                                                     
\object{$\eta$ Ser}     & \bf{4915}             & \bf{E}    & \bf{3.07}             & \bf{I}   & \bf{1.14} & \bf{--0.21}             & \bf{$<$-2.00} & \bf{--0.08} & \bf{+0.25} & \bf{+0.06} & \bf{+0.05} & \\
                        & \bf{4935}             & \bf{E}    & \bf{[3.00]}           & \bf{A}   & \bf{1.17} & \bf{--0.24}             & \bf{$<$-1.90} & \bf{--0.07} & \bf{+0.28} & \bf{+0.06} & \bf{+0.10} & \\
                        & 4850                  & E         & 2.96                  &   I      & 1.04      & --0.11                  & --0.91 & --0.18 & ... & ... & --0.05 & 24\\
                        & 4972                  & E         & 3.12                  &   I      & 1.17      & --0.18                  & ... & --0.03 & ... & ... & +0.05 & 25\tablefootmark{c}\\
                        & 4955                  & E         & 3.20                  &   I      & 1.33      & --0.15                  & ... & ... & ... & ... & ... & 26\\
                        & 4944                  & E         & 3.12                  &   I      & 1.25      & --0.17                  & ... & ... & ... & ... & ... & 27\\
\cline{2-13}                                                                                                
                        & 4890                  & P         & 3.21                  & LMT      & 2.1       & --0.25                  & ... & ... & ... & ... & ... & 28\\
                        & 4809                  & LDR       &  ...                  & ...      & ...       & ...                     & ... & ... & ... & ... & ... & 29\\
                        & 4975                  & S         & 3.29                  &   S      & 0.85      & --0.05                  & ... & ... & ... & ... & --0.05 & 30\\
                        & 4917                  & B         & 3.06                  &   I      & 1.14      & --0.20                  & ... & ... & ... & ... & ... & 31\\
                        & 4936                  & P         & 3.04                  & LMT      & 1.1       & --0.27                  & ... & ... & ... & ... & ... & 32\\
\hline 
\object{$\epsilon$ Oph} & \bf{4935}             & \bf{E}    & \bf{2.66}             & \bf{I}   & \bf{1.42} & \bf{--0.03}             & \bf{$<$--0.80} & \bf{--0.27} & \bf{+0.30} & \bf{--0.08} & \bf{+0.09} & \\
                        & \bf{4940}             & \bf{E}    & \bf{[2.64]}           & \bf{A}   & \bf{1.43} & \bf{--0.04}             & \bf{$<$--0.80} & \bf{--0.26} & \bf{+0.31} & \bf{--0.08} & \bf{+0.10} \\
                        & 4931                  & E         & 2.69                  &   I      & 1.34      & --0.07                  & ... & --0.27 & ... & ... & +0.05 & 25\tablefootmark{c}\\
                        & 4970                  & E         & 2.90                  &   I      & 1.52      & --0.07                  & ... & ... & ... & ... & ... & 26\\
                        & 5021                  & E         & 3.02                  &   I      & 1.54      & --0.01                  & $<$--1.26 & --0.28 & +0.38 & +0.06 & +0.07 & 33\\
\cline{2-13}                                                                                                
                        & 4850                  & P         & 3.03                  & LMT      & 2.2       & --0.08                  & ... & ... & ... & ... & ... & 28\\
                        & 4861                  & P         & 2.77                  & LMT      & 1.4       & --0.08                  & ... & ... & ... & --0.01\tablefootmark{b} & +0.12 & 34\\
\hline                                                                                                     
\object{$\xi$ Hya}      & \bf{5080}             & \bf{E}    & \bf{2.96}             & \bf{I}   & \bf{1.32} & \bf{+0.13}              & \bf{+0.07} & \bf{--0.25} & \bf{+0.43} & \bf{--0.14} & \bf{+0.24} & \\
                        & \bf{5095}             & \bf{E}    & \bf{[2.88]}           & \bf{A}   & \bf{1.34} & \bf{+0.10}              & \bf{+0.09} & \bf{--0.24} & \bf{+0.47} & \bf{--0.15} & \bf{+0.28} & \\
                        & 5045                  & E         & 2.76                  &   I      & 1.20      &  +0.21                  & +0.39 & --0.45 & ... & --0.02 & +0.23 & 24\\
\cline{2-13}                                                                                                
                        & 4996                  & E         & 2.53                  & LMT      & 1.39      &  +0.08                  & ... & ... & ... & ... & ... & 23\\
                        & 5010                  & P         & 2.93                  & LMT      & 2.1       &  +0.13                  & ... & ... & ... & ... & ... & 28\\
\hline                                                                                                     
\object{$\beta$ Aql}    & \bf{5100}             & \bf{E}    & \bf{3.56}             & \bf{I}   & \bf{0.97} & \bf{--0.21}             & \bf{$<$--1.20} & \bf{+0.02} & \bf{--0.13} & \bf{+0.01} & \bf{+0.03} & \\
                        & \bf{5110}             & \bf{E}    & \bf{[3.53]}           & \bf{A}   & \bf{0.99} & \bf{--0.22}             & \bf{$<$--1.15} & \bf{+0.02} & \bf{--0.12} & \bf{+0.00} & \bf{+0.04} & \\
                        & 5030                  & E         & 3.49                  &   I      & 0.88      & --0.21                  & --0.73 & --0.09 & ... & ... & +0.03 & 24\\
                        & 5160                  & E         & 3.68                  &   I      & 0.92      & --0.12                  & $<$--0.47 & --0.04 & ... & +0.06 & ... & 35\\
\cline{2-13}                                                                                                
                        & 5062                  & E         & 3.54                  & LMT      & 0.97      & --0.19                  & ... & ... & ... & +0.13\tablefootmark{b} & ... & 21\\
                        & 5100                  & P         & 3.60                  & LMT      & 1.8       & --0.13                  & ... & ... & ... & ... & ... & 28\\
                        & 5163                  & S         & 3.79                  &   S      & 0.85       & --0.10                  & ... & ... & ... & ... & --0.05 & 30\\
                        & 5111                  & P         & 3.52                  & LMT      & 1.2       & --0.28                  & ... & ... & ... & ... & ... & 32\\
                        & 5106                  & P         & 3.54                  & LMT      & 1.15      & --0.15                  & ... & --0.23 & ... & --0.03 & ... & 36\\
                        & 5110                  & B         & 3.60                  &   I      & 0.92      & --0.17                  & ... & ... & ... & ... & ... & 37\\
\hline
\end{tabular}
\tablefoot{The rows in boldface show the results of this study, either using the spectroscopic or the seismic \logg \ (given in square brackets). The abundances were rescaled to our adopted solar values when appropriate and whenever these were not quoted in the original paper. {\bf E}: from excitation equilibrium of the \ion{Fe}{i} lines; {\bf I}: from ionisation equilibrium of Fe; {\bf P}: from photometric data; {\bf LMT}: from estimates of the luminosity, mass, and effective temperature; {\bf S}: from spectral synthesis; {\bf F}: from fitting of the spectral energy distribution; {\bf LDR}: from line-depth ratios; {\bf B}: from fitting the Balmer line wings; {\bf A}: from asteroseismology. References. (1) Hill \cite{hill97}; (2) Gonzalez \& Wallerstein \cite{gonzalez_wallerstein98}; (3) Tomkin \& Lambert \cite{tomkin_lambert99}; (4) Smith \etal \cite{smith00}; (5) Mishenina \& Kovtyukh \cite{mishenina01}; (6) Reddy \etal \cite{reddy02}; (7) Zoccali \etal \cite{zoccali04}; (8) Mortier \etal \cite{mortier13} (using Kurucz models and the 'TS13' line list); (9) Gratton \& Sneden \cite{gratton_sneden87}; (10) Brown \etal \cite{brown89}; (11) Fern\'{a}ndez-Villaca\~nas \etal \cite{fernandez90}; (12) Brown \& Wallerstein \cite{brown_wallerstein92}; (13) Peterson \etal \cite{peterson93}; (14) McWilliam \& Rich \cite{mcwilliam_rich94}; (15) Griffin \& Lynas-Gray \cite{griffin_lynas_gray99}; (16) Decin \etal \cite{decin03}; (17) Mel\'endez \etal \cite{melendez03}; (18) Recio-Blanco \etal \cite{recio_blanco06}; (19) Lecureur \etal \cite{lecureur07}; (20) Ram{\'{\i}}rez \& Allende Prieto \cite{ramirez11}; (21) Ram{\'{\i}}rez \etal \cite{ramirez13}; (22) Smith \etal \cite{smith13}; (23) Fulbright \etal \cite{fulbright06}; (24) Bruntt \etal \cite{bruntt10}; (25) Takeda \etal \cite{takeda08}; (26) Hekker \& Mel\'endez \cite{hekker_melendez07}; (27) Ghezzi \etal \cite{ghezzi10}; (28) McWilliam \cite{mcwilliam90}; (29) Kovtyukh \etal \cite{kovtyukh06}; (30) Valenti \& Fischer \cite{valenti_fischer05}; (31) Fuhrmann \& Chini \cite{fuhrmann12}; (32) Matrozis \etal \cite{matrozis13}; (33) Luck \& Heiter \cite{luck_heiter07}; (34) Liu \etal \cite{liu07}; (35) Luck \& Heiter \cite{luck_heiter06}; (36) Allende Prieto \etal \cite{allende04}; (37) Fuhrmann \cite{fuhrmann04}. \tablefoottext{a}{Also based on Hinkle \etal (\cite{hinkle00}) atlas.} \tablefoottext{b}{NLTE value.} \tablefootmark{c}{As Takeda \etal caution, their oxygen abundances may not be reliable.}}
\end{table*}

\onltab{2}{
\begin{table*}
\caption{EW measurements (the full table is available in electronic form). Only values eventually retained for the analysis are listed.}
\scriptsize
\hspace*{-1.6cm}
\label{tab_EWs}
\tabcolsep=0.11cm
\begin{tabular}{lrrrrrrrrrrrrrrrrrrrrrrrr} 
\hline\hline
Line & \multicolumn{15}{c}{EW [m\AA]}\\
     & HD 40726 & HD 42911 & HD 43023 & HD 45398 & HD 49429 & HD 49566 & HD 50890 & HD 169370 & HD 169751 & HD 170008 & HD 170031 & HD 171427 & HD 175294 & HD 175679 & HD 178484 & HD 181907 & HD 170053 & HD 170174 & HD 170231 & $\alpha$ Boo & $\eta$ Ser & $\epsilon$ Oph & $\xi$ Hya & $\beta$ Aql\\ 
\hline
\ion{Fe}{i}  $\lambda$5543.937 &  91.2 &  95.5 & 83.3 & 101.2 & 82.2 & 81.7 & 110.6 & 84.7 & 83.4 & 63.6 & 96.7 & 118.9 & 105.8 & 92.1 & 90.1 & 94.7 & 104.5 & 89.4 & 86.7 & 86.9 & 75.1 & 89.5 & 88.6 & 69.6\\ 
\ion{Fe}{i}  $\lambda$5638.262 & 109.0 & 114.5 & 101.2 &   ... & 96.7 & 96.6 & 134.3 & 104.8 & 100.3 & 75.0 &   ... & 141.9 & 124.3 & 112.2 & 110.0 & 112.5 &   ... & 110.7 & 106.6 & 104.0 & 89.0 & 107.1 & 106.6 & 83.2\\ 
\ion{Fe}{i}  $\lambda$5679.025 &  81.2 &  87.7 & 76.2 & 85.3 & 74.9 & 72.9 & 102.5 & 73.8 & 77.2 & 55.9 & 86.6 & 100.9 & 96.0 & 84.9 &   ... & 83.8 & 91.2 & 79.6 & 76.6 & 72.5 & 67.9 & 80.8 & 81.6 & 63.3\\ 
\ion{Fe}{i}  $\lambda$5732.275 &   ... &   ... & 29.3 &   ... &   ... &   ... &   ... &   ... &   ... &   ... & 40.0 & 41.3 & 51.0 &   ... & 31.5 & 34.1 & 41.1 &   ... &   ... & 21.5 &   ... & 32.3 &   ... &   ... \\ 
\ion{Fe}{i}  $\lambda$5775.080 &  87.6 &  94.7 & 82.8 & 99.6 & 78.6 & 78.6 & 108.6 & 83.9 & 85.8 & 61.1 & 95.8 & 117.7 & 105.9 & 92.1 & 92.0 & 93.5 & 102.5 & 88.2 & 83.7 & 84.0 & 72.7 & 86.0 & 88.3 & 66.1 \\ 
\ion{Fe}{i}  $\lambda$5806.717 &  80.9 &  88.9 & 74.1 & 88.5 & 73.6 & 71.8 & 108.1 & 76.1 & 78.0 & 52.6 & 87.6 & 100.9 & 101.0 & 82.9 & 81.1 & 87.0 & 94.6 & 81.7 & 78.8 & 70.3 & 65.1 & 81.7 & 82.1 & 58.9 \\ 
\ion{Fe}{i}  $\lambda$5848.123 &   ... &  86.1 & 70.2 &   ... &   ... &   ... & 100.3 & 76.8 & 76.3 &   ... & 93.7 &   ... & 93.7 &   ... & 82.9 & 82.0 & 98.4 & 73.8 &   ... & 70.5 & 62.2 & 72.5 & 77.4 &   ... \\ 
\ion{Fe}{i}  $\lambda$5855.091 &  40.3 &  51.2 & 38.4 & 53.5 & 36.9 & 37.0 & 51.5 & 40.2 & 42.4 & 23.1 & 51.6 & 51.8 & 55.6 & 44.1 & 43.7 & 48.3 & 55.0 & 42.9 & 41.3 & 35.4 & 33.6 & 43.2 & 45.7 & 28.5 \\
\ion{Fe}{i}  $\lambda$5902.473 &   ... &  37.5 &   ... &   ... & 24.8 &   ... &   ... & 29.6 &   ... & 13.9 & 41.4 &   ... &   ... & 33.6 & 34.4 & 37.3 & 43.8 & 29.1 & 29.5 & 23.7 &   ... &   ... &   ... &   ... \\
\ion{Fe}{i}  $\lambda$5905.689 &  79.8 &  86.8 & 74.0 &   ... & 74.1 & 72.1 & 102.8 &   ... &   ... & 55.1 & 87.0 &   ... &   ... & 84.6 & 80.5 & 91.4 & 93.0 & 81.0 & 76.9 & 71.6 & 67.8 &   ... &   ... &   ... \\ 
\ion{Fe}{i}  $\lambda$5909.970 &   ... &   ... &   ... &   ... &   ... &   ... &   ... & 83.3 & 74.4 &   ... &   ... & 119.1 & 99.4 &   ... &   ... & 93.7 &   ... &   ... & 76.7 & 89.4 &   ... &   ... & 78.5 & 48.7 \\
\ion{Fe}{i}  $\lambda$5916.247 &   ... & 110.1 & 93.2 & 132.8 & 91.4 & 90.1 &   ... & 109.6 & 98.7 &   ... & 119.5 & 147.0 & 128.1 & 109.7 & 118.1 & 121.8 & 139.1 & 102.6 & 97.7 & 118.5 & 86.5 & 107.8 & 101.7 & 73.9 \\ 
\ion{Fe}{i}  $\lambda$5927.786 &  65.1 &  71.0 &   ... & 74.1 &   ... & 58.2 &   ... & 60.8 & 63.9 & 42.1 & 73.2 & 80.7 & 73.2 &   ... & 65.6 & 66.4 & 77.5 & 65.7 & 62.3 & 55.7 & 53.3 & 66.2 & 65.5 & 48.3 \\
\ion{Fe}{i}  $\lambda$5929.667 &  64.2 &  71.2 & 61.2 & 73.1 & 56.5 & 57.3 & 84.1 & 61.3 & 64.1 & 40.2 & 72.2 & 83.6 &   ... & 66.1 & 64.2 & 68.5 & 78.6 & 64.7 & 61.8 & 56.2 & 52.7 & 64.5 & 66.0 & 47.5 \\
\ion{Fe}{i}  $\lambda$5930.173 & 112.9 & 115.9 & 105.4 & 114.5 & 101.3 & 100.3 & 136.4 & 100.8 & 106.1 & 81.4 & 116.4 & 143.6 & 130.4 & 116.5 & 108.8 & 114.9 & 124.4 & 111.7 & 108.4 & 100.9 & 93.2 & 109.2 & 111.7 & 87.6 \\ 
\ion{Fe}{i}  $\lambda$5947.503 &   ... &   ... &   ... &   ... &   ... & 19.2 &   ... &   ... &   ... & 10.3 & 34.2 &   ... &   ... & 26.9 & 25.7 &   ... &   ... &   ... &   ... & 18.5 &   ... & 29.3 &   ... &   ... \\
\ion{Fe}{i}  $\lambda$6027.050 & 102.2 & 101.8 & 89.6 & 106.3 & 87.0 & 85.9 & 132.0 & 90.8 & 89.9 & 66.0 & 101.5 & 132.7 & 114.1 & 102.3 & 97.1 & 103.1 & 112.0 & 96.3 & 93.1 & 92.0 & 79.3 & 94.2 & 93.6 & 71.9 \\
\ion{Fe}{i}  $\lambda$6078.491 & 101.1 & 104.4 & 91.9 & 102.5 & 89.8 & 89.2 & 123.7 & 89.9 & 94.3 & 71.4 & 103.8 & 128.7 & 114.9 & 101.7 & 95.7 & 100.0 & 110.5 & 99.2 & 94.6 & 85.0 & 81.9 & 96.6 & 100.5 & 77.0 \\ 
\ion{Fe}{i}  $\lambda$6078.999 &  69.8 & 75.3 & 64.3 & 75.6 & 62.1 & 61.7 & 94.0 & 64.3 & 67.1 & 45.8 & 75.9 & 91.9 & 83.9 & 72.3 & 68.7 & 72.8 & 81.5 & 71.1 & 66.4 & 59.4 & 56.5 & 68.8 & 71.3 & 51.5 \\
\ion{Fe}{i}  $\lambda$6093.644 &  51.1 & 61.2 & 49.5 & 62.3 & 47.0 & 46.1 & 74.3 & 51.4 & 52.0 & 31.5 & 62.5 & 69.1 & 68.1 & 55.0 & 54.4 & 57.0 & 66.7 & 53.9 & 51.5 & 44.5 & 42.6 & 53.1 & 55.2 & 37.7 \\
\ion{Fe}{i}  $\lambda$6094.364 &  35.1 & 47.1 & 34.7 & 49.7 & 32.6 & 32.2 & 62.2 & 38.9 & 39.6 & 21.0 & 49.1 & 51.3 & 55.8 & 40.4 & 40.4 & 43.3 & 52.8 & 39.0 & 37.7 & 29.7 & 29.9 & 39.4 & 42.0 & 25.8 \\ 
\ion{Fe}{i}  $\lambda$6096.665 &  64.2 & 75.6 & 62.0 & 80.5 & 59.3 & 59.0 & 96.0 & 67.0 & 67.5 & 41.2 & 78.5 & 86.9 & 82.7 & 71.2 & 72.4 & 73.6 & 85.4 & 66.8 & 63.6 & 64.5 & 56.2 & 68.2 & 68.6 & 48.5 \\ 
\ion{Fe}{i}  $\lambda$6098.280 &  36.7 & 45.1 & 34.1 & 52.2 & 33.2 & 30.8 &   ... & 37.1 & 38.6 & 18.6 & 48.9 &   ... & 55.0 & 41.7 & 41.0 & 44.8 & 55.3 & 39.1 & 37.0 & 29.7 & 27.9 & 40.5 & 39.7 & 23.3 \\
\ion{Fe}{i}  $\lambda$6120.249 &  28.2 & 49.4 & 31.6 & 90.7 & 28.6 & 25.1 & 74.5 & 57.8 & 42.4 & 15.6 & 67.4 & 57.8 & 54.1 & 32.8 & 66.3 & 52.6 & 86.0 & 35.6 & 30.6 &   ... & 29.6 & 40.7 & 37.4 & 19.4 \\
\ion{Fe}{i}  $\lambda$6151.617 &  91.5 & 104.0 & 89.5 & 135.4 & 85.5 & 84.0 & 134.0 & 103.7 & 94.8 & 64.2 & 115.3 & 136.7 & 112.4 & 95.5 & 115.0 & 108.9 & 136.6 & 97.4 & 89.2 & 118.9 & 81.8 & 97.0 & 93.6 & 69.6 \\ 
\ion{Fe}{i}  $\lambda$6165.361 &  73.6 & 81.1 & 67.9 & 87.5 & 67.1 & 66.0 & 104.1 & 72.0 & 72.9 & 47.6 & 84.1 & 98.3 & 88.6 & 76.4 & 78.8 & 81.2 & 92.7 & 73.9 & 71.3 & 70.0 & 61.3 & 74.4 & 75.7 & 53.8 \\
\ion{Fe}{i}  $\lambda$6187.987 &  78.6 & 90.2 & 74.2 & 96.1 & 71.2 & 70.6 & 100.5 & 78.7 & 79.3 & 51.2 & 91.2 & 107.7 & 96.9 & 83.8 & 85.6 & 89.7 & 101.8 & 80.1 & 76.0 & 77.6 & 66.5 & 80.5 & 81.8 & 58.9 \\
\ion{Fe}{i}  $\lambda$6219.279 & 141.8 & 155.2 & 132.7 &   ... & 128.9 & 128.1 & 191.5 & 149.1 & 141.7 & 100.4 & 170.5 & 207.9 & 170.3 & 148.9 & 168.4 & 166.1 &   ... & 146.7 & 138.7 & 172.7 & 122.4 & 144.7 & 144.0 & 108.4 \\
\ion{Fe}{i}  $\lambda$6240.646 &  91.2 & 106.2 & 88.9 & 130.7 & 84.6 & 83.3 &   ... & 101.4 & 96.6 & 62.6 & 114.4 & 139.4 & 114.2 & 97.0 & 113.8 & 109.2 & 135.3 & 97.3 & 89.0 & 117.6 & 81.0 & 97.1 & 94.8 & 69.1 \\
\ion{Fe}{i}  $\lambda$6252.554 & 167.6 & 188.9 & 162.9 &   ... & 156.8 &   ... & 215.0 & 186.7 &   ... & 129.6 & 207.2 &   ... & 201.8 & 174.7 &   ... & 194.8 &   ... & 177.5 & 168.1 &   ... & 155.6 & 173.8 &   ... & 141.3 \\
\ion{Fe}{i}  $\lambda$6322.690 & 120.0 & 134.7 &   ... & 158.2 & 110.1 & 108.8 & 169.7 & 127.2 & 122.1 & 85.9 & 143.4 & 176.6 & 145.2 & 122.2 & 140.1 & 137.4 & 162.6 & 125.3 & 116.5 & 141.1 & 105.8 & 122.6 & 123.1 & 93.4 \\ 
\ion{Fe}{i}  $\lambda$6335.328 & 147.8 & 165.3 & 142.7 &   ... & 137.9 & 137.4 &   ... & 163.3 & 149.1 & 108.5 &   ... &   ... & 182.4 & 153.8 &   ... & 175.0 &   ... & 157.7 & 147.2 &   ... & 134.3 & 152.4 &   ... & 117.5 \\
\ion{Fe}{i}  $\lambda$6336.823 & 137.0 &   ... & 125.6 & 150.2 & 122.3 & 122.8 &   ... & 131.1 & 128.0 & 100.6 & 149.9 & 189.0 & 165.9 & 139.3 & 141.2 & 149.5 & 162.8 & 139.2 & 135.2 & 134.4 & 116.9 & 135.8 & 138.6 & 109.3 \\
\ion{Fe}{i}  $\lambda$6436.411 &   ... & 39.6 & 25.7 & 45.8 & 22.5 & 23.7 &   ... & 29.0 & 30.7 & 11.6 & 42.0 & 42.8 & 47.2 & 32.5 &   ... & 35.1 & 47.1 & 29.5 & 30.1 & 23.8 & 20.6 & 30.3 & 32.9 & 15.6 \\
\ion{Fe}{i}  $\lambda$6469.213 &   ... &   ... &   ... &   ... &   ... &   ... &   ... &   ... &   ... &   ... &   ... & 118.6 &   ... &   ... &   ... &   ... &   ... &   ... & 88.2 &   ... &   ... &   ... &   ... &   ... \\
\ion{Fe}{i}  $\lambda$6498.939 & 105.9 & 122.9 & 101.8 &   ... & 97.6 & 92.5 & 166.6 & 126.5 & 110.6 & 69.7 & 142.0 & 164.9 & 134.1 & 110.8 & 146.2 & 132.3 &   ... & 112.1 & 101.8 & 155.4 & 93.9 & 116.5 & 106.9 & 76.9 \\
\ion{Fe}{i}  $\lambda$6574.228 &  82.1 & 103.2 & 82.7 & 147.8 &   ... & 73.3 &   ... & 107.5 & 91.6 & 53.0 & 120.5 & 130.9 & 115.8 & 88.9 & 122.9 & 115.9 & 148.6 & 92.3 & 80.8 & 129.8 & 77.2 & 95.6 & 88.7 & 62.2 \\
\ion{Fe}{i}  $\lambda$6593.871 & 137.2 & 152.7 & 127.7 & 179.3 &   ... &   ... &   ... & 144.9 & 138.4 &   ... & 161.2 & 199.6 & 167.6 & 145.0 & 159.6 & 161.0 & 186.5 & 141.0 & 131.8 & 160.0 & 118.9 & 138.2 & 138.2 & 105.6 \\
\ion{Fe}{i}  $\lambda$6699.162 &   ... & 30.4 & 18.2 & 32.9 & 19.0 & 16.7 &   ... & 21.3 & 27.0 &   ... & 32.6 & 38.6 & 36.4 & 23.7 & 25.5 & 26.3 & 35.2 & 22.0 & 21.1 & 17.8 & 14.5 & 22.3 & 25.2 & 11.5 \\
\ion{Fe}{i}  $\lambda$6703.567 &  76.2 & 88.8 & 73.0 & 105.8 & 70.1 & 68.6 &   ... & 82.6 & 78.5 & 48.5 & 95.2 & 112.0 & 96.9 & 82.9 & 92.7 & 92.3 & 110.2 & 81.2 & 73.3 & 91.6 & 65.3 & 81.1 & 80.1 & 55.3 \\
\ion{Fe}{i}  $\lambda$6713.771 &  39.1 & 49.2 & 36.7 & 49.1 &   ... &   ... &   ... & 37.5 & 41.5 & 20.6 & 50.0 & 53.0 & 58.7 & 43.2 & 41.7 & 46.8 & 53.2 & 42.4 & 39.6 & 29.4 & 31.2 & 40.8 & 44.6 & 26.2 \\
\ion{Fe}{i}  $\lambda$6725.353 &  36.4 & 49.1 & 36.0 & 53.6 & 35.6 & 33.0 & 62.9 & 41.5 & 42.6 & 21.1 & 52.6 & 50.8 & 56.1 & 42.9 & 45.3 & 48.6 & 57.6 & 39.2 & 38.0 & 36.4 & 32.3 & 41.0 & 43.5 & 25.9 \\ 
\ion{Fe}{i}  $\lambda$6726.661 &  67.4 & 78.0 & 65.8 & 78.5 & 63.8 & 62.3 & 96.2 & 67.1 & 69.8 & 45.8 & 78.6 & 88.4 & 88.3 & 73.8 & 71.8 & 76.3 & 83.0 & 71.2 & 67.5 & 60.7 & 58.5 & 70.4 & 72.4 & 52.9 \\
\ion{Fe}{i}  $\lambda$6733.151 &  48.8 & 60.6 & 46.0 & 61.1 & 43.9 & 43.3 & 72.3 & 47.4 & 51.6 & 27.6 & 60.3 & 67.2 & 67.3 & 55.0 & 52.7 & 58.3 & 64.8 & 53.0 & 48.5 & 38.8 & 39.6 & 50.0 & 53.9 & 33.1 \\ 
\ion{Fe}{i}  $\lambda$6745.090 &   ... & 29.9 & 20.0 & 32.7 & 19.4 & 17.6 &   ... & 22.2 & 24.6 &   ... & 31.6 & 34.3 & 37.2 &   ... & 21.8 & 28.0 & 34.3 & 23.7 & 21.7 & 15.4 & 15.8 &   ... & 26.4 & 12.3 \\ 
\ion{Fe}{i}  $\lambda$6750.150 & 122.7 & 136.7 & 114.3 & 160.1 &   ... & 110.8 &   ... & 128.0 & 123.3 & 86.5 & 146.1 & 180.0 & 151.9 & 130.9 & 142.8 & 143.7 & 170.5 & 130.7 & 118.1 & 142.9 & 105.5 & 124.2 & 125.5 & 92.0 \\
\ion{Fe}{i}  $\lambda$6806.847 &  74.6 & 88.3 & 71.7 & 107.3 & 67.6 & 66.1 & 119.1 & 82.1 & 79.1 & 47.2 & 95.4 & 112.4 & 100.9 & 80.7 & 93.0 & 90.7 & 112.0 & 79.4 & 73.7 & 89.4 & 64.4 &   ... & 79.4 & 53.4 \\
\ion{Fe}{i}  $\lambda$6810.257 &  73.7 & 82.8 & 70.0 & 84.0 & 66.7 & 67.2 & 103.6 & 68.6 & 73.2 & 49.0 & 83.6 & 96.3 & 91.9 & 77.7 & 76.2 & 78.0 & 89.5 & 76.9 & 71.2 & 65.5 & 61.3 &   ... & 76.4 & 55.4 \\
\ion{Fe}{i}  $\lambda$6820.369 &  66.8 & 79.1 & 61.9 &   ... & 61.2 & 59.4 & 94.6 & 65.7 & 68.7 & 41.2 & 80.9 & 89.4 & 86.1 & 70.6 & 72.4 & 75.8 &   ... & 67.9 & 65.1 & 61.1 & 55.0 &   ... & 71.1 & 48.1 \\
\ion{Fe}{i}  $\lambda$6843.648 &  89.5 & 97.8 & 80.9 & 96.9 & 80.3 & 80.2 & 114.4 & 84.8 & 89.0 & 60.3 & 97.5 & 118.2 & 108.8 & 91.1 & 89.9 & 95.3 & 106.3 & 91.0 & 86.7 & 77.4 & 71.8 &   ... & 89.5 & 65.8 \\
\ion{Fe}{i}  $\lambda$6857.243 &  45.4 & 58.5 & 42.7 & 63.4 & 43.5 & 40.9 & 65.5 & 48.1 & 48.7 & 24.8 & 59.8 & 61.6 & 62.4 & 48.9 & 52.5 & 55.2 & 64.3 & 47.9 & 45.6 & 41.5 & 37.8 &   ... & 51.4 & 31.4 \\
\ion{Fe}{i}  $\lambda$6862.492 &  55.4 & 66.3 & 49.0 & 67.2 & 51.0 & 49.1 & 75.4 & 51.6 & 55.9 & 31.3 & 64.7 & 76.2 &   ... & 68.2 & 57.3 & 69.4 & 71.0 & 55.8 & 54.2 & 45.0 & 42.4 &   ... & 60.6 & 36.8 \\
\ion{Fe}{i}  $\lambda$6988.523 &   ... &   ... &   ... &   ... &   ... &   ... &   ... &   ... &   ... &   ... &   ... &   ... &   ... &   ... &   ... & 99.7 &   ... &   ... &   ... & 96.0 &   ... &   ... &   ... &   ... \\
\ion{Fe}{i}  $\lambda$7022.953 &   ... &   ... &   ... &   ... &   ... &   ... &   ... &   ... &   ... &   ... &   ... &   ... &   ... &   ... &   ... &   ... &   ... &   ... &   ... & 88.3 &   ... &   ... &   ... &   ... \\
\ion{Fe}{i}  $\lambda$7219.678 &   ... &   ... &   ... &   ... &   ... &   ... &   ... &   ... &   ... &   ... &   ... &   ... &   ... &   ... &   ... & 84.4 &   ... &   ... &   ... & 73.1 &   ... &   ... &   ... &   ... \\
\ion{Fe}{i}  $\lambda$7306.556 &   ... &   ... &   ... &   ... &   ... &   ... &   ... &   ... &   ... &   ... &   ... &   ... &   ... &   ... &   ... &   ... &   ... &   ... &   ... & 67.4 &   ... &   ... &   ... &   ... \\
\ion{Fe}{i}  $\lambda$7547.896 &   ... &   ... &   ... &   ... &   ... &   ... &   ... &   ... &   ... &   ... &   ... &   ... & 55.4 & 43.2 &   ... & 40.8 &   ... &   ... &   ... & 27.8 &   ... &   ... &   ... &   ... \\
\ion{Fe}{i}  $\lambda$7723.208 &   ... &   ... &   ... &   ... &   ... &   ... &   ... &   ... &   ... &   ... &   ... &   ... & 119.7 & 95.8 &   ... & 110.4 &   ... &   ... &   ... & 108.7 &   ... &   ... &   ... &   ... \\
\ion{Fe}{i}  $\lambda$7746.587 &   ... &   ... &   ... &   ... &   ... &   ... &   ... &   ... &   ... &   ... &   ... &   ... & 58.4 &   ... &   ... & 42.5 &   ... &   ... &   ... & 26.8 &   ... &   ... &   ... &   ... \\
\ion{Fe}{i}  $\lambda$7748.274 &   ... &   ... &   ... &   ... &   ... &   ... &   ... &   ... &   ... &   ... &   ... &   ... &   ... &   ... &   ... &   ... &   ... &   ... &   ... & 170.8 &   ... &   ... &   ... &   ... \\
\ion{Fe}{i}  $\lambda$7751.137 &   ... &   ... &   ... &   ... &   ... &   ... &   ... &   ... &   ... &   ... &   ... &   ... & 95.1 & 78.2 &   ... & 81.9 &   ... &   ... &   ... & 57.8 &   ... &   ... &   ... &   ... \\
\ion{Fe}{i}  $\lambda$7780.552 &   ... &   ... &   ... &   ... &   ... &   ... &   ... &   ... &   ... &   ... &   ... &   ... & 164.9 & 146.4 &   ... & 145.0 &   ... &   ... &   ... & 129.4 &   ... &   ... &   ... &   ... \\ 
\ion{Fe}{i}  $\lambda$7802.473 &   ... &   ... &   ... &   ... &   ... &   ... &   ... &   ... &   ... &   ... &   ... &   ... & 44.9 & 31.6 &   ... & 31.1 &   ... &   ... &   ... & 18.5 &   ... &   ... &   ... &   ... \\
\ion{Fe}{i}  $\lambda$7807.952 &   ... &   ... &   ... &   ... &   ... &   ... &   ... &   ... &   ... &   ... &   ... &   ... & 105.9 & 93.5 &   ... & 87.3 &   ... &   ... &   ... & 67.4 &   ... &   ... &   ... &   ... \\
\ion{Fe}{i}  $\lambda$8239.127 &   ... &   ... &   ... &   ... &   ... &   ... &   ... &   ... &   ... &   ... &   ... &   ... &   ... &   ... &   ... &   ... &   ... &   ... &   ... & 114.4 &   ... &   ... &   ... &   ... \\
\ion{Fe}{i}  $\lambda$8922.643 &   ... &   ... &   ... &   ... &   ... &   ... &   ... &   ... &   ... &   ... &   ... &   ... &   ... &   ... &   ... &   ... &   ... &   ... &   ... & 18.6 &   ... &   ... &   ... &   ... \\
\ion{Fe}{ii} $\lambda$5264.812 &   ... & 68.8 & 66.2 & 55.1 & 65.9 & 67.1 &   ... & 50.4 & 62.1 & 43.8 & 59.2 & 104.2 & 77.3 & 76.3 & 58.8 & 64.0 & 62.9 & 73.8 & 73.1 & 47.1 & 51.1 & 67.4 & 69.7 & 45.3 \\
\ion{Fe}{ii} $\lambda$5425.257 &  76.4 & 62.2 & 58.2 & 50.4 & 58.2 & 59.2 &   ... & 45.2 & 56.8 & 37.7 & 52.6 & 97.8 & 70.1 & 71.9 & 52.0 & 60.2 & 57.0 & 67.1 & 64.2 & 44.1 & 45.3 &   ... & 63.2 & 39.8 \\
\ion{Fe}{ii} $\lambda$5991.376 &  68.5 &   ... & 50.1 &   ... &   ... & 52.3 &   ... &   ... & 49.4 & 29.0 &   ... & 90.2 &   ... &   ... & 45.7 &   ... &   ... & 58.6 & 59.8 & 34.2 & 35.8 & 54.6 & 56.2 &   ... \\
\ion{Fe}{ii} $\lambda$6149.258 &  70.6 & 54.6 & 49.2 &   ... & 49.6 & 51.8 &   ... & 34.8 & 47.3 & 32.2 & 42.7 & 86.8 & 63.0 & 62.0 & 41.4 & 48.9 &   ... &   ... & 57.0 & 32.1 & 35.5 & 51.9 & 54.9 & 31.2 \\
\ion{Fe}{ii} $\lambda$6247.557 &  92.4 & 70.3 & 67.6 & 45.8 & 65.4 & 69.4 & 94.7 & 46.2 & 64.4 & 45.9 & 52.7 & 112.4 & 76.8 & 80.5 & 51.9 & 60.2 & 56.2 & 74.3 & 74.8 & 42.3 & 49.3 & 68.2 & 72.6 & 45.5 \\
\ion{Fe}{ii} $\lambda$6369.462 &  49.9 & 40.8 & 34.8 & 31.2 & 35.2 & 36.9 & 57.1 & 24.8 & 36.0 & 18.2 & 32.1 & 67.3 &   ... & 43.3 & 31.7 & 34.8 & 37.1 & 43.9 & 41.5 & 23.4 & 23.9 & 39.1 & 40.3 & 19.4 \\
\ion{Fe}{ii} $\lambda$6416.919 &  74.1 &   ... & 52.8 &   ... & 52.1 & 54.8 & 80.6 &   ... & 53.8 & 34.1 &   ... & 84.8 & 65.4 & 63.6 &   ... & 55.4 &   ... & 60.7 & 61.4 &   ... &   ... &   ... & 59.4 &   ... \\
\ion{Fe}{ii} $\lambda$6432.680 &   ... & 67.2 & 60.3 & 50.5 & 60.8 &   ... &   ... & 47.0 & 60.3 & 38.5 &   ... & 101.4 & 74.6 & 76.8 &   ... & 61.2 & 60.4 & 70.5 & 69.4 & 45.0 & 46.1 & 68.0 & 66.9 & 39.2 \\
\ion{Fe}{ii} $\lambda$6456.383 & 104.6 & 80.1 & 78.2 & 55.2 &   ... & 81.6 &   ... & 55.7 & 75.1 & 54.0 & 62.8 & 130.5 & 93.1 & 92.2 & 63.6 & 76.0 & 69.0 &   ... &   ... & 52.2 & 59.0 & 77.4 & 83.5 & 54.2 \\
\ion{Fe}{ii} $\lambda$7711.723 &   ... &   ... &   ... &   ... &   ... &   ... &   ... &   ... &   ... &   ... &   ... &   ... &   ... & 77.1 &   ... &   ... &   ... &   ... &   ... & 37.0 &   ... &   ... &   ... &   ... \\
\ion{O}{i}   $\lambda$7771.944 &   ... &   ... &   ... &   ... &   ... &   ... &   ... &   ... &   ... &   ... &   ... &   ... & 60.1 & 70.4 &   ... & 41.0 &   ... &   ... &   ... & 22.1 &   ... &   ... &   ... &   ... \\
\ion{O}{i}   $\lambda$7774.166 &   ... &   ... &   ... &   ... &   ... &   ... &   ... &   ... &   ... &   ... &   ... &   ... & 56.7 & 59.4 &   ... &   ... &   ... &   ... &   ... & 21.2 &   ... &   ... &   ... &   ... \\
\ion{O}{i}   $\lambda$7775.388 &   ... &   ... &   ... &   ... &   ... &   ... &   ... &   ... &   ... &   ... &   ... &   ... & 51.5 & 45.8 &   ... & 34.5 &   ... &   ... &   ... & 17.0 &   ... &   ... &   ... &   ... \\
\ion{Na}{i}  $\lambda$6154.226 &  76.7 & 89.9 & 63.9 & 111.3 & 63.2 & 63.0 & 129.4 & 70.5 & 70.9 & 37.3 & 97.8 & 106.8 & 108.8 & 79.3 & 86.1 & 83.5 & 111.3 & 76.5 & 71.0 & 71.0 & 54.8 & 70.9 & 83.9 & 46.8 \\ 
\ion{Na}{i}  $\lambda$6160.747 &  97.6 & 111.5 & 86.3 & 134.0 & 85.4 & 84.0 & 138.7 & 93.4 & 94.2 & 57.8 & 121.4 & 127.9 & 125.3 & 99.9 & 109.3 & 101.6 & 135.8 & 99.0 & 92.8 & 94.4 & 77.7 & 91.5 & 106.5 & 68.0 \\
\ion{Mg}{i}  $\lambda$5711.088 & 130.4 & 139.7 & 122.7 & 153.7 & 119.8 & 120.8 & 163.7 & 138.7 & 127.0 & 104.5 & 150.4 & 156.3 & 155.4 & 138.3 & 143.0 & 146.0 & 156.3 & 131.9 & 124.9 & 152.4 & 120.4 & 128.7 & 129.1 & 114.6 \\
\ion{Al}{i}  $\lambda$6696.023 &   ... & 78.3 & 59.4 & 106.5 & 57.6 & 54.2 &   ... & 82.8 & 67.4 & 43.1 & 94.7 & 87.1 & 92.8 & 68.9 & 83.1 & 79.1 & 97.7 & 64.0 & 60.0 & 85.4 & 57.4 &   ... & 68.2 & 49.9 \\
\ion{Al}{i}  $\lambda$6698.673 &  38.2 & 55.2 & 37.7 & 78.4 & 36.1 & 34.4 & 78.2 & 58.4 & 46.9 & 25.8 & 70.8 & 62.5 & 67.5 & 46.0 & 62.0 & 56.6 & 70.8 & 42.7 & 39.2 & 60.3 & 37.3 & 44.8 & 47.3 & 30.8 \\
\ion{Al}{i}  $\lambda$7835.309 &   ... &   ... &   ... &   ... &   ... &   ... &   ... &   ... &   ... &   ... &   ... &   ... & 83.2 & 61.7 &   ... & 64.3 &   ... &   ... &   ... & 64.6 &   ... &   ... &   ... &   ... \\
\multicolumn{1}{c}{...} & ... & ... & ... & ... & ... & ... & ... & ... & ... & ... & ... & ... & ... & ... & ... & ... & ... & ... & ... & ... & ... & ... & ... &...\\  
\hline
\end{tabular}
\end{table*}}

\onltab{3}{
\begin{table*}
\centering
\caption{Line list and atomic data.}
\label{tab_atomic_data}
\begin{tabular}{lccrcc|lccrcc} 
\hline\hline
Ion          & $\lambda$& LEP    & \loggf  & $\log C_6$              & $E_\gamma^{5/2}$          & Ion           & $\lambda$& LEP    & \loggf  & $\log C_6$               & $E_\gamma^{5/2}$       \\
             & [\AA]    & [eV]   &         &                         &                          &              & [\AA]    & [eV]   &         &                          &                        \\
\hline
\ion{Fe}{i}  & 5543.937 &  4.218 & --1.089 & --30.523                & \multicolumn{1}{c|}{...} & \ion{Fe}{i}  & 7746.587 &  5.064 & --1.330 & \multicolumn{1}{c}{...}  & 3.07                   \\                  
\ion{Fe}{i}  & 5638.262 &  4.221 & --0.841 & --30.541                & \multicolumn{1}{c|}{...} & \ion{Fe}{i}  & 7748.274 &  2.949 & --1.816 & --31.321                 & \multicolumn{1}{c}{...}\\ 
\ion{Fe}{i}  & 5679.025 &  4.652 & --0.802 & --30.090                & \multicolumn{1}{c|}{...} & \ion{Fe}{i}  & 7751.137 &  4.992 & --0.799 & \multicolumn{1}{c}{...}  & 3.07                   \\                  
\ion{Fe}{i}  & 5732.275 &  4.992 & --1.233 & --30.731                & \multicolumn{1}{c|}{...} & \ion{Fe}{i}  & 7780.552 &  4.474 & --0.154 & \multicolumn{1}{c}{...}  & 3.07                   \\                  
\ion{Fe}{i}  & 5775.080 &  4.221 & --1.141 & \multicolumn{1}{c}{...} & 3.07                     & \ion{Fe}{i}  & 7802.473 &  5.086 & --1.297 & \multicolumn{1}{c}{...}  & 3.07                   \\                   
\ion{Fe}{i}  & 5806.717 &  4.608 & --0.913 & --30.216                & \multicolumn{1}{c|}{...} & \ion{Fe}{i}  & 7807.952 &  4.992 & --0.527 & \multicolumn{1}{c}{...}  & 3.07                   \\                  
\ion{Fe}{i}  & 5848.123 &  4.608 & --0.900 & \multicolumn{1}{c}{...} & 3.07                     & \ion{Fe}{i}  & 8239.127 &  2.424 & --3.306 & \multicolumn{1}{c}{...}  & 3.07                   \\                   
\ion{Fe}{i}  & 5855.091 &  4.608 & --1.612 & --30.241                & \multicolumn{1}{c|}{...} & \ion{Fe}{i}  & 8922.643 &  4.992 & --1.442 & \multicolumn{1}{c}{...}  & 3.07                   \\                  
\ion{Fe}{i}  & 5902.473 &  4.593 & --1.806 & --31.809                & \multicolumn{1}{c|}{...} & \ion{Fe}{ii} & 5264.812 &  3.231 & --3.093 & \multicolumn{1}{c}{...}  & 3.07                   \\                  
\ion{Fe}{i}  & 5905.689 &  4.652 & --0.792 & --30.206                & \multicolumn{1}{c|}{...} & \ion{Fe}{ii} & 5425.257 &  3.200 & --3.271 & \multicolumn{1}{c}{...}  & 3.07                   \\                  
\ion{Fe}{i}  & 5909.970 &  3.211 & --2.551 & --30.562                & \multicolumn{1}{c|}{...} & \ion{Fe}{ii} & 5991.376 &  3.153 & --3.595 & --32.110                 & \multicolumn{1}{c}{...}\\  
\ion{Fe}{i}  & 5916.247 &  2.454 & --2.923 & --31.367                & \multicolumn{1}{c|}{...} & \ion{Fe}{ii} & 6149.258 &  3.889 & --2.782 & --32.025                 & \multicolumn{1}{c}{...}\\ 
\ion{Fe}{i}  & 5927.786 &  4.652 & --1.121 & --30.217                & \multicolumn{1}{c|}{...} & \ion{Fe}{ii} & 6247.557 &  3.892 & --2.367 & --32.025                 & \multicolumn{1}{c}{...}\\ 
\ion{Fe}{i}  & 5929.667 &  4.549 & --1.251 & --30.358                & \multicolumn{1}{c|}{...} & \ion{Fe}{ii} & 6369.462 &  2.891 & --4.169 & --32.129                 & \multicolumn{1}{c}{...}\\ 
\ion{Fe}{i}  & 5930.173 &  4.652 & --0.356 & --30.218                & \multicolumn{1}{c|}{...} & \ion{Fe}{ii} & 6416.919 &  3.892 & --2.738 & --32.031                 & \multicolumn{1}{c}{...}\\ 
\ion{Fe}{i}  & 5947.503 &  4.607 & --1.928 & \multicolumn{1}{c}{...} & 3.07                     & \ion{Fe}{ii} & 6432.680 &  2.891 & --3.606 & --32.129                 & \multicolumn{1}{c}{...}\\ 
\ion{Fe}{i}  & 6027.050 &  4.076 & --1.166 & \multicolumn{1}{c}{...} & 3.07                     & \ion{Fe}{ii} & 6456.383 &  3.904 & --2.185 & --32.031                 & \multicolumn{1}{c}{...}\\ 
\ion{Fe}{i}  & 6078.491 &  4.796 & --0.388 & \multicolumn{1}{c}{...} & 3.07                     & \ion{Fe}{ii} & 7711.723 &  3.904 & --2.555 & --32.031                 & \multicolumn{1}{c}{...}\\ 
\ion{Fe}{i}  & 6078.999 &  4.652 & --1.050 & --30.290                & \multicolumn{1}{c|}{...} & \ion{O}{i}   & 7771.944 &  9.147 &   0.339 & --31.059                 & \multicolumn{1}{c}{...}\\ 
\ion{Fe}{i}  & 6093.644 &  4.608 & --1.378 & --30.356                & \multicolumn{1}{c|}{...} & \ion{O}{i}   & 7774.166 &  9.147 &   0.169 & --31.059                 & \multicolumn{1}{c}{...}\\ 
\ion{Fe}{i}  & 6094.364 &  4.652 & --1.622 & --30.297                & \multicolumn{1}{c|}{...} & \ion{O}{i}   & 7775.388 &  9.147 & --0.065 & --31.059                 & \multicolumn{1}{c}{...}\\ 
\ion{Fe}{i}  & 6096.665 &  3.984 & --1.855 & --30.240                & \multicolumn{1}{c|}{...} & \ion{Na}{i}  & 6154.226 &  2.102 & --1.581 & \multicolumn{1}{c}{...}  & 5.20                   \\                   
\ion{Fe}{i}  & 6098.280 &  4.559 & --1.777 & --30.446                & \multicolumn{1}{c|}{...} & \ion{Na}{i}  & 6160.747 &  2.105 & --1.323 & \multicolumn{1}{c}{...}  & 5.20                   \\                   
\ion{Fe}{i}  & 6120.249 &  0.915 & --5.950 & \multicolumn{1}{c}{...} & 3.07                     & \ion{Mg}{i}  & 5711.088 &  4.346 & --1.852 & \multicolumn{1}{c}{...}  & 9.88                   \\                  
\ion{Fe}{i}  & 6151.617 &  2.176 & --3.350 & --31.593                & \multicolumn{1}{c|}{...} & \ion{Al}{i}  & 6696.023 &  3.143 & --1.612 & \multicolumn{1}{c}{...}  & 9.88                   \\                   
\ion{Fe}{i}  & 6165.361 &  4.143 & --1.522 & \multicolumn{1}{c}{...} & 3.07                     & \ion{Al}{i}  & 6698.673 &  3.143 & --1.878 & \multicolumn{1}{c}{...}  & 9.88                   \\                  
\ion{Fe}{i}  & 6187.987 &  3.944 & --1.688 & --30.310                & \multicolumn{1}{c|}{...} & \ion{Al}{i}  & 7835.309 &  4.022 & --0.731 & \multicolumn{1}{c}{...}  & 9.88                   \\                    
\ion{Fe}{i}  & 6219.279 &  2.198 & --2.468 & --31.589                & \multicolumn{1}{c|}{...} & \ion{Si}{i}  & 5793.073 &  4.930 & --1.947 & \multicolumn{1}{c}{...}  & 1.93                   \\                   
\ion{Fe}{i}  & 6240.646 &  2.223 & --3.376 & --31.503                & \multicolumn{1}{c|}{...} & \ion{Si}{i}  & 5948.541 &  5.083 & --1.243 & --29.660\tablefootmark{a}& \multicolumn{1}{c}{...}\\              
\ion{Fe}{i}  & 6252.554 &  2.404 & --1.891 & --31.416                & \multicolumn{1}{c|}{...} & \ion{Si}{i}  & 6029.869 &  5.984 & --1.581 & \multicolumn{1}{c}{...}  & 1.93                   \\                   
\ion{Fe}{i}  & 6322.690 &  2.588 & --2.435 & --31.355                & \multicolumn{1}{c|}{...} & \ion{Si}{i}  & 6155.134 &  5.620 & --0.840 & \multicolumn{1}{c}{...}  & 1.93                   \\                   
\ion{Fe}{i}  & 6335.328 &  2.198 & --2.427 & --31.601                & \multicolumn{1}{c|}{...} & \ion{Si}{i}  & 6721.848 &  5.863 & --1.147 & \multicolumn{1}{c}{...}  & 1.93                   \\                   
\ion{Fe}{i}  & 6336.823 &  3.687 & --1.049 & --30.382                & \multicolumn{1}{c|}{...} & \ion{Si}{i}  & 7034.901 &  5.871 & --0.829 & \multicolumn{1}{c}{...}  & 1.93                   \\                   
\ion{Fe}{i}  & 6436.411 &  4.187 & --2.380 & \multicolumn{1}{c}{...} & 3.07                     & \ion{Si}{i}  & 7680.266 &  5.863 & --0.697 & \multicolumn{1}{c}{...}  & 1.93                   \\                  
\ion{Fe}{i}  & 6469.213 &  4.835 & --0.616 & --30.219                & \multicolumn{1}{c|}{...} & \ion{Si}{i}  & 7760.628 &  6.206 & --1.354 & \multicolumn{1}{c}{...}  & 1.93                   \\                   
\ion{Fe}{i}  & 6498.939 &  0.958 & --4.662 & --31.814                & \multicolumn{1}{c|}{...} & \ion{Si}{i}  & 8742.446 &  5.871 & --0.531 & \multicolumn{1}{c}{...}  & 1.93                   \\                   
\ion{Fe}{i}  & 6574.228 &  0.990 & --5.011 & \multicolumn{1}{c}{...} & 3.07                     & \ion{Si}{i}  & 8892.720 &  5.984 & --0.689 & \multicolumn{1}{c}{...}  & 1.93                   \\                  
\ion{Fe}{i}  & 6593.871 &  2.433 & --2.296 & --31.433                & \multicolumn{1}{c|}{...} & \ion{Ca}{i}  & 6166.439 &  2.521 & --1.196 & --30.226                 & \multicolumn{1}{c}{...}\\ 
\ion{Fe}{i}  & 6699.162 &  4.593 & --2.059 & --31.517                & \multicolumn{1}{c|}{...} & \ion{Ca}{i}  & 6455.598 &  2.523 & --1.383 & --31.293                 & \multicolumn{1}{c}{...}\\ 
\ion{Fe}{i}  & 6703.567 &  2.759 & --3.080 & --31.436                & \multicolumn{1}{c|}{...} & \ion{Ca}{i}  & 6499.650 &  2.523 & --0.866 & --31.296                 & \multicolumn{1}{c}{...}\\ 
\ion{Fe}{i}  & 6713.771 &  4.796 & --1.540 & --30.367                & \multicolumn{1}{c|}{...} & \ion{Sc}{ii} & 6320.851 &  1.500 & --1.826 & \multicolumn{1}{c}{...}  & 6.91                   \\                  
\ion{Fe}{i}  & 6725.353 &  4.104 & --2.221 & --30.317                & \multicolumn{1}{c|}{...} & \ion{Ti}{i}  & 5766.330 &  3.294 &   0.345 & \multicolumn{1}{c}{...}  & 4.98                   \\                  
\ion{Fe}{i}  & 6726.661 &  4.607 & --1.132 & \multicolumn{1}{c}{...} & 3.07                     & \ion{Cr}{i}  & 5787.965 &  3.323 & --0.210 & --30.099                 & \multicolumn{1}{c}{...}\\ 
\ion{Fe}{i}  & 6733.151 &  4.639 & --1.454 & --30.468                & \multicolumn{1}{c|}{...} & \ion{Cr}{i}  & 6882.475 &  3.438 & --0.293 & --30.382                 & \multicolumn{1}{c}{...}\\ 
\ion{Fe}{i}  & 6745.090 &  4.580 & --2.058 & --31.679                & \multicolumn{1}{c|}{...} & \ion{Cr}{i}  & 6882.996 &  3.438 & --0.355 & --30.382                 & \multicolumn{1}{c}{...}\\ 
\ion{Fe}{i}  & 6750.150 &  2.424 & --2.610 & --31.387                & \multicolumn{1}{c|}{...} & \ion{Cr}{i}  & 6925.202 &  3.450 & --0.282 & --30.382                 & \multicolumn{1}{c}{...}\\ 
\ion{Fe}{i}  & 6806.847 &  2.728 & --3.140 & --31.460                & \multicolumn{1}{c|}{...} & \ion{Co}{i}  & 6454.990 &  3.632 & --0.262 & --30.423                 & \multicolumn{1}{c}{...}\\ 
\ion{Fe}{i}  & 6810.257 &  4.607 & --1.035 & --30.347                & \multicolumn{1}{c|}{...} & \ion{Ni}{i}  & 5593.733 &  3.899 & --0.823 & --30.403                 & \multicolumn{1}{c}{...}\\ 
\ion{Fe}{i}  & 6820.369 &  4.639 & --1.199 & --30.311                & \multicolumn{1}{c|}{...} & \ion{Ni}{i}  & 5805.213 &  4.168 & --0.572 & --30.387                 & \multicolumn{1}{c}{...}\\ 
\ion{Fe}{i}  & 6843.648 &  4.549 & --0.891 & --30.532                & \multicolumn{1}{c|}{...} & \ion{Ni}{i}  & 6111.066 &  4.088 & --0.873 & --30.382                 & \multicolumn{1}{c}{...}\\ 
\ion{Fe}{i}  & 6857.243 &  4.076 & --2.052 & \multicolumn{1}{c}{...} & 3.07                     & \ion{Ni}{i}  & 6176.807 &  4.088 & --0.287 & --30.407                 & \multicolumn{1}{c}{...}\\ 
\ion{Fe}{i}  & 6862.492 &  4.559 & --1.392 & --30.436                & \multicolumn{1}{c|}{...} & \ion{Ni}{i}  & 6186.709 &  4.106 & --0.882 & --30.392                 & \multicolumn{1}{c}{...}\\ 
\ion{Fe}{i}  & 6988.523 &  2.404 & --3.485 & --31.467                & \multicolumn{1}{c|}{...} & \ion{Ni}{i}  & 6204.600 &  4.088 & --1.128 & --30.557                 & \multicolumn{1}{c}{...}\\ 
\ion{Fe}{i}  & 7022.953 &  4.191 & --1.183 & --30.299                & \multicolumn{1}{c|}{...} & \ion{Ni}{i}  & 6223.981 &  4.106 & --0.909 & --30.405                 & \multicolumn{1}{c}{...}\\ 
\ion{Fe}{i}  & 7219.678 &  4.076 & --1.607 & \multicolumn{1}{c}{...} & 3.07                     & \ion{Ni}{i}  & 6772.313 &  3.658 & --0.983 & --30.448                 & \multicolumn{1}{c}{...}\\ 
\ion{Fe}{i}  & 7306.556 &  4.178 & --1.574 & \multicolumn{1}{c}{...} & 3.07                     & \ion{Ni}{i}  & 7555.598 &  3.848 & --0.109 & \multicolumn{1}{c}{...}  & 4.04                   \\                  
\ion{Fe}{i}  & 7547.896 &  5.100 & --1.158 & \multicolumn{1}{c}{...} & 3.07                     & \ion{Ni}{i}  & 7797.586 &  3.899 & --0.292 & \multicolumn{1}{c}{...}  & 4.04                   \\                  
\ion{Fe}{i}  & 7723.208 &  2.279 & --3.500 & --31.492                & \multicolumn{1}{c|}{...} & \ion{Ba}{ii} & 5853.668 &  0.604 & --0.921 & --31.293                 & \multicolumn{1}{c}{...}\\                      
\hline
\end{tabular}
\tablefoot{Wavelength of the transition, lower excitation potential, calibrated oscillator strength, interaction constant for van der Waals interaction, $C_6$, or empirical enhancement factor to the line width parameter, $E_\gamma$, when the latter is not available. The interaction constants were derived from the line broadening cross sections computed by Barklem \etal (\cite{barklem00}) and Barklem \& Aspelund-Johansson (\cite{barklem05}). \tablefoottext{a}{From Barbuy \etal (\cite{barbuy06}).}}
\end{table*}}

\onltab{4}{
\begin{table*}
\centering
\caption{Abundance data with 1-$\sigma$ uncertainties.}
\scriptsize
\label{tab_abundances}
\begin{tabular}{lrrrrrrrrrr} 
\hline\hline
              & \multicolumn{1}{c}{[Mg/Fe]} & \multicolumn{1}{c}{[Al/Fe]} & \multicolumn{1}{c}{[Si/Fe]} & \multicolumn{1}{c}{[Ca/Fe]} & \multicolumn{1}{c}{[Sc/Fe]} & \multicolumn{1}{c}{[Ti/Fe]} & \multicolumn{1}{c}{[Cr/Fe]} & \multicolumn{1}{c}{[Co/Fe]} & \multicolumn{1}{c}{[Ni/Fe]} & \multicolumn{1}{c}{[Ba/Fe]}\\
\hline
Sun                     &  \multicolumn{1}{c}{7.58} &   \multicolumn{1}{c}{6.47}         &   \multicolumn{1}{c}{7.55}         &      \multicolumn{1}{c}{6.36}   &   \multicolumn{1}{c}{3.17}                      &   \multicolumn{1}{c}{5.02}           &    \multicolumn{1}{c}{5.67}          &    \multicolumn{1}{c}{4.92}       &   \multicolumn{1}{c}{6.25}           &     \multicolumn{1}{c}{2.13}         \\
\hline
HD 40726       &   0.09\p0.08 &   0.02\p0.08 & 0.09\p0.08 &   0.07\p0.07 & --0.14\p0.13            & --0.08\p0.08 & --0.07\p0.08 &   0.05\p0.08 & --0.05\p0.05 &   0.37\p0.11\\
HD 42911       &   0.01\p0.11 &   0.07\p0.09 & 0.17\p0.11 &   0.02\p0.08 & --0.05\p0.14            & --0.04\p0.11 & --0.03\p0.09 &   0.12\p0.08 &   0.02\p0.09 &   0.06\p0.12\\
HD 43023       &   0.04\p0.09 &   0.06\p0.07 & 0.08\p0.08 &   0.03\p0.06 & --0.05\p0.14            &   0.00\p0.09 &   0.04\p0.08 &   0.02\p0.08 & --0.04\p0.06 &   0.23\p0.12\\
HD 45398       &   0.22\p0.12 &   0.20\p0.16 & 0.24\p0.07 &   0.08\p0.11 & --0.04\p0.12            & --0.03\p0.15 &   0.02\p0.09 &   0.25\p0.08 &   0.05\p0.07 &   0.18\p0.09\\
HD 49429       &   0.02\p0.10 &   0.06\p0.07 & 0.08\p0.08 &   0.06\p0.05 & --0.06\p0.12            & --0.07\p0.09 & --0.02\p0.08 & --0.02\p0.08 & --0.05\p0.05 &   0.27\p0.11\\
HD 49566       &   0.06\p0.09 &   0.04\p0.06 & 0.06\p0.08 &   0.07\p0.05 & --0.07\p0.12            & --0.04\p0.08 & --0.05\p0.08 &   0.02\p0.08 & --0.05\p0.06 &   0.32\p0.11\\
               &   0.12\p0.08 &   0.09\p0.06 & 0.09\p0.06 &   0.12\p0.05 & --0.09\p0.08            &   0.01\p0.08 &   0.00\p0.08 &   0.06\p0.08 & --0.02\p0.06 &   0.31\p0.10\\
HD 50890       &   0.21\p0.10 &   0.30\p0.09 & 0.28\p0.13 &   0.06\p0.09 & \multicolumn{1}{c}{...} &   0.05\p0.10 & --0.09\p0.08 &   0.13\p0.09 & --0.03\p0.09 &   0.23\p0.12\\
               &   0.10\p0.08 &   0.20\p0.09 & 0.25\p0.13 & --0.04\p0.09 & \multicolumn{1}{c}{...} & --0.05\p0.10 & --0.19\p0.08 &   0.07\p0.08 & --0.08\p0.09 &   0.23\p0.10\\
HD 169370      &   0.23\p0.13 &   0.26\p0.11 & 0.16\p0.07 &   0.09\p0.09 &   0.05\p0.12            &   0.05\p0.13 & --0.06\p0.09 &   0.22\p0.08 &   0.00\p0.07 & --0.08\p0.11\\
               &   0.22\p0.08 &   0.25\p0.09 & 0.15\p0.05 &   0.08\p0.06 &   0.05\p0.08            &   0.04\p0.10 & --0.07\p0.07 &   0.22\p0.08 & --0.01\p0.07 & --0.09\p0.09\\
HD 169751      &   0.02\p0.10 &   0.07\p0.07 & 0.12\p0.10 &   0.05\p0.07 & --0.10\p0.14            & --0.05\p0.10 & --0.06\p0.06 &   0.04\p0.08 & --0.03\p0.08 &   0.18\p0.12\\
               &   0.05\p0.08 &   0.09\p0.05 & 0.13\p0.09 &   0.08\p0.06 & --0.10\p0.09            & --0.02\p0.09 & --0.03\p0.05 &   0.05\p0.08 & --0.01\p0.08 &   0.18\p0.10\\
HD 170008      &   0.09\p0.11 &   0.15\p0.06 & 0.09\p0.08 &   0.11\p0.05 & --0.03\p0.11            &   0.01\p0.09 & --0.08\p0.10 &   0.05\p0.08 & --0.01\p0.05 &   0.10\p0.13\\
               &   0.08\p0.08 &   0.14\p0.06 & 0.08\p0.06 &   0.10\p0.05 & --0.03\p0.08            &   0.00\p0.09 & --0.09\p0.08 &   0.04\p0.08 & --0.01\p0.05 &   0.10\p0.10\\
HD 170031      &   0.14\p0.15 &   0.17\p0.13 & 0.14\p0.05 &   0.07\p0.11 &   0.00\p0.11            & --0.02\p0.16 & --0.05\p0.09 &   0.24\p0.08 &   0.01\p0.08 &   0.04\p0.12\\
               &   0.08\p0.09 &   0.11\p0.10 & 0.12\p0.05 &   0.02\p0.08 & --0.02\p0.08            & --0.09\p0.13 & --0.11\p0.07 &   0.21\p0.08 & --0.02\p0.08 &   0.03\p0.09\\
HD 171427      &   0.05\p0.08 &   0.19\p0.05 & 0.20\p0.12 &   0.01\p0.08 & --0.03\p0.16            &   0.02\p0.09 & --0.09\p0.07 &   0.03\p0.09 & --0.08\p0.06 &   0.65\p0.12\\
HD 175294      &   0.05\p0.11 &   0.09\p0.11 & 0.11\p0.10 & --0.03\p0.10 & --0.01\p0.14            & --0.01\p0.12 & --0.09\p0.08 &   0.12\p0.08 &   0.02\p0.08 &   0.05\p0.14\\
HD 175679      &   0.09\p0.09 &   0.04\p0.08 & 0.05\p0.10 &   0.01\p0.06 & --0.07\p0.13            & --0.09\p0.09 & --0.04\p0.15 &   0.03\p0.08 & --0.01\p0.05 &   0.28\p0.11\\
               &   0.23\p0.08 &   0.16\p0.07 & 0.11\p0.10 &   0.12\p0.05 & --0.11\p0.10            &   0.03\p0.08 &   0.07\p0.15 &   0.11\p0.08 &   0.05\p0.05 &   0.25\p0.10\\
HD 178484      &   0.26\p0.12 &   0.26\p0.07 & 0.25\p0.11 &   0.10\p0.08 &   0.01\p0.13            &   0.01\p0.12 & --0.06\p0.08 &   0.20\p0.08 &   0.01\p0.07 &   0.12\p0.09\\
               &   0.22\p0.09 &   0.22\p0.07 & 0.24\p0.11 &   0.06\p0.07 &   0.01\p0.09            & --0.03\p0.12 & --0.10\p0.08 &   0.18\p0.08 &   0.00\p0.07 &   0.11\p0.09\\
HD 181907      &   0.17\p0.12 &   0.12\p0.10 & 0.19\p0.11 &   0.00\p0.09 & --0.02\p0.14            & --0.04\p0.13 & --0.09\p0.10 &   0.11\p0.08 &   0.04\p0.07 &   0.14\p0.12\\
               &   0.22\p0.08 &   0.18\p0.08 & 0.20\p0.10 &   0.05\p0.05 & --0.03\p0.08            &   0.02\p0.09 & --0.04\p0.08 &   0.14\p0.07 &   0.06\p0.06 &   0.14\p0.10\\
HD 170053      &   0.18\p0.11 &   0.11\p0.12 & 0.21\p0.07 &   0.01\p0.11 & --0.05\p0.12            & --0.08\p0.13 & --0.09\p0.09 &   0.18\p0.08 & --0.05\p0.10 &   0.32\p0.10\\
               &   0.08\p0.09 & --0.01\p0.12 & 0.18\p0.07 & --0.13\p0.11 & --0.06\p0.08            & --0.21\p0.13 & --0.22\p0.09 &   0.13\p0.08 & --0.10\p0.10 &   0.30\p0.10\\
HD 170174      &   0.08\p0.09 &   0.07\p0.06 & 0.12\p0.09 &   0.06\p0.07 & --0.04\p0.13            & --0.01\p0.09 & --0.05\p0.06 &   0.03\p0.08 & --0.05\p0.06 &   0.28\p0.11\\
               &   0.16\p0.07 &   0.15\p0.05 & 0.16\p0.09 &   0.13\p0.06 & --0.07\p0.09            &   0.07\p0.08 &   0.02\p0.06 &   0.08\p0.08 &   0.00\p0.06 &   0.27\p0.11\\
HD 170231      &   0.01\p0.10 &   0.04\p0.07 & 0.06\p0.09 &   0.07\p0.07 & --0.04\p0.12            &   0.01\p0.09 & --0.05\p0.05 &   0.03\p0.08 & --0.06\p0.05 &   0.38\p0.11\\
               &   0.12\p0.08 &   0.12\p0.06 & 0.11\p0.08 &   0.15\p0.05 & --0.08\p0.09            &   0.10\p0.08 &   0.04\p0.05 &   0.10\p0.08 &   0.00\p0.05 &   0.36\p0.11\\
$\alpha$ Boo   &   0.58\p0.13 &   0.45\p0.09 & 0.42\p0.11 &   0.19\p0.08 &   0.12\p0.11            &   0.22\p0.12 & --0.02\p0.09 &   0.31\p0.08 &   0.04\p0.07 & --0.23\p0.10\\
               &   0.60\p0.08 &   0.47\p0.08 & 0.42\p0.11 &   0.22\p0.07 &   0.12\p0.08            &   0.24\p0.12 &   0.00\p0.09 &   0.32\p0.08 &   0.05\p0.07 & --0.22\p0.09\\
$\eta$ Ser     &   0.09\p0.11 &   0.12\p0.08 & 0.09\p0.08 &   0.08\p0.05 &   0.00\p0.13            & --0.01\p0.10 & --0.05\p0.06 &   0.10\p0.08 &   0.00\p0.06 &   0.08\p0.14\\
               &   0.14\p0.08 &   0.17\p0.06 & 0.10\p0.07 &   0.13\p0.05 &   0.00\p0.08            &   0.04\p0.09 & --0.01\p0.05 &   0.12\p0.08 &   0.02\p0.06 &   0.07\p0.09\\
$\epsilon$ Oph &   0.04\p0.10 &   0.05\p0.08 & 0.11\p0.09 &   0.03\p0.06 & --0.05\p0.15            &   0.00\p0.09 & --0.02\p0.09 &   0.05\p0.08 & --0.03\p0.06 &   0.24\p0.12\\
               &   0.05\p0.08 &   0.06\p0.08 & 0.11\p0.07 &   0.05\p0.05 & --0.05\p0.09            &   0.01\p0.09 & --0.01\p0.08 &   0.06\p0.08 & --0.03\p0.05 &   0.24\p0.10\\
$\xi$ Hya      & --0.02\p0.10 &   0.04\p0.06 & 0.10\p0.08 &   0.04\p0.06 & --0.08\p0.12            & --0.05\p0.09 & --0.04\p0.05 &   0.06\p0.08 & --0.02\p0.09 &   0.19\p0.11\\
               &   0.03\p0.08 &   0.08\p0.05 & 0.12\p0.07 &   0.08\p0.06 & --0.10\p0.08            & --0.01\p0.09 &   0.00\p0.05 &   0.08\p0.08 &   0.00\p0.09 &   0.18\p0.09\\
$\beta$ Aql    &   0.06\p0.13 &   0.10\p0.08 & 0.06\p0.08 &   0.10\p0.07 & --0.03\p0.11            & --0.02\p0.10 & --0.05\p0.06 &   0.04\p0.08 &   0.01\p0.05 &   0.08\p0.15\\
               &   0.08\p0.08 &   0.12\p0.06 & 0.06\p0.06 &   0.11\p0.05 & --0.04\p0.08            &   0.00\p0.09 & --0.04\p0.05 &   0.05\p0.07 &   0.01\p0.05 &   0.07\p0.10\\
\hline
\end{tabular}
\tablefoot{When available, the second row shows the results with the surface gravity fixed to the seismic value for each star. We use the usual notation [X/Fe] = $[\log \epsilon($X$)-\log \epsilon($Fe$)]-[\log \epsilon($X$)-\log \epsilon($Fe$)]_{\sun}$ with $\log \epsilon$(X) = 12+$\log [{\cal N}($X$)/{\cal N}($H$)]$ (${\cal N}$ is the number density of the species). The adopted solar abundances ($\log \epsilon_{\sun}$) are given in the first row (Grevesse \& Sauval \cite{grevesse_sauval98}).}
\end{table*}}

\onltab{5}{
\begin{table*}
\centering
\caption{Abundance data with 1-$\sigma$ uncertainties for the mixing indicators.}
\scriptsize
\label{tab_abundances_mixing}
\begin{tabular}{lccrcrrrrrrc} 
\hline\hline
              & \multicolumn{2}{c}{[Li/H]} & \multicolumn{1}{c}{[C/Fe]} & \multicolumn{1}{c}{[C/Fe]$_{\rm corr}$} & \multicolumn{1}{c}{[N/Fe]} & \multicolumn{1}{c}{[\ion{O}{i} 6300/Fe]} & \multicolumn{1}{c}{[\ion{O}{i} 6300/Fe]$_{\rm corr}$} & \multicolumn{1}{c}{[O triplet/Fe]} & \multicolumn{1}{c}{[Na/Fe]}    & \multicolumn{1}{c}{[Na/Fe]$_{\rm corr}$} & \iso \\
&  \multicolumn{1}{c}{LTE} & \multicolumn{1}{c}{NLTE} &  \multicolumn{1}{c}{LTE} & \multicolumn{1}{c}{LTE} & \multicolumn{1}{c}{LTE} & \multicolumn{1}{c}{LTE} & \multicolumn{1}{c}{LTE} & \multicolumn{1}{c}{LTE} &  \multicolumn{1}{c}{LTE} & \multicolumn{1}{c}{LTE} & \multicolumn{1}{c}{LTE}\\
\hline
Sun            &     1.09   &     1.13           &  8.52     &   8.52   &  7.92      &    8.83    &     8.83 &    \multicolumn{1}{c}{8.83}    &  \multicolumn{1}{c}{6.33}    &  \multicolumn{1}{c}{6.33}  &               \\
\hline
HD 40726         &  +0.34\p0.13 &  +0.43\p0.13 & --0.30 & --0.28 &   0.45 & --0.15 & --0.12 & \multicolumn{1}{c}{...} & 0.25\p0.05 &   0.26\p0.05 & ...  \\
HD 42911         & $<$--1.15    & $<$--0.99    & --0.21 & --0.16 &   0.36 & --0.13 & --0.07 & \multicolumn{1}{c}{...} & 0.18\p0.08 &   0.20\p0.08 & ...  \\
HD 43023         & $<$--1.40    & $<$--1.29    & --0.23 & --0.25 &   0.39 & --0.02 & --0.04 & \multicolumn{1}{c}{...} & 0.10\p0.06 &   0.09\p0.06 & ...  \\
HD 45398         & --0.82\p0.13 & --0.50\p0.13 & --0.11 & --0.17 &   0.16 & --0.03 & --0.10 & \multicolumn{1}{c}{...} & 0.24\p0.14 &   0.22\p0.14 & ...  \\
HD 49429         & $<$--1.00    & $<$--0.90    & --0.23 & --0.25 &   0.32 & --0.02 & --0.05 & \multicolumn{1}{c}{...} & 0.12\p0.06 &   0.11\p0.06 & ...  \\
HD 49566         & --0.27\p0.13 & --0.18\p0.13 & --0.18 & --0.20 &   0.38 & --0.04 & --0.06 & \multicolumn{1}{c}{...} & 0.14\p0.06 &   0.13\p0.06 & ...  \\
                 & --0.25\p0.12 & --0.16\p0.12 & --0.15 & --0.18 &   0.42 & --0.05 & --0.09 & \multicolumn{1}{c}{...} & 0.19\p0.05 &   0.18\p0.05 & ...  \\
HD 50890         & --0.07\p0.16 &  +0.15\p0.16 & --0.40 & --0.41 &   0.58 & --0.24 & --0.25 & \multicolumn{1}{c}{...} & 0.51\p0.11 &   0.51\p0.11 & ...  \\
                 & --0.13\p0.14 &  +0.09\p0.14 & --0.43 & --0.41 &   0.57 & --0.20 & --0.17 & \multicolumn{1}{c}{...} & 0.41\p0.11 &   0.42\p0.11 & ...  \\
HD 169370        & --1.41\p0.13 & --1.19\p0.13 & --0.03 & --0.14 &   0.06 &   0.08 & --0.05 & \multicolumn{1}{c}{...} & 0.03\p0.12 & --0.01\p0.12 & ...  \\
                 & --1.41\p0.12 & --1.19\p0.12 & --0.04 & --0.14 &   0.05 &   0.07 & --0.06 & \multicolumn{1}{c}{...} & 0.02\p0.10 & --0.01\p0.10 & ...  \\
HD 169751        & --0.52\p0.13 & --0.38\p0.13 & --0.26 & --0.26 &   0.29 & --0.10 & --0.10 & \multicolumn{1}{c}{...} & 0.08\p0.09 &   0.08\p0.09 & ...  \\
                 & --0.50\p0.12 & --0.36\p0.12 & --0.25 & --0.26 &   0.31 & --0.10 & --0.11 & \multicolumn{1}{c}{...} & 0.11\p0.08 &   0.11\p0.08 & ...  \\
HD 170008        & $<$--0.90    & $<$--0.83    & --0.03 & --0.17 &   0.02 &   0.03 & --0.14 & \multicolumn{1}{c}{...} & 0.03\p0.07 & --0.02\p0.07 & ...  \\
                 & $<$--0.90    & $<$--0.83    & --0.03 & --0.16 &   0.01 &   0.03 & --0.14 & \multicolumn{1}{c}{...} & 0.02\p0.07 & --0.02\p0.07 & ...  \\
HD 170031        & --1.39\p0.13 & --1.15\p0.13 & --0.11 & --0.11 &   0.24 & --0.03 & --0.03 & \multicolumn{1}{c}{...} & 0.17\p0.15 &   0.17\p0.15 & ...  \\
                 & --1.42\p0.12 & --1.18\p0.12 & --0.13 & --0.11 &   0.20 & --0.05 & --0.03 & \multicolumn{1}{c}{...} & 0.11\p0.12 &   0.12\p0.12 & ...  \\
HD 171427        & --0.38\p0.13 & --0.20\p0.13 & --0.22 & --0.23 &   0.59 & --0.06 & --0.07 & \multicolumn{1}{c}{...} & 0.33\p0.05 &   0.33\p0.05 & ...  \\
HD 175294        & $<$--1.20    & $<$--1.03    & --0.25 & --0.15 &   0.43 & --0.12 &   0.00 & 0.14\p0.26              & 0.26\p0.08 &   0.29\p0.08 & 16\p2\\
HD 175679        &  +0.15\p0.13 &  +0.26\p0.13 & --0.24 & --0.20 &   0.46 & --0.05 &   0.00 & 0.05\p0.15              & 0.17\p0.05 &   0.18\p0.05 & 17\p5\\
                 &  +0.20\p0.12 &  +0.31\p0.12 & --0.19 & --0.18 &   0.56 & --0.09 & --0.08 & --0.02\p0.14            & 0.29\p0.05 &   0.29\p0.05 & 17\p5\\
HD 178484        & --0.90\p0.13 & --0.65\p0.13 &   0.00 & --0.12 &   0.17 &   0.08 & --0.08 & \multicolumn{1}{c}{...} & 0.23\p0.10 &   0.19\p0.10 & ...  \\ 
                 & --0.92\p0.12 & --0.67\p0.12 & --0.02 & --0.13 &   0.17 &   0.07 & --0.07 & \multicolumn{1}{c}{...} & 0.19\p0.10 &   0.15\p0.10 & ...  \\ 
HD 181907        & $<$--1.30    & $<$--1.11    & --0.14 & --0.18 &   0.26 &   0.05 &   0.00 & 0.29\p0.27              & 0.12\p0.09 &   0.11\p0.09 & 9\p1 \\
                 & $<$--1.25    & $<$--1.06    & --0.13 & --0.19 &   0.28 &   0.05 & --0.02 & 0.25\p0.23              & 0.17\p0.05 &   0.15\p0.05 & 9\p1 \\
HD 170053        &  +0.11\p0.13 &  +0.36\p0.13 & --0.15 & --0.20 &   0.42 & --0.04 & --0.11 & \multicolumn{1}{c}{...} & 0.27\p0.13 &   0.25\p0.13 &  ... \\
                 &  +0.06\p0.12 &  +0.31\p0.12 & --0.18 & --0.19 &   0.43 & --0.06 & --0.07 & \multicolumn{1}{c}{...} & 0.13\p0.13 &   0.13\p0.13 &  ... \\
HD 170174        & --0.48\p0.13 & --0.36\p0.13 & --0.20 & --0.20 &   0.46 & --0.03 & --0.03 & \multicolumn{1}{c}{...} & 0.20\p0.06 &   0.20\p0.06 &  ... \\
                 & --0.45\p0.12 & --0.33\p0.12 & --0.17 & --0.20 &   0.52 & --0.05 & --0.08 & \multicolumn{1}{c}{...} & 0.28\p0.06 &   0.27\p0.06 &  ... \\
HD 170231        &  +0.24\p0.13 &  +0.34\p0.13 & --0.27 & --0.25 &   0.46 & --0.09 & --0.07 & \multicolumn{1}{c}{...} & 0.16\p0.07 &   0.17\p0.07 &  ... \\ 
                 &  +0.26\p0.12 &  +0.36\p0.12 & --0.23 & --0.24 &   0.54 & --0.12 & --0.13 & \multicolumn{1}{c}{...} & 0.25\p0.06 &   0.25\p0.06 &  ... \\ 
$\alpha$ Boo     & $<$--2.50    & $<$--2.24    &   0.18 & --0.21 &   0.22 &   0.43 & --0.02 & 0.69\p0.20              & 0.19\p0.10 &   0.07\p0.10 & 8\p1 \\
                 & $<$--2.45    & $<$--2.19    &   0.19 & --0.21 &   0.22 &   0.44 & --0.02 & 0.68\p0.20              & 0.21\p0.09 &   0.09\p0.09 & 8\p1 \\
$\eta$ Ser       & $<$--2.00    & $<$--1.88    & --0.08 & --0.16 &   0.25 &   0.06 & --0.04 & \multicolumn{1}{c}{...} & 0.05\p0.09 &   0.02\p0.09 & ...  \\
                 & $<$--1.90    & $<$--1.78    & --0.07 & --0.16 &   0.28 &   0.06 & --0.06 & \multicolumn{1}{c}{...} & 0.10\p0.08 &   0.07\p0.08 &  ... \\
$\epsilon$ Oph   & $<$--0.80    & $<$--0.66    & --0.27 & --0.28 &   0.30 & --0.08 & --0.09 & \multicolumn{1}{c}{...} & 0.09\p0.06 &   0.09\p0.06 &  ... \\
                 & $<$--0.80    & $<$--0.66    & --0.26 & --0.28 &   0.31 & --0.08 & --0.10 & \multicolumn{1}{c}{...} & 0.10\p0.05 &   0.09\p0.05 &  ... \\
$\xi$ Hya        &  +0.07\p0.13 &  +0.19\p0.13 & --0.25 & --0.20 &   0.43 & --0.14 & --0.07 & \multicolumn{1}{c}{...} & 0.24\p0.08 &   0.26\p0.08 &  ... \\
                 &  +0.09\p0.12 &  +0.21\p0.12 & --0.24 & --0.20 &   0.47 & --0.15 & --0.10 & \multicolumn{1}{c}{...} & 0.28\p0.07 &   0.29\p0.07 &  ... \\
$\beta$ Aql      & $<$--1.20    & $<$--1.12    &   0.02 & --0.06 & --0.13 &   0.01 & --0.09 & \multicolumn{1}{c}{...} & 0.03\p0.09 &   0.00\p0.09 &  ... \\
                 & $<$--1.15    & $<$--1.07    &   0.02 & --0.07 & --0.12 &   0.00 & --0.11 & \multicolumn{1}{c}{...} & 0.04\p0.06 &   0.01\p0.06 &  ... \\
\hline
\end{tabular}
\tablefoot{When available, the second row shows the results with the surface gravity fixed to the seismic value for each star. We use the usual notation [X/Fe] = $[\log \epsilon($X$)-\log \epsilon($Fe$)]-[\log \epsilon($X$)-\log \epsilon($Fe$)]_{\sun}$ with $\log \epsilon$(X) = 12+$\log [{\cal N}($X$)/{\cal N}($H$)]$ (${\cal N}$ is the number density of the species). The adopted solar abundances ($\log \epsilon_{\sun}$) are given in the first row and are taken from Grevesse \& Sauval (\cite{grevesse_sauval98}), except for Li for which we adopt $\log \epsilon({\rm Li})_{\sun}$ = 1.09 and 1.13 in LTE and NLTE, respectively (Sect.~\ref{spectral_synthesis_atomic_data}). For lithium, the following notation is used: [Li/H] = $\log {\cal N}({\rm Li})-\log {\cal N}({\rm Li})_{\sun}$. Lithium abundances corrected for departures from LTE (Lind \etal \cite{lind09}) are also provided. The values with the ``corr'' subscript were corrected for the chemical evolution of the Galaxy (Sect.~\ref{sect_correction_chemical_evolution}). The 1-$\sigma$ uncertainties for the C, N, and the O abundances derived from \ion{[O}{i]} $\lambda$6300 are 0.09, 0.13, and 0.14 dex, respectively. With the gravity fixed to the seismic value, this translates to 0.09, 0.13, and 0.10 dex. Due to the difficulty in fitting the CNO features in HD 50890, the uncertainties for the C, N, and O abundances are 0.12, 0.14, and 0.16 dex (0.12, 0.14, and 0.12 dex for the gravity fixed to the seismic value).}
\end{table*}}

\onltab{6}{
\begin{table*}
\centering
\caption{Logarithmic abundance ratios of C, N, and O.}
\label{tab_abundances_ratios}
\begin{tabular}{lrrrrrr} 
\hline\hline
              & \multicolumn{1}{c}{[N/C]}  & \multicolumn{1}{c}{[N/C]$_{\rm corr}$} & \multicolumn{1}{c}{[N/O]}  & \multicolumn{1}{c}{[N/O]$_{\rm corr}$} & \multicolumn{1}{c}{[C/O]}  & \multicolumn{1}{c}{[C/O]$_{\rm corr}$}\\
\hline						      			    	          
\object{HD 40726}        &   0.75 &   0.73 &   0.60 &   0.57 & --0.15 & --0.16\\
\object{HD 42911}        &   0.57 &   0.52 &   0.49 &   0.43 & --0.08 & --0.09\\
\object{HD 43023}        &   0.62 &   0.64 &   0.41 &   0.43 & --0.21 & --0.20\\
\object{HD 45398}        &   0.27 &   0.33 &   0.19 &   0.26 & --0.08 & --0.06\\ 
\object{HD 49429}        &   0.55 &   0.57 &   0.34 &   0.37 & --0.21 & --0.20\\
\object{HD 49566}        &   0.56 &   0.58 &   0.42 &   0.44 & --0.14 & --0.14\\
                         &   0.57 &   0.60 &   0.47 &   0.51 & --0.10 & --0.09\\
\object{HD 50890}        &   0.98 &   0.99 &   0.82 &   0.83 & --0.16 & --0.16\\
                         &   1.00 &   0.98 &   0.77 &   0.74 & --0.23 & --0.24\\
\object{HD 169370}       &   0.09 &   0.20 & --0.02 &   0.11 & --0.11 & --0.08\\
                         &   0.09 &   0.19 & --0.02 &   0.11 & --0.11 & --0.08\\
\object{HD 169751}       &   0.55 &   0.55 &   0.39 &   0.39 & --0.16 & --0.16\\
                         &   0.56 &   0.57 &   0.41 &   0.42 & --0.15 & --0.15\\
\object{HD 170008}       &   0.05 &   0.19 & --0.01 &   0.16 & --0.06 & --0.02\\
                         &   0.04 &   0.17 & --0.02 &   0.15 & --0.06 & --0.02\\
\object{HD 170031}       &   0.35 &   0.35 &   0.27 &   0.27 & --0.08 & --0.08\\
                         &   0.33 &   0.31 &   0.25 &   0.23 & --0.08 & --0.08\\
\object{HD 171427}       &   0.81 &   0.82 &   0.65 &   0.66 & --0.16 & --0.16\\
\object{HD 175294}       &   0.68 &   0.58 &   0.55 &   0.42 & --0.13 & --0.16\\
\object{HD 175679}       &   0.70 &   0.66 &   0.51 &   0.45 & --0.19 & --0.20\\
                         &   0.75 &   0.74 &   0.65 &   0.64 & --0.10 & --0.10\\
\object{HD 178484}       &   0.17 &   0.29 &   0.09 &   0.25 & --0.08 & --0.04\\
                         &   0.19 &   0.30 &   0.10 &   0.24 & --0.09 & --0.06\\
\object{HD 181907}       &   0.40 &   0.44 &   0.21 &   0.26 & --0.19 & --0.18\\
                         &   0.41 &   0.47 &   0.23 &   0.30 & --0.18 & --0.16\\
\object{HD 170053}       &   0.57 &   0.62 &   0.46 &   0.53 & --0.11 & --0.09\\
                         &   0.61 &   0.62 &   0.49 &   0.50 & --0.12 & --0.12\\
\object{HD 170174}       &   0.66 &   0.66 &   0.49 &   0.49 & --0.17 & --0.17\\
                         &   0.69 &   0.72 &   0.57 &   0.60 & --0.12 & --0.11\\
\object{HD 170231}       &   0.73 &   0.71 &   0.55 &   0.53 & --0.18 & --0.18\\ 
                         &   0.77 &   0.78 &   0.66 &   0.67 & --0.11 & --0.11\\ 
\object{$\alpha$ Boo}    &   0.04 &   0.43 & --0.21 &   0.24 & --0.25 & --0.19\\
                         &   0.03 &   0.43 & --0.22 &   0.24 & --0.25 & --0.18\\
\object{$\eta$ Ser}      &   0.33 &   0.41 &   0.19 &   0.29 & --0.14 & --0.12\\
                         &   0.35 &   0.44 &   0.22 &   0.34 & --0.13 & --0.10\\
\object{$\epsilon$ Oph}  &   0.57 &   0.58 &   0.38 &   0.39 & --0.19 & --0.19\\
                         &   0.57 &   0.59 &   0.39 &   0.41 & --0.18 & --0.18\\
\object{$\xi$ Hya}       &   0.68 &   0.63 &   0.57 &   0.50 & --0.11 & --0.12\\
                         &   0.71 &   0.67 &   0.62 &   0.57 & --0.09 & --0.10\\
\object{$\beta$ Aql}     & --0.15 & --0.07 & --0.14 & --0.03 &   0.01 &   0.03\\
                         & --0.14 & --0.05 & --0.12 & --0.01 &   0.02 &   0.04\\
\hline    
\end{tabular}
\tablefoot{When available, the second row shows the results with the surface gravity fixed to the seismic value for each star. The values with the ``corr'' subscript were corrected for the chemical evolution of the Galaxy (Sect.~\ref{sect_correction_chemical_evolution}). The 1-$\sigma$ uncertainties for [N/C], [N/O], and [C/O] are 0.10, 0.15, and 0.12 dex, respectively. With the surface gravity fixed to the seismic value, this translates to 0.08, 0.13, and 0.09 dex. Due to the difficulty in fitting the CNO features in HD 50890, the uncertainties for [N/C], [N/O], and [C/O] are 0.13, 0.17, and 0.14 dex (0.11, 0.15, and 0.12 dex for the gravity fixed to the seismic value).}
\end{table*}}

\end{document}